\documentclass[12pt]{article}

    \usepackage{times}

    \usepackage[breakable]{tcolorbox}
    \usepackage{parskip} 

    \usepackage{graphicx}
    
    \usepackage{float}
    \usepackage{xcolor} 
    \usepackage{enumerate} 
    \usepackage{geometry} 
    \usepackage{amsmath} 
    \usepackage{amssymb} 
    \usepackage{textcomp} 
    \AtBeginDocument{%
        \def\PYZsq{\textquotesingle}
    }
    \usepackage{upquote} 
    \usepackage{eurosym} 

    \usepackage{iftex}
    \ifPDFTeX
        \usepackage[T1]{fontenc}
    \else
        \usepackage{fontspec}
        \usepackage{unicode-math}
    \fi

    \usepackage{fancyvrb} 
    \usepackage{grffile} 
    \makeatletter 
    \@ifpackagelater{grffile}{2019/11/01}
    {
    }
    {
      \def\Gread@@xetex#1{%
        \IfFileExists{"\Gin@base".bb}%
        {\Gread@eps{\Gin@base.bb}}%
        {\Gread@@xetex@aux#1}%
      }
    }
    \makeatother
    \usepackage[Export]{adjustbox} 
    \adjustboxset{max size={0.9\linewidth}{0.9\paperheight}}

    \usepackage{hyperref}
    \usepackage{titling}
    \usepackage{longtable} 
    \usepackage{booktabs}  
    \usepackage{array}     
    \usepackage{calc}      
    \usepackage[inline]{enumitem} 
    \usepackage[normalem]{ulem} 
    \usepackage{mathrsfs}

    \definecolor{urlcolor}{rgb}{0,.145,.698}
    \definecolor{linkcolor}{rgb}{.71,0.21,0.01}
    \definecolor{citecolor}{rgb}{.12,.54,.11}

    \definecolor{ansi-black}{HTML}{3E424D}
    \definecolor{ansi-black-intense}{HTML}{282C36}
    \definecolor{ansi-red}{HTML}{E75C58}
    \definecolor{ansi-red-intense}{HTML}{B22B31}
    \definecolor{ansi-green}{HTML}{00A250}
    \definecolor{ansi-green-intense}{HTML}{007427}
    \definecolor{ansi-yellow}{HTML}{DDB62B}
    \definecolor{ansi-yellow-intense}{HTML}{B27D12}
    \definecolor{ansi-blue}{HTML}{208FFB}
    \definecolor{ansi-blue-intense}{HTML}{0065CA}
    \definecolor{ansi-magenta}{HTML}{D160C4}
    \definecolor{ansi-magenta-intense}{HTML}{A03196}
    \definecolor{ansi-cyan}{HTML}{60C6C8}
    \definecolor{ansi-cyan-intense}{HTML}{258F8F}
    \definecolor{ansi-white}{HTML}{C5C1B4}
    \definecolor{ansi-white-intense}{HTML}{A1A6B2}
    \definecolor{ansi-default-inverse-fg}{HTML}{FFFFFF}
    \definecolor{ansi-default-inverse-bg}{HTML}{000000}

    \definecolor{outerrorbackground}{HTML}{FFDFDF}

    \DefineVerbatimEnvironment{Highlighting}{Verbatim}{commandchars=\\\{\}}

    \let\Oldtex\TeX
    \let\Oldlatex\LaTeX
    \renewcommand{\TeX}{\textrm{\Oldtex}}
    \renewcommand{\LaTeX}{\textrm{\Oldlatex}}
    \title{How To Program Your Own Quantum Computer \\
    or\\
    QUBE: QUantum computing for BEginners
}    
\author{Martin N. P. Nilsson, \\
	RISE Research Institutes of Sweden, 
	POB 1263, SE-164 29 Kista, Sweden\\
Email: {\tt martin.nilsson@ri.se} \\
ORCiD: 0000-0002-7504-0328}

\makeatletter
\def\PY@reset{\let\PY@it=\relax \let\PY@bf=\relax%
    \let\PY@ul=\relax \let\PY@tc=\relax%
    \let\PY@bc=\relax \let\PY@ff=\relax}
\def\PY@tok#1{\csname PY@tok@#1\endcsname}
\def\PY@toks#1+{\ifx\relax#1\empty\else%
    \PY@tok{#1}\expandafter\PY@toks\fi}
\def\PY@do#1{\PY@bc{\PY@tc{\PY@ul{%
    \PY@it{\PY@bf{\PY@ff{#1}}}}}}}
\def\PY#1#2{\PY@reset\PY@toks#1+\relax+\PY@do{#2}}

\@namedef{PY@tok@w}{\def\PY@tc##1{\textcolor[rgb]{0.73,0.73,0.73}{##1}}}
\@namedef{PY@tok@c}{\let\PY@it=\textit\def\PY@tc##1{\textcolor[rgb]{0.24,0.48,0.48}{##1}}}
\@namedef{PY@tok@cp}{\def\PY@tc##1{\textcolor[rgb]{0.61,0.40,0.00}{##1}}}
\@namedef{PY@tok@k}{\let\PY@bf=\textbf\def\PY@tc##1{\textcolor[rgb]{0.00,0.50,0.00}{##1}}}
\@namedef{PY@tok@kp}{\def\PY@tc##1{\textcolor[rgb]{0.00,0.50,0.00}{##1}}}
\@namedef{PY@tok@kt}{\def\PY@tc##1{\textcolor[rgb]{0.69,0.00,0.25}{##1}}}
\@namedef{PY@tok@o}{\def\PY@tc##1{\textcolor[rgb]{0.40,0.40,0.40}{##1}}}
\@namedef{PY@tok@ow}{\let\PY@bf=\textbf\def\PY@tc##1{\textcolor[rgb]{0.67,0.13,1.00}{##1}}}
\@namedef{PY@tok@nb}{\def\PY@tc##1{\textcolor[rgb]{0.00,0.50,0.00}{##1}}}
\@namedef{PY@tok@nf}{\def\PY@tc##1{\textcolor[rgb]{0.00,0.00,1.00}{##1}}}
\@namedef{PY@tok@nc}{\let\PY@bf=\textbf\def\PY@tc##1{\textcolor[rgb]{0.00,0.00,1.00}{##1}}}
\@namedef{PY@tok@nn}{\let\PY@bf=\textbf\def\PY@tc##1{\textcolor[rgb]{0.00,0.00,1.00}{##1}}}
\@namedef{PY@tok@ne}{\let\PY@bf=\textbf\def\PY@tc##1{\textcolor[rgb]{0.80,0.25,0.22}{##1}}}
\@namedef{PY@tok@nv}{\def\PY@tc##1{\textcolor[rgb]{0.10,0.09,0.49}{##1}}}
\@namedef{PY@tok@no}{\def\PY@tc##1{\textcolor[rgb]{0.53,0.00,0.00}{##1}}}
\@namedef{PY@tok@nl}{\def\PY@tc##1{\textcolor[rgb]{0.46,0.46,0.00}{##1}}}
\@namedef{PY@tok@ni}{\let\PY@bf=\textbf\def\PY@tc##1{\textcolor[rgb]{0.44,0.44,0.44}{##1}}}
\@namedef{PY@tok@na}{\def\PY@tc##1{\textcolor[rgb]{0.41,0.47,0.13}{##1}}}
\@namedef{PY@tok@nt}{\let\PY@bf=\textbf\def\PY@tc##1{\textcolor[rgb]{0.00,0.50,0.00}{##1}}}
\@namedef{PY@tok@nd}{\def\PY@tc##1{\textcolor[rgb]{0.67,0.13,1.00}{##1}}}
\@namedef{PY@tok@s}{\def\PY@tc##1{\textcolor[rgb]{0.73,0.13,0.13}{##1}}}
\@namedef{PY@tok@sd}{\let\PY@it=\textit\def\PY@tc##1{\textcolor[rgb]{0.73,0.13,0.13}{##1}}}
\@namedef{PY@tok@si}{\let\PY@bf=\textbf\def\PY@tc##1{\textcolor[rgb]{0.64,0.35,0.47}{##1}}}
\@namedef{PY@tok@se}{\let\PY@bf=\textbf\def\PY@tc##1{\textcolor[rgb]{0.67,0.36,0.12}{##1}}}
\@namedef{PY@tok@sr}{\def\PY@tc##1{\textcolor[rgb]{0.64,0.35,0.47}{##1}}}
\@namedef{PY@tok@ss}{\def\PY@tc##1{\textcolor[rgb]{0.10,0.09,0.49}{##1}}}
\@namedef{PY@tok@sx}{\def\PY@tc##1{\textcolor[rgb]{0.00,0.50,0.00}{##1}}}
\@namedef{PY@tok@m}{\def\PY@tc##1{\textcolor[rgb]{0.40,0.40,0.40}{##1}}}
\@namedef{PY@tok@gh}{\let\PY@bf=\textbf\def\PY@tc##1{\textcolor[rgb]{0.00,0.00,0.50}{##1}}}
\@namedef{PY@tok@gu}{\let\PY@bf=\textbf\def\PY@tc##1{\textcolor[rgb]{0.50,0.00,0.50}{##1}}}
\@namedef{PY@tok@gd}{\def\PY@tc##1{\textcolor[rgb]{0.63,0.00,0.00}{##1}}}
\@namedef{PY@tok@gi}{\def\PY@tc##1{\textcolor[rgb]{0.00,0.52,0.00}{##1}}}
\@namedef{PY@tok@gr}{\def\PY@tc##1{\textcolor[rgb]{0.89,0.00,0.00}{##1}}}
\@namedef{PY@tok@ge}{\let\PY@it=\textit}
\@namedef{PY@tok@gs}{\let\PY@bf=\textbf}
\@namedef{PY@tok@gp}{\let\PY@bf=\textbf\def\PY@tc##1{\textcolor[rgb]{0.00,0.00,0.50}{##1}}}
\@namedef{PY@tok@go}{\def\PY@tc##1{\textcolor[rgb]{0.44,0.44,0.44}{##1}}}
\@namedef{PY@tok@gt}{\def\PY@tc##1{\textcolor[rgb]{0.00,0.27,0.87}{##1}}}
\@namedef{PY@tok@err}{\def\PY@bc##1{{\setlength{\fboxsep}{\string -\fboxrule}\fcolorbox[rgb]{1.00,0.00,0.00}{1,1,1}{\strut ##1}}}}
\@namedef{PY@tok@kc}{\let\PY@bf=\textbf\def\PY@tc##1{\textcolor[rgb]{0.00,0.50,0.00}{##1}}}
\@namedef{PY@tok@kd}{\let\PY@bf=\textbf\def\PY@tc##1{\textcolor[rgb]{0.00,0.50,0.00}{##1}}}
\@namedef{PY@tok@kn}{\let\PY@bf=\textbf\def\PY@tc##1{\textcolor[rgb]{0.00,0.50,0.00}{##1}}}
\@namedef{PY@tok@kr}{\let\PY@bf=\textbf\def\PY@tc##1{\textcolor[rgb]{0.00,0.50,0.00}{##1}}}
\@namedef{PY@tok@bp}{\def\PY@tc##1{\textcolor[rgb]{0.00,0.50,0.00}{##1}}}
\@namedef{PY@tok@fm}{\def\PY@tc##1{\textcolor[rgb]{0.00,0.00,1.00}{##1}}}
\@namedef{PY@tok@vc}{\def\PY@tc##1{\textcolor[rgb]{0.10,0.09,0.49}{##1}}}
\@namedef{PY@tok@vg}{\def\PY@tc##1{\textcolor[rgb]{0.10,0.09,0.49}{##1}}}
\@namedef{PY@tok@vi}{\def\PY@tc##1{\textcolor[rgb]{0.10,0.09,0.49}{##1}}}
\@namedef{PY@tok@vm}{\def\PY@tc##1{\textcolor[rgb]{0.10,0.09,0.49}{##1}}}
\@namedef{PY@tok@sa}{\def\PY@tc##1{\textcolor[rgb]{0.73,0.13,0.13}{##1}}}
\@namedef{PY@tok@sb}{\def\PY@tc##1{\textcolor[rgb]{0.73,0.13,0.13}{##1}}}
\@namedef{PY@tok@sc}{\def\PY@tc##1{\textcolor[rgb]{0.73,0.13,0.13}{##1}}}
\@namedef{PY@tok@dl}{\def\PY@tc##1{\textcolor[rgb]{0.73,0.13,0.13}{##1}}}
\@namedef{PY@tok@s2}{\def\PY@tc##1{\textcolor[rgb]{0.73,0.13,0.13}{##1}}}
\@namedef{PY@tok@sh}{\def\PY@tc##1{\textcolor[rgb]{0.73,0.13,0.13}{##1}}}
\@namedef{PY@tok@s1}{\def\PY@tc##1{\textcolor[rgb]{0.73,0.13,0.13}{##1}}}
\@namedef{PY@tok@mb}{\def\PY@tc##1{\textcolor[rgb]{0.40,0.40,0.40}{##1}}}
\@namedef{PY@tok@mf}{\def\PY@tc##1{\textcolor[rgb]{0.40,0.40,0.40}{##1}}}
\@namedef{PY@tok@mh}{\def\PY@tc##1{\textcolor[rgb]{0.40,0.40,0.40}{##1}}}
\@namedef{PY@tok@mi}{\def\PY@tc##1{\textcolor[rgb]{0.40,0.40,0.40}{##1}}}
\@namedef{PY@tok@il}{\def\PY@tc##1{\textcolor[rgb]{0.40,0.40,0.40}{##1}}}
\@namedef{PY@tok@mo}{\def\PY@tc##1{\textcolor[rgb]{0.40,0.40,0.40}{##1}}}
\@namedef{PY@tok@ch}{\let\PY@it=\textit\def\PY@tc##1{\textcolor[rgb]{0.24,0.48,0.48}{##1}}}
\@namedef{PY@tok@cm}{\let\PY@it=\textit\def\PY@tc##1{\textcolor[rgb]{0.24,0.48,0.48}{##1}}}
\@namedef{PY@tok@cpf}{\let\PY@it=\textit\def\PY@tc##1{\textcolor[rgb]{0.24,0.48,0.48}{##1}}}
\@namedef{PY@tok@c1}{\let\PY@it=\textit\def\PY@tc##1{\textcolor[rgb]{0.24,0.48,0.48}{##1}}}
\@namedef{PY@tok@cs}{\let\PY@it=\textit\def\PY@tc##1{\textcolor[rgb]{0.24,0.48,0.48}{##1}}}

\def\PYZbs{\char`\\}
\def\PYZus{\char`\_}

\def\PYZca{\char`\^}

\def\PYZgt{\char`\>}
\def\PYZsh{\char`\#}
\def\PYZpc{\char`\%}

\def\PYZhy{\char`\-}
\def\PYZsq{\char`\'}
\def\PYZdq{\char`\"}

\makeatother

    \makeatletter
        \newbox\Wrappedcontinuationbox
        \newbox\Wrappedvisiblespacebox
        \newcommand*\Wrappedvisiblespace {\textcolor{red}{\textvisiblespace}}
        \newcommand*\Wrappedcontinuationsymbol {\textcolor{red}{\llap{\tiny$\m@th\hookrightarrow$}}}
        \newcommand*\Wrappedcontinuationindent {3ex }
        \newcommand*\Wrappedafterbreak {\kern\Wrappedcontinuationindent\copy\Wrappedcontinuationbox}
        \newcommand*\Wrappedbreaksatspecials {%
            \def\PYGZus{\discretionary{\char`\_}{\Wrappedafterbreak}{\char`\_}}%
            \def\PYGZob{\discretionary{}{\Wrappedafterbreak\char`\{}{\char`\{}}%
            \def\PYGZcb{\discretionary{\char`\}}{\Wrappedafterbreak}{\char`\}}}%
            \def\PYGZca{\discretionary{\char`\^}{\Wrappedafterbreak}{\char`\^}}%
            \def\PYGZam{\discretionary{\char`\&}{\Wrappedafterbreak}{\char`\&}}%
            \def\PYGZlt{\discretionary{}{\Wrappedafterbreak\char`\<}{\char`\<}}%
            \def\PYGZgt{\discretionary{\char`\>}{\Wrappedafterbreak}{\char`\>}}%
            \def\PYGZsh{\discretionary{}{\Wrappedafterbreak\char`\#}{\char`\#}}%
            \def\PYGZpc{\discretionary{}{\Wrappedafterbreak\char`\%}{\char`\%}}%
            \def\PYGZdl{\discretionary{}{\Wrappedafterbreak\char`\$}{\char`\$}}%
            \def\PYGZhy{\discretionary{\char`\-}{\Wrappedafterbreak}{\char`\-}}%
            \def\PYGZsq{\discretionary{}{\Wrappedafterbreak\textquotesingle}{\textquotesingle}}%
            \def\PYGZdq{\discretionary{}{\Wrappedafterbreak\char`\"}{\char`\"}}%
            \def\PYGZti{\discretionary{\char`\~}{\Wrappedafterbreak}{\char`\~}}%
        }
        \newcommand*\Wrappedbreaksatpunct {%
            \lccode`\~`\.\lowercase{\def~}{\discretionary{\hbox{\char`\.}}{\Wrappedafterbreak}{\hbox{\char`\.}}}%
            \lccode`\~`\,\lowercase{\def~}{\discretionary{\hbox{\char`\,}}{\Wrappedafterbreak}{\hbox{\char`\,}}}%
            \lccode`\~`\;\lowercase{\def~}{\discretionary{\hbox{\char`\;}}{\Wrappedafterbreak}{\hbox{\char`\;}}}%
            \lccode`\~`\:\lowercase{\def~}{\discretionary{\hbox{\char`\:}}{\Wrappedafterbreak}{\hbox{\char`\:}}}%
            \lccode`\~`\?\lowercase{\def~}{\discretionary{\hbox{\char`\?}}{\Wrappedafterbreak}{\hbox{\char`\?}}}%
            \lccode`\~`\!\lowercase{\def~}{\discretionary{\hbox{\char`\!}}{\Wrappedafterbreak}{\hbox{\char`\!}}}%
            \lccode`\~`\/\lowercase{\def~}{\discretionary{\hbox{\char`\/}}{\Wrappedafterbreak}{\hbox{\char`\/}}}%
            \catcode`\.\active
            \catcode`\,\active
            \catcode`\;\active
            \catcode`\:\active
            \catcode`\?\active
            \catcode`\!\active
            \catcode`\/\active
            \lccode`\~`\~
        }
    \makeatother

    \let\OriginalVerbatim=\Verbatim
    \makeatletter
    \renewcommand{\Verbatim}[1][1]{%
        \sbox\Wrappedcontinuationbox {\Wrappedcontinuationsymbol}%
        \sbox\Wrappedvisiblespacebox {\FV@SetupFont\Wrappedvisiblespace}%
        \def\FancyVerbFormatLine ##1{\hsize\linewidth
            \vtop{\raggedright\hyphenpenalty\z@\exhyphenpenalty\z@
                \doublehyphendemerits\z@\finalhyphendemerits\z@
                \strut ##1\strut}%
        }%
        \def\FV@Space {%
            \nobreak\hskip\z@ plus\fontdimen3\font minus\fontdimen4\font
            \discretionary{\copy\Wrappedvisiblespacebox}{\Wrappedafterbreak}
            {\kern\fontdimen2\font}%
        }%

        \Wrappedbreaksatspecials
        \OriginalVerbatim[#1,codes*=\Wrappedbreaksatpunct]%
    }
    \makeatother

    \definecolor{incolor}{HTML}{303F9F}
    \definecolor{outcolor}{HTML}{D84315}
    \definecolor{cellborder}{HTML}{CFCFCF}
    \definecolor{cellbackground}{HTML}{F7F7F7}

    \makeatletter
    \newcommand{\boxspacing}{\kern\kvtcb@left@rule\kern\kvtcb@boxsep}
    \makeatother
    \newcommand{\prompt}[4]{}

    \hypersetup{
      breaklinks=true,  
      colorlinks=true,
      urlcolor=urlcolor,
      linkcolor=linkcolor,
      citecolor=citecolor,
      }
    
\begin{document}
    
    \maketitle




\textbf{Do you think you need to know quantum physics to understand how
a quantum computer works? Nope, no worries there. You don't need a deep
dive into physics or mathematics, just a bit of familiarity with vectors
and matrix multiplication. That's really it. A good handle on Python
programming and a few numpy functions will do the trick, specifically:}
\href{https://numpy.org/doc/stable/reference/generated/numpy.reshape.html}{\textbf{\texttt{reshape}}},
\href{https://numpy.org/doc/stable/reference/generated/numpy.kron.html}{\textbf{\texttt{kron}}},
\href{https://numpy.org/doc/stable/reference/generated/numpy.matmul.html}{\textbf{\texttt{matmul}}},
\href{https://numpy.org/doc/stable/reference/generated/numpy.swapaxes.html}{\textbf{\texttt{swapaxes}}},
\href{https://numpy.org/doc/stable/reference/generated/numpy.linalg.norm.html}{\textbf{\texttt{linalg.norm}}},
\textbf{and}
\href{https://numpy.org/doc/stable/reference/generated/numpy.random.choice.norm.html}{\textbf{\texttt{random.choice}}}.
\textbf{In fact, appendix B shows that twelve lines of Python code suffice to define a complete simulator.}

\textbf{The whole point of this article is to give you an informal,
brief, hopefully digestible and educational description of \emph{how you
can easily implement your own quantum computer simulator}. It's
\emph{not} about `Yet Another Quantum Computer Simulator' (YAQCS?),
which are a dime a dozen {[}1{]}, but about \emph{how to build your
own}. And, honestly, there's probably no better way to learn how a
quantum computer works!}

\section{A Quantum Computer is Like a Rubik's Cube}
\label{a-quantum-computer-is-like-a-rubiks-cube}

\begin{figure}[!hbt] 
	\centering{\includegraphics[width=0.5 \textwidth]{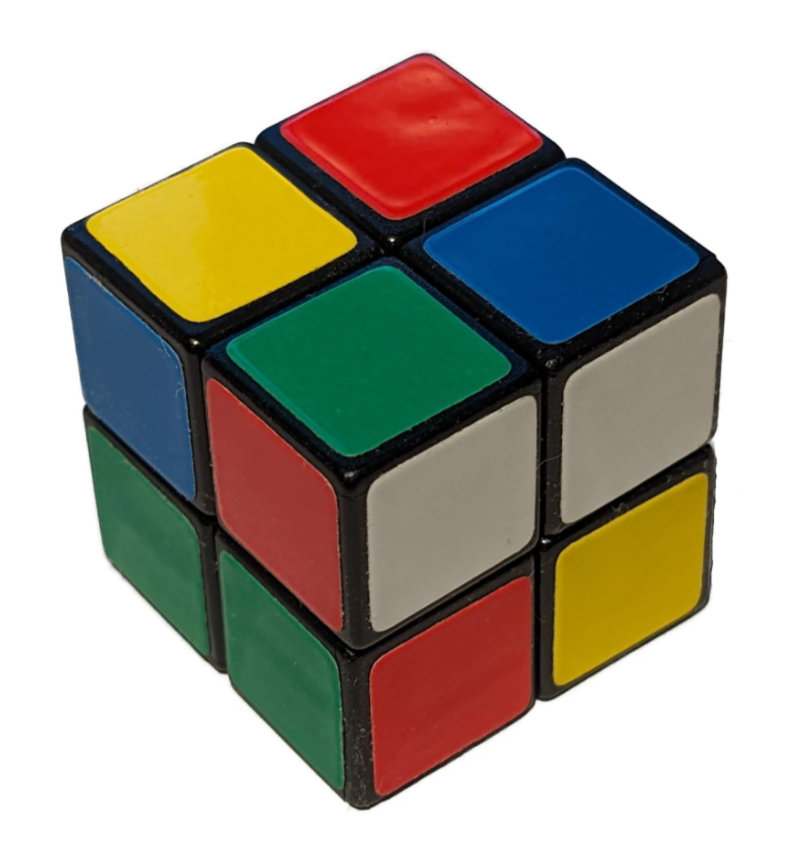}}
	\caption{A quantum computer has similarities with a high-dimensional Rubik's mini-cube.}
\end{figure}

    Imagine a quantum computer as a high-dimensional Rubik's mini-cube (Fig.~1).
This cube is formed by a bunch of \textbf{quantum bits} or
`qubits', each representing a coordinate axis. So, three qubits define a
three-dimensional cube (that's \(2 \times 2 \times 2\)), and \(N\)
qubits define an \(N\)-dimensional cube (we're talking
\(2 \times 2 \times \ldots \times 2\) here). The cube consists of
\(2^N\) smaller cubes, forming the quantum computer's exponential
`workspace'. Each small cube is like a drawer, holding a weight, which
is either a real or complex number. But, for simplicity, let's mostly
talk about real numbers here.

In this text, I'll primarily use computer science terms. However, since
physics jargon is pretty common online and in literature, let me briefly
describe some basic physics terms and how they map to the concepts I'm
using. Physicists call the small cubes \textbf{eigenstates}. The
workspace is known as the \textbf{state vector}, comprising a weighted
sum of these \(2^N\) eigenstates. Mathematicians would call such a sum a
\textbf{linear combination}, while physicists call it a
\textbf{superposition}. The weights are known as \textbf{amplitudes}. A
small cube is represented in \textbf{Dirac} or \textbf{bra-ket notation}
as \(|k\rangle\), where \(k\) is the cube's binary index. For example,
cube number 5 is \(|101\rangle\), equivalent to the unit vector
\(\boldsymbol{e}_5\) in traditional math notation. There are some
special symbols too. For instance, \(|+\rangle\) typically means the
same as \((\boldsymbol{e}_0 + \boldsymbol{e}_1)/\sqrt{2}\), which can
also be written as \((\sqrt{1/2},\sqrt{1/2})\), while \(|-\rangle\)
represents
\((\boldsymbol{e}_0 - \boldsymbol{e}_1)/\sqrt{2} = (\sqrt{1/2},-\sqrt{1/2})\).
The set of small cubes
\(\{\boldsymbol{e}_0,\boldsymbol{e}_1,\ldots,\boldsymbol{e}_{(2^N-1)}\}\)
forms a \textbf{basis} for the state vector.

A physical quantum computer with \(N\) qubits has a size of \(O(N)\),
but to simulate it on a classical computer, we need \(O(2^N)\) memory.
So, we can't simulate huge quantum computers. But the size we can
simulate is enough to understand how a quantum computer works from a
programmer's perspective.

There are three main steps in quantum computing: 
\begin{enumerate}
	\item  Building up the workspace.
	\item Performing operations on the workspace.
	\item Reading from the workspace.
\end{enumerate}

There are some rules and limitations from physical laws, but they're not
too weird. We have an exponentially large workspace, and the cool thing
is we can perform certain primitive operations (like matrix
multiplications) on the entire memory in constant time on a physical
quantum computer. Despite the limitations, there are clever combinations
of operations that are very useful, even if finding them can be tricky.

First, I'll describe how you can build a quantum computer simulator as a
\textbf{stack machine}. Then, I'll show with a slightly larger example
how to use a quantum computer to solve a `practical' problem, which
isn't immediately obvious. Finally, I'll demonstrate running the quantum
computer simulator on a GPU, which offers great potential for speed-up.
The complete code is compiled in Appendix A, and Appendix B contains a
minimalist yet fully functional and complete quantum computer simulator
that's just twelve lines long.

    \section{The Quantum Computer as a Stack
Machine}\label{the-quantum-computer-as-a-stack-machine}

\begin{figure}[!hbt] 
	\centering{\includegraphics[width=\textwidth]{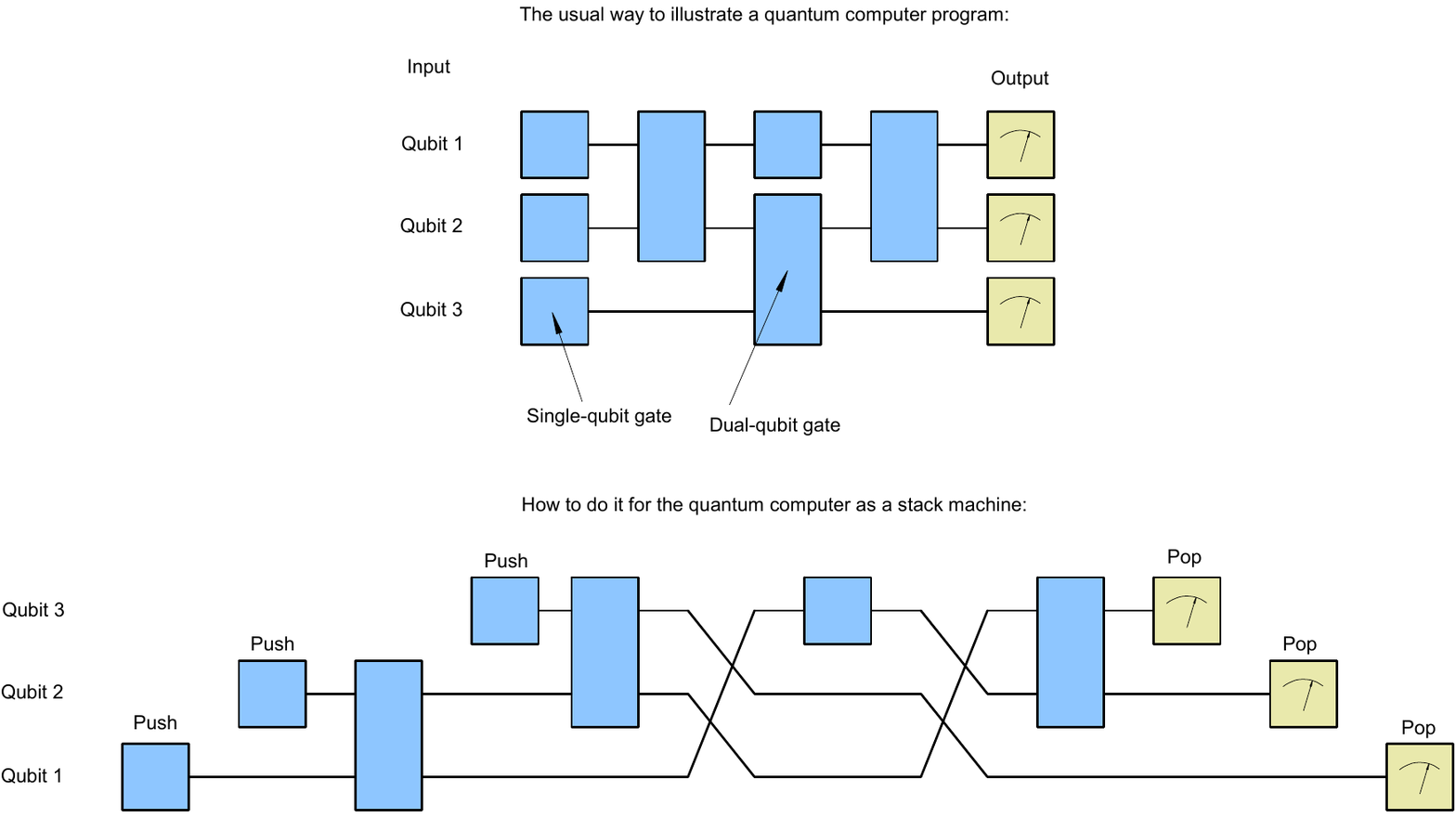}}
	\caption{A quantum computer can be seen as a stack machine.}
\end{figure}

Think of a quantum computer as a stack machine (Fig. 2). For me,
this was a real `aha!' moment! The workspace functions like a stack and
can be represented as an array that's built up by \textbf{pushing}
qubits onto it. This is done by calculating the \textbf{Kronecker
product} `\(\otimes\)' (also known as the tensor product) of the
workspace \(M\), represented as a vector with \(2^N\) elements, and the
new qubit \(q\), represented as a vector with two elements, so that
\(M \leftarrow M \otimes q\). Here's a little example of how the
Kronecker product works:
\begin{align*}
(a, b, c, d) \otimes (\alpha, \beta) &= 
\left(a \cdot (\alpha, \beta), b \cdot (\alpha, \beta), c \cdot (\alpha, \beta), d \cdot (\alpha, \beta)\right) \\
&= (a\alpha, a\beta, b\alpha, b\beta, c\alpha, c\beta, d\alpha, d\beta) .
\end{align*}
The difference from a traditional stack is that the whole quantum stack
changes when we push a qubit, not just the top. The Kronecker product
can be computed using the \texttt{kron} function in numpy. The qubit
gets, sort of, `woven' into the entire stack.

A \textbf{gate operation} is performed on the top qubits in the stack,
kind of like twisting layers on a Rubik's cube. Such an operation is
simply done with a matrix multiplication, \texttt{matmul} in numpy.

We also need an operation to \textbf{move a qubit} in the stack to the
top position (TOS, `Top Of Stack'). For the workspace, this means
permuting the array's axes, which we achieve with numpy's
\texttt{swapaxes} procedure.

Finally, we need to \textbf{pop} qubits from the stack and read them
off. This is done by calculating the probability that a qubit will be
zero or one using numpy's \texttt{linalg.norm} function, and then
randomly drawing a value according to these probabilities. (The
remaining stack, i.e., the workspace, can be associated with the
\href{https://en.wikipedia.org/wiki/Conditional_probability}{conditional
probabilities} given that the popped qubit had the drawn value. You
don't need to know exactly what this means, but it will become clear how
this works in the code later.)

First, I'll go through the implementation of the four basic instructions
\emph{push}, \emph{gate operation}, \emph{move}, and \emph{pop} in
detail. All the instructions can be implemented by first transforming
the workspace with numpy's \texttt{reshape} and then applying a numpy
function (\texttt{kron}, \texttt{matmul}, \texttt{swapaxes}, or
\texttt{linalg.norm}) to this array. If you reshape the workspace into a
vector, you can think of it as a table containing a weight for each
small cube. The transformation is a permutation of the array's indices
but never changes the content.
        
    \subsection{Instruction 1: Push a Qubit onto the
Stack}\label{instruction-1-pushing-a-qubit-onto-the-stack}

Every time you push a new qubit onto the stack, the size of the
workspace \textbf{doubles}. Imagine a `fresh' qubit represented by a
weight vector $(\alpha,\beta)$, where $\alpha$ and $\beta$ are
weights that have to satisfy the condition
$|\alpha|^2 + |\beta|^2 = 1$.
You can think of $|\alpha|^2$ as the probability
that the qubit is zero, and $|\beta|^2$ as the
probability that it's one. A qubit that's definitely zero can be
represented by the vector $(1,0)$, while one that's certainly one is
$(0,1)$. And a qubit that's equally likely to be zero or one can be
represented with $(\sqrt{1/2},\sqrt{1/2})$.

To implement this operation, we first reshape the workspace into a $(1
\times 2^N)$-matrix (that's a row vector) and then use numpy's
\texttt{kron} function to \textbf{expand the memory to the Kronecker
product of the workspace and the new qubit's weight vector}.

    \begin{tcolorbox}[breakable, size=fbox, boxrule=1pt, pad at break*=1mm,colback=cellbackground, colframe=cellborder]
\prompt{In}{incolor}{51}{\boxspacing}
\begin{Verbatim}[commandchars=\\\{\}]
\PY{k+kn}{import} \PY{n+nn}{numpy} \PY{k}{as} \PY{n+nn}{np}                           

\PY{k}{def} \PY{n+nf}{pushQubit}\PY{p}{(}\PY{n}{weights}\PY{p}{)}\PY{p}{:}
    \PY{k}{global} \PY{n}{workspace}
    \PY{n}{workspace} \PY{o}{=} \PY{n}{np}\PY{o}{.}\PY{n}{reshape}\PY{p}{(}\PY{n}{workspace}\PY{p}{,}\PY{p}{(}\PY{l+m+mi}{1}\PY{p}{,}\PY{o}{\PYZhy{}}\PY{l+m+mi}{1}\PY{p}{)}\PY{p}{)}  
    \PY{n}{workspace} \PY{o}{=} \PY{n}{np}\PY{o}{.}\PY{n}{kron}\PY{p}{(}\PY{n}{workspace}\PY{p}{,}\PY{n}{weights}\PY{p}{)}    
\end{Verbatim}
\end{tcolorbox}

    The parameter \texttt{(1,-1)} in the \texttt{reshape} function means
that we're reshaping the array into a matrix with one row (that's what
the one in \texttt{(1,-1)} is telling us) and as many columns as needed
(and that's what the minus one in \texttt{(1,-1)} is for).

Here comes a code example. We'll start by creating an empty workspace.

    \begin{tcolorbox}[breakable, size=fbox, boxrule=1pt, pad at break*=1mm,colback=cellbackground, colframe=cellborder]
\prompt{In}{incolor}{52}{\boxspacing}
\begin{Verbatim}[commandchars=\\\{\}]
\PY{n}{workspace} \PY{o}{=} \PY{n}{np}\PY{o}{.}\PY{n}{array}\PY{p}{(}\PY{p}{[}\PY{p}{[}\PY{l+m+mf}{1.}\PY{p}{]}\PY{p}{]}\PY{p}{)}        \PY{c+c1}{\PYZsh{} create empty qubit stack}
\PY{n}{pushQubit}\PY{p}{(}\PY{p}{[}\PY{l+m+mi}{1}\PY{p}{,}\PY{l+m+mi}{0}\PY{p}{]}\PY{p}{)}
\PY{n+nb}{print}\PY{p}{(}\PY{n}{workspace}\PY{p}{)}
\PY{n}{pushQubit}\PY{p}{(}\PY{p}{[}\PY{l+m+mi}{3}\PY{o}{/}\PY{l+m+mi}{5}\PY{p}{,}\PY{l+m+mi}{4}\PY{o}{/}\PY{l+m+mi}{5}\PY{p}{]}\PY{p}{)}                \PY{c+c1}{\PYZsh{} push a 2nd qubit}
\PY{n+nb}{print}\PY{p}{(}\PY{n}{workspace}\PY{p}{)}
\end{Verbatim}
\end{tcolorbox}

    \begin{Verbatim}[commandchars=\\\{\}]
[[1. 0.]]
[[0.6 0.8 0.  0. ]]
    \end{Verbatim}

    The only place in the code where we explicitly specify the data type is
when we create the quantum stack \texttt{workspace}. An empty quantum
stack is represented by a matrix that consists of exactly one row and
one column containing the number one.

\subsection{Instruction 2: Perform a Gate
Operation}\label{instruction-2-perform-a-gate-operation}

Now, let's do operations on the qubits. These are called \textbf{gates}
and are carried out through matrix multiplications that involve the
whole workspace. You can't manipulate individual small cubes; you have
to manipulate the entire workspace at once. It's a bit like how you
can't just turn a single cube in a Rubik's Cube; you have to turn an
entire layer.

Gate operations never remove or add any qubits. A gate applied to \(M\)
qubits corresponds to a matrix multiplication of the \textbf{entire}
workspace with a \((2^M \times 2^M)\)-matrix \(U\). We don't really need
to worry about the formal requirements for the matrix \(U\), as almost
always only a handful of predefined matrices are used. But it's good to
know that the matrix \(U\) must be \textbf{unitary}, which means that
the inverse \(U^{-1}\) is the same as the transposed and
complex-conjugated matrix \(U^H\). Other than this, the matrix can look
any way you like. Of course, it can be tricky to practically realize
irregular matrices on a physical quantum computer.

Usually, gates are used on one to three qubits (\(1 \le M \le 3\)). It
can be shown that all gates with \(M>2\) can be expressed as sequences
of gates with \(M \le 2\).

We implement the gate operation by first reshaping the workspace into a
\((2^{N-M}\times 2^M)\)-matrix, which we then multiply with the gate
matrix, having dimensions \(2^M \times 2^M\).

    \begin{tcolorbox}[breakable, size=fbox, boxrule=1pt, pad at break*=1mm,colback=cellbackground, colframe=cellborder]
\prompt{In}{incolor}{53}{\boxspacing}
\begin{Verbatim}[commandchars=\\\{\}]
\PY{k}{def} \PY{n+nf}{applyGate}\PY{p}{(}\PY{n}{gate}\PY{p}{)}\PY{p}{:}
    \PY{k}{global} \PY{n}{workspace}
    \PY{n}{workspace} \PY{o}{=} \PY{n}{np}\PY{o}{.}\PY{n}{reshape}\PY{p}{(}\PY{n}{workspace}\PY{p}{,}\PY{p}{(}\PY{o}{\PYZhy{}}\PY{l+m+mi}{1}\PY{p}{,}\PY{n}{gate}\PY{o}{.}\PY{n}{shape}\PY{p}{[}\PY{l+m+mi}{0}\PY{p}{]}\PY{p}{)}\PY{p}{)} 
    \PY{n}{np}\PY{o}{.}\PY{n}{matmul}\PY{p}{(}\PY{n}{workspace}\PY{p}{,}\PY{n}{gate}\PY{o}{.}\PY{n}{T}\PY{p}{,}\PY{n}{out}\PY{o}{=}\PY{n}{workspace}\PY{p}{)} 
\end{Verbatim}
\end{tcolorbox}

    \paragraph{\texorpdfstring{Example of
\texttt{applyGate}}{Example of applyGate}}\label{example-of-applygate}

A common gate is the X-gate:

    \begin{tcolorbox}[breakable, size=fbox, boxrule=1pt, pad at break*=1mm,colback=cellbackground, colframe=cellborder]
\prompt{In}{incolor}{54}{\boxspacing}
\begin{Verbatim}[commandchars=\\\{\}]
\PY{n}{X\PYZus{}gate} \PY{o}{=} \PY{n}{np}\PY{o}{.}\PY{n}{array}\PY{p}{(}\PY{p}{[}\PY{p}{[}\PY{l+m+mi}{0}\PY{p}{,} \PY{l+m+mi}{1}\PY{p}{]}\PY{p}{,}                    \PY{c+c1}{\PYZsh{} Pauli X gate}
                   \PY{p}{[}\PY{l+m+mi}{1}\PY{p}{,} \PY{l+m+mi}{0}\PY{p}{]}\PY{p}{]}\PY{p}{)}                   \PY{c+c1}{\PYZsh{} = NOT gate}
\end{Verbatim}
\end{tcolorbox}

    This gate switches the probabilities for a qubit being zero or one,
which you can see by performing the multiplication
\[\begin{pmatrix}0 & 1\\ 1 & 0\end{pmatrix} \begin{pmatrix}a \\ b\end{pmatrix} = \begin{pmatrix}b \\ a\end{pmatrix}.\]
That's why the X-gate is also called a NOT gate. Here's what a run looks
like on a qubit that is zero, i.e., represented by the weights
\((1, 0)\):

    \begin{tcolorbox}[breakable, size=fbox, boxrule=1pt, pad at break*=1mm,colback=cellbackground, colframe=cellborder]
\prompt{In}{incolor}{55}{\boxspacing}
\begin{Verbatim}[commandchars=\\\{\}]
\PY{n}{np}\PY{o}{.}\PY{n}{matmul}\PY{p}{(}\PY{n}{X\PYZus{}gate}\PY{p}{,}\PY{p}{[}\PY{l+m+mi}{1}\PY{p}{,}\PY{l+m+mi}{0}\PY{p}{]}\PY{p}{)}
\end{Verbatim}
\end{tcolorbox}

            \begin{tcolorbox}[breakable, size=fbox, boxrule=.5pt, pad at break*=1mm, opacityfill=0]
\prompt{Out}{outcolor}{55}{\boxspacing}
\begin{Verbatim}[commandchars=\\\{\}]
array([0, 1])
\end{Verbatim}
\end{tcolorbox}
        
    Or as a sample run using our new instructions \texttt{pushQubit} and
\texttt{applyGate}:

    \begin{tcolorbox}[breakable, size=fbox, boxrule=1pt, pad at break*=1mm,colback=cellbackground, colframe=cellborder]
\prompt{In}{incolor}{56}{\boxspacing}
\begin{Verbatim}[commandchars=\\\{\}]
\PY{n}{workspace} \PY{o}{=} \PY{n}{np}\PY{o}{.}\PY{n}{array}\PY{p}{(}\PY{p}{[}\PY{p}{[}\PY{l+m+mf}{1.}\PY{p}{]}\PY{p}{]}\PY{p}{)}              \PY{c+c1}{\PYZsh{} reset workspace}
\PY{n}{pushQubit}\PY{p}{(}\PY{p}{[}\PY{l+m+mi}{1}\PY{p}{,}\PY{l+m+mi}{0}\PY{p}{]}\PY{p}{)}
\PY{n+nb}{print}\PY{p}{(}\PY{l+s+s2}{\PYZdq{}}\PY{l+s+s2}{input}\PY{l+s+s2}{\PYZdq{}}\PY{p}{,}\PY{n}{workspace}\PY{p}{)}
\PY{n}{applyGate}\PY{p}{(}\PY{n}{X\PYZus{}gate}\PY{p}{)}                         \PY{c+c1}{\PYZsh{} = NOT}
\PY{n+nb}{print}\PY{p}{(}\PY{l+s+s2}{\PYZdq{}}\PY{l+s+s2}{output}\PY{l+s+s2}{\PYZdq{}}\PY{p}{,}\PY{n}{workspace}\PY{p}{)}
\end{Verbatim}
\end{tcolorbox}

    \begin{Verbatim}[commandchars=\\\{\}]
input [[1. 0.]]
output [[0. 1.]]
    \end{Verbatim}

    The Hadamard gate is another common and important gate:

    \begin{tcolorbox}[breakable, size=fbox, boxrule=1pt, pad at break*=1mm,colback=cellbackground, colframe=cellborder]
\prompt{In}{incolor}{57}{\boxspacing}
\begin{Verbatim}[commandchars=\\\{\}]
\PY{n}{H\PYZus{}gate} \PY{o}{=} \PY{n}{np}\PY{o}{.}\PY{n}{array}\PY{p}{(}\PY{p}{[}\PY{p}{[}\PY{l+m+mi}{1}\PY{p}{,} \PY{l+m+mi}{1}\PY{p}{]}\PY{p}{,}                \PY{c+c1}{\PYZsh{} Hadamard gate}
                   \PY{p}{[}\PY{l+m+mi}{1}\PY{p}{,}\PY{o}{\PYZhy{}}\PY{l+m+mi}{1}\PY{p}{]}\PY{p}{]}\PY{p}{)} \PY{o}{*} \PY{n}{np}\PY{o}{.}\PY{n}{sqrt}\PY{p}{(}\PY{l+m+mi}{1}\PY{o}{/}\PY{l+m+mi}{2}\PY{p}{)}
\end{Verbatim}
\end{tcolorbox}

    When applied to a qubit that is zero, represented by \$(1, 0)\$, it
changes the qubit so that it has an equal probability of being zero or
one. You can see this through the following matrix multiplication:
\[\frac{1}{\sqrt{2}}\begin{pmatrix}1 & 1\\ 1 & -1\end{pmatrix} \begin{pmatrix}1 \\ 0\end{pmatrix} = \frac{1}{\sqrt{2}}\begin{pmatrix}1 \\ 1\end{pmatrix}.\]
Some code examples:

    \begin{tcolorbox}[breakable, size=fbox, boxrule=1pt, pad at break*=1mm,colback=cellbackground, colframe=cellborder]
\prompt{In}{incolor}{58}{\boxspacing}
\begin{Verbatim}[commandchars=\\\{\}]
\PY{n}{workspace} \PY{o}{=} \PY{n}{np}\PY{o}{.}\PY{n}{array}\PY{p}{(}\PY{p}{[}\PY{p}{[}\PY{l+m+mf}{1.}\PY{p}{]}\PY{p}{]}\PY{p}{)}
\PY{n}{pushQubit}\PY{p}{(}\PY{p}{[}\PY{l+m+mi}{1}\PY{p}{,}\PY{l+m+mi}{0}\PY{p}{]}\PY{p}{)}
\PY{n+nb}{print}\PY{p}{(}\PY{l+s+s2}{\PYZdq{}}\PY{l+s+s2}{input}\PY{l+s+s2}{\PYZdq{}}\PY{p}{,}\PY{n}{workspace}\PY{p}{)}
\PY{n}{applyGate}\PY{p}{(}\PY{n}{H\PYZus{}gate}\PY{p}{)} 
\PY{n+nb}{print}\PY{p}{(}\PY{l+s+s2}{\PYZdq{}}\PY{l+s+s2}{output}\PY{l+s+s2}{\PYZdq{}}\PY{p}{,}\PY{n}{workspace}\PY{p}{)}
\end{Verbatim}
\end{tcolorbox}

    \begin{Verbatim}[commandchars=\\\{\}]
input [[1. 0.]]
output [[0.70710678 0.70710678]]
    \end{Verbatim}

    If we use complex weights, for example when we're using T-gates, we need
to initialize \texttt{workspace} with a complex data type:

    \begin{tcolorbox}[breakable, size=fbox, boxrule=1pt, pad at break*=1mm,colback=cellbackground, colframe=cellborder]
\prompt{In}{incolor}{59}{\boxspacing}
\begin{Verbatim}[commandchars=\\\{\}]
\PY{n}{T\PYZus{}gate} \PY{o}{=} \PY{n}{np}\PY{o}{.}\PY{n}{array}\PY{p}{(}\PY{p}{[}\PY{p}{[}\PY{l+m+mi}{1}\PY{p}{,}                \PY{l+m+mi}{0}\PY{p}{]}\PY{p}{,} 
                   \PY{p}{[}\PY{l+m+mi}{0}\PY{p}{,}\PY{n}{np}\PY{o}{.}\PY{n}{exp}\PY{p}{(}\PY{n}{np}\PY{o}{.}\PY{n}{pi}\PY{o}{/}\PY{o}{\PYZhy{}}\PY{l+m+mi}{4}\PY{n}{j}\PY{p}{)}\PY{p}{]}\PY{p}{]}\PY{p}{)}

\PY{n}{workspace} \PY{o}{=} \PY{n}{np}\PY{o}{.}\PY{n}{array}\PY{p}{(}\PY{p}{[}\PY{p}{[}\PY{l+m+mf}{1.}\PY{o}{+}\PY{l+m+mi}{0}\PY{n}{j}\PY{p}{]}\PY{p}{]}\PY{p}{)}      \PY{c+c1}{\PYZsh{} set complex workspace}
\PY{n}{pushQubit}\PY{p}{(}\PY{p}{[}\PY{l+m+mf}{.6}\PY{p}{,}\PY{l+m+mf}{.8}\PY{p}{]}\PY{p}{)}
\PY{n+nb}{print}\PY{p}{(}\PY{l+s+s2}{\PYZdq{}}\PY{l+s+s2}{input}\PY{l+s+s2}{\PYZdq{}}\PY{p}{,}\PY{n}{workspace}\PY{p}{)}
\PY{n}{applyGate}\PY{p}{(}\PY{n}{T\PYZus{}gate}\PY{p}{)}
\PY{n+nb}{print}\PY{p}{(}\PY{l+s+s2}{\PYZdq{}}\PY{l+s+s2}{output}\PY{l+s+s2}{\PYZdq{}}\PY{p}{,}\PY{n}{workspace}\PY{p}{)}
\end{Verbatim}
\end{tcolorbox}

    \begin{Verbatim}[commandchars=\\\{\}]
input [[0.6+0.j 0.8+0.j]]
output [[0.6       +0.j         0.56568542+0.56568542j]]
    \end{Verbatim}

    In this article, however, most examples use only real numbers.

A gate that can be applied to two qubits is the SWAP gate:

    \begin{tcolorbox}[breakable, size=fbox, boxrule=1pt, pad at break*=1mm,colback=cellbackground, colframe=cellborder]
\prompt{In}{incolor}{60}{\boxspacing}
\begin{Verbatim}[commandchars=\\\{\}]
\PY{n}{SWAP\PYZus{}gate} \PY{o}{=} \PY{n}{np}\PY{o}{.}\PY{n}{array}\PY{p}{(}\PY{p}{[}\PY{p}{[}\PY{l+m+mi}{1}\PY{p}{,} \PY{l+m+mi}{0}\PY{p}{,} \PY{l+m+mi}{0}\PY{p}{,} \PY{l+m+mi}{0}\PY{p}{]}\PY{p}{,}       \PY{c+c1}{\PYZsh{} Swap gate}
                      \PY{p}{[}\PY{l+m+mi}{0}\PY{p}{,} \PY{l+m+mi}{0}\PY{p}{,} \PY{l+m+mi}{1}\PY{p}{,} \PY{l+m+mi}{0}\PY{p}{]}\PY{p}{,}
                      \PY{p}{[}\PY{l+m+mi}{0}\PY{p}{,} \PY{l+m+mi}{1}\PY{p}{,} \PY{l+m+mi}{0}\PY{p}{,} \PY{l+m+mi}{0}\PY{p}{]}\PY{p}{,}
                      \PY{p}{[}\PY{l+m+mi}{0}\PY{p}{,} \PY{l+m+mi}{0}\PY{p}{,} \PY{l+m+mi}{0}\PY{p}{,} \PY{l+m+mi}{1}\PY{p}{]}\PY{p}{]}\PY{p}{)}
\end{Verbatim}
\end{tcolorbox}

    It changes the qubit order:

    \begin{tcolorbox}[breakable, size=fbox, boxrule=1pt, pad at break*=1mm,colback=cellbackground, colframe=cellborder]
\prompt{In}{incolor}{61}{\boxspacing}
\begin{Verbatim}[commandchars=\\\{\}]
\PY{n}{workspace} \PY{o}{=} \PY{n}{np}\PY{o}{.}\PY{n}{array}\PY{p}{(}\PY{p}{[}\PY{p}{[}\PY{l+m+mf}{1.}\PY{p}{]}\PY{p}{]}\PY{p}{)}
\PY{n}{pushQubit}\PY{p}{(}\PY{p}{[}\PY{l+m+mi}{1}\PY{p}{,}\PY{l+m+mi}{0}\PY{p}{]}\PY{p}{)}                          \PY{c+c1}{\PYZsh{} qubit 1}
\PY{n}{pushQubit}\PY{p}{(}\PY{p}{[}\PY{l+m+mf}{0.6}\PY{p}{,}\PY{l+m+mf}{0.8}\PY{p}{]}\PY{p}{)}                      \PY{c+c1}{\PYZsh{} qubit 2}
\PY{n+nb}{print}\PY{p}{(}\PY{n}{workspace}\PY{p}{)}
\PY{n}{applyGate}\PY{p}{(}\PY{n}{SWAP\PYZus{}gate}\PY{p}{)}
\PY{n+nb}{print}\PY{p}{(}\PY{n}{workspace}\PY{p}{)}
\end{Verbatim}
\end{tcolorbox}

    \begin{Verbatim}[commandchars=\\\{\}]
[[0.6 0.8 0.  0. ]]
[[0.6 0.  0.8 0. ]]
    \end{Verbatim}

    After a bit of thought, you'll see that the first row in the output
corresponds to qubit 1 \(\otimes\) qubit 2, while the second one
corresponds to qubit 2 \(\otimes\) qubit 1.

Here's a list of most of the common gates:

    \begin{tcolorbox}[breakable, size=fbox, boxrule=1pt, pad at break*=1mm,colback=cellbackground, colframe=cellborder]
\prompt{In}{incolor}{62}{\boxspacing}
\begin{Verbatim}[commandchars=\\\{\}]
\PY{n}{X\PYZus{}gate} \PY{o}{=} \PY{n}{np}\PY{o}{.}\PY{n}{array}\PY{p}{(}\PY{p}{[}\PY{p}{[}\PY{l+m+mi}{0}\PY{p}{,} \PY{l+m+mi}{1}\PY{p}{]}\PY{p}{,}                \PY{c+c1}{\PYZsh{} Pauli X gate}
                   \PY{p}{[}\PY{l+m+mi}{1}\PY{p}{,} \PY{l+m+mi}{0}\PY{p}{]}\PY{p}{]}\PY{p}{)}               \PY{c+c1}{\PYZsh{} = NOT gate}

\PY{n}{Y\PYZus{}gate} \PY{o}{=} \PY{n}{np}\PY{o}{.}\PY{n}{array}\PY{p}{(}\PY{p}{[}\PY{p}{[} \PY{l+m+mi}{0}\PY{p}{,}\PY{o}{\PYZhy{}}\PY{l+m+mi}{1}\PY{n}{j}\PY{p}{]}\PY{p}{,}              \PY{c+c1}{\PYZsh{} Pauli Y gate}
                   \PY{p}{[}\PY{l+m+mi}{1}\PY{n}{j}\PY{p}{,}  \PY{l+m+mi}{0}\PY{p}{]}\PY{p}{]}\PY{p}{)}             \PY{c+c1}{\PYZsh{} = SHZHZS}

\PY{n}{Z\PYZus{}gate} \PY{o}{=} \PY{n}{np}\PY{o}{.}\PY{n}{array}\PY{p}{(}\PY{p}{[}\PY{p}{[}\PY{l+m+mi}{1}\PY{p}{,} \PY{l+m+mi}{0}\PY{p}{]}\PY{p}{,}                \PY{c+c1}{\PYZsh{} Pauli Z gate}
                   \PY{p}{[}\PY{l+m+mi}{0}\PY{p}{,}\PY{o}{\PYZhy{}}\PY{l+m+mi}{1}\PY{p}{]}\PY{p}{]}\PY{p}{)}               \PY{c+c1}{\PYZsh{} = P(pi) = S\PYZca{}2}
                                          \PY{c+c1}{\PYZsh{} = HXH}
\PY{n}{H\PYZus{}gate} \PY{o}{=} \PY{n}{np}\PY{o}{.}\PY{n}{array}\PY{p}{(}\PY{p}{[}\PY{p}{[}\PY{l+m+mi}{1}\PY{p}{,} \PY{l+m+mi}{1}\PY{p}{]}\PY{p}{,}                \PY{c+c1}{\PYZsh{} Hadamard gate}
                   \PY{p}{[}\PY{l+m+mi}{1}\PY{p}{,}\PY{o}{\PYZhy{}}\PY{l+m+mi}{1}\PY{p}{]}\PY{p}{]}\PY{p}{)} \PY{o}{*} \PY{n}{np}\PY{o}{.}\PY{n}{sqrt}\PY{p}{(}\PY{l+m+mi}{1}\PY{o}{/}\PY{l+m+mi}{2}\PY{p}{)}

\PY{n}{S\PYZus{}gate} \PY{o}{=} \PY{n}{np}\PY{o}{.}\PY{n}{array}\PY{p}{(}\PY{p}{[}\PY{p}{[}\PY{l+m+mi}{1}\PY{p}{,} \PY{l+m+mi}{0}\PY{p}{]}\PY{p}{,}                \PY{c+c1}{\PYZsh{} Phase gate}
                   \PY{p}{[}\PY{l+m+mi}{0}\PY{p}{,}\PY{l+m+mi}{1}\PY{n}{j}\PY{p}{]}\PY{p}{]}\PY{p}{)}               \PY{c+c1}{\PYZsh{} = P(pi/2) = T\PYZca{}2}

\PY{n}{T\PYZus{}gate} \PY{o}{=} \PY{n}{np}\PY{o}{.}\PY{n}{array}\PY{p}{(}\PY{p}{[}\PY{p}{[}\PY{l+m+mi}{1}\PY{p}{,}                \PY{l+m+mi}{0}\PY{p}{]}\PY{p}{,} \PY{c+c1}{\PYZsh{} = P(pi/4)}
                   \PY{p}{[}\PY{l+m+mi}{0}\PY{p}{,}\PY{n}{np}\PY{o}{.}\PY{n}{exp}\PY{p}{(}\PY{n}{np}\PY{o}{.}\PY{n}{pi}\PY{o}{/}\PY{o}{\PYZhy{}}\PY{l+m+mi}{4}\PY{n}{j}\PY{p}{)}\PY{p}{]}\PY{p}{]}\PY{p}{)}

\PY{n}{Tinv\PYZus{}gate} \PY{o}{=} \PY{n}{np}\PY{o}{.}\PY{n}{array}\PY{p}{(}\PY{p}{[}\PY{p}{[}\PY{l+m+mi}{1}\PY{p}{,}               \PY{l+m+mi}{0}\PY{p}{]}\PY{p}{,} \PY{c+c1}{\PYZsh{} = P(\PYZhy{}pi/4)}
                      \PY{p}{[}\PY{l+m+mi}{0}\PY{p}{,}\PY{n}{np}\PY{o}{.}\PY{n}{exp}\PY{p}{(}\PY{n}{np}\PY{o}{.}\PY{n}{pi}\PY{o}{/}\PY{l+m+mi}{4}\PY{n}{j}\PY{p}{)}\PY{p}{]}\PY{p}{]}\PY{p}{)} \PY{c+c1}{\PYZsh{} = T\PYZca{}\PYZhy{}1}

\PY{k}{def} \PY{n+nf}{P\PYZus{}gate}\PY{p}{(}\PY{n}{phi}\PY{p}{)}\PY{p}{:}                          \PY{c+c1}{\PYZsh{} Phase shift gate}
    \PY{k}{return} \PY{n}{np}\PY{o}{.}\PY{n}{array}\PY{p}{(}\PY{p}{[}\PY{p}{[}\PY{l+m+mi}{1}\PY{p}{,}             \PY{l+m+mi}{0}\PY{p}{]}\PY{p}{,}
                     \PY{p}{[}\PY{l+m+mi}{0}\PY{p}{,}\PY{n}{np}\PY{o}{.}\PY{n}{exp}\PY{p}{(}\PY{n}{phi}\PY{o}{*}\PY{l+m+mi}{1}\PY{n}{j}\PY{p}{)}\PY{p}{]}\PY{p}{]}\PY{p}{)}

\PY{k}{def} \PY{n+nf}{Rx\PYZus{}gate}\PY{p}{(}\PY{n}{theta}\PY{p}{)}\PY{p}{:}                       \PY{c+c1}{\PYZsh{} X rotation gate}
    \PY{k}{return} \PY{n}{np}\PY{o}{.}\PY{n}{array}\PY{p}{(}\PY{p}{[}\PY{p}{[}\PY{n}{np}\PY{o}{.}\PY{n}{cos}\PY{p}{(}\PY{n}{theta}\PY{o}{/}\PY{l+m+mi}{2}\PY{p}{)}\PY{p}{,}\PY{o}{\PYZhy{}}\PY{l+m+mi}{1}\PY{n}{j}\PY{o}{*}\PY{n}{np}\PY{o}{.}\PY{n}{sin}\PY{p}{(}\PY{n}{theta}\PY{o}{/}\PY{l+m+mi}{2}\PY{p}{)}\PY{p}{]}\PY{p}{,}
                     \PY{p}{[}\PY{o}{\PYZhy{}}\PY{l+m+mi}{1}\PY{n}{j}\PY{o}{*}\PY{n}{np}\PY{o}{.}\PY{n}{sin}\PY{p}{(}\PY{n}{theta}\PY{o}{/}\PY{l+m+mi}{2}\PY{p}{)}\PY{p}{,}\PY{n}{np}\PY{o}{.}\PY{n}{cos}\PY{p}{(}\PY{n}{theta}\PY{o}{/}\PY{l+m+mi}{2}\PY{p}{)}\PY{p}{]}\PY{p}{]}\PY{p}{)}

\PY{k}{def} \PY{n+nf}{Ry\PYZus{}gate}\PY{p}{(}\PY{n}{theta}\PY{p}{)}\PY{p}{:}                       \PY{c+c1}{\PYZsh{} Y rotation gate}
    \PY{k}{return} \PY{n}{np}\PY{o}{.}\PY{n}{array}\PY{p}{(}\PY{p}{[}\PY{p}{[}\PY{n}{np}\PY{o}{.}\PY{n}{cos}\PY{p}{(}\PY{n}{theta}\PY{o}{/}\PY{l+m+mi}{2}\PY{p}{)}\PY{p}{,}\PY{o}{\PYZhy{}}\PY{n}{np}\PY{o}{.}\PY{n}{sin}\PY{p}{(}\PY{n}{theta}\PY{o}{/}\PY{l+m+mi}{2}\PY{p}{)}\PY{p}{]}\PY{p}{,}
                     \PY{p}{[}\PY{n}{np}\PY{o}{.}\PY{n}{sin}\PY{p}{(}\PY{n}{theta}\PY{o}{/}\PY{l+m+mi}{2}\PY{p}{)}\PY{p}{,} \PY{n}{np}\PY{o}{.}\PY{n}{cos}\PY{p}{(}\PY{n}{theta}\PY{o}{/}\PY{l+m+mi}{2}\PY{p}{)}\PY{p}{]}\PY{p}{]}\PY{p}{)}

\PY{k}{def} \PY{n+nf}{Rz\PYZus{}gate}\PY{p}{(}\PY{n}{theta}\PY{p}{)}\PY{p}{:}                       \PY{c+c1}{\PYZsh{} Z rotation gate}
    \PY{k}{return} \PY{n}{np}\PY{o}{.}\PY{n}{array}\PY{p}{(}\PY{p}{[}\PY{p}{[}\PY{n}{np}\PY{o}{.}\PY{n}{exp}\PY{p}{(}\PY{o}{\PYZhy{}}\PY{l+m+mi}{1}\PY{n}{j}\PY{o}{*}\PY{n}{theta}\PY{o}{/}\PY{l+m+mi}{2}\PY{p}{)}\PY{p}{,}                 \PY{l+m+mi}{0}\PY{p}{]}\PY{p}{,}
                     \PY{p}{[}                  \PY{l+m+mi}{0}\PY{p}{,}\PY{n}{np}\PY{o}{.}\PY{n}{exp}\PY{p}{(}\PY{l+m+mi}{1}\PY{n}{j}\PY{o}{*}\PY{n}{theta}\PY{o}{/}\PY{l+m+mi}{2}\PY{p}{)}\PY{p}{]}\PY{p}{]}\PY{p}{)}

\PY{n}{CNOT\PYZus{}gate} \PY{o}{=} \PY{n}{np}\PY{o}{.}\PY{n}{array}\PY{p}{(}\PY{p}{[}\PY{p}{[}\PY{l+m+mi}{1}\PY{p}{,} \PY{l+m+mi}{0}\PY{p}{,} \PY{l+m+mi}{0}\PY{p}{,} \PY{l+m+mi}{0}\PY{p}{]}\PY{p}{,}       \PY{c+c1}{\PYZsh{} Ctled NOT gate}
                      \PY{p}{[}\PY{l+m+mi}{0}\PY{p}{,} \PY{l+m+mi}{1}\PY{p}{,} \PY{l+m+mi}{0}\PY{p}{,} \PY{l+m+mi}{0}\PY{p}{]}\PY{p}{,}       \PY{c+c1}{\PYZsh{} = XOR gate}
                      \PY{p}{[}\PY{l+m+mi}{0}\PY{p}{,} \PY{l+m+mi}{0}\PY{p}{,} \PY{l+m+mi}{0}\PY{p}{,} \PY{l+m+mi}{1}\PY{p}{]}\PY{p}{,}
                      \PY{p}{[}\PY{l+m+mi}{0}\PY{p}{,} \PY{l+m+mi}{0}\PY{p}{,} \PY{l+m+mi}{1}\PY{p}{,} \PY{l+m+mi}{0}\PY{p}{]}\PY{p}{]}\PY{p}{)}

\PY{n}{CZ\PYZus{}gate} \PY{o}{=} \PY{n}{np}\PY{o}{.}\PY{n}{array}\PY{p}{(}\PY{p}{[}\PY{p}{[}\PY{l+m+mi}{1}\PY{p}{,} \PY{l+m+mi}{0}\PY{p}{,} \PY{l+m+mi}{0}\PY{p}{,} \PY{l+m+mi}{0}\PY{p}{]}\PY{p}{,}         \PY{c+c1}{\PYZsh{} Ctled Z gate}
                    \PY{p}{[}\PY{l+m+mi}{0}\PY{p}{,} \PY{l+m+mi}{1}\PY{p}{,} \PY{l+m+mi}{0}\PY{p}{,} \PY{l+m+mi}{0}\PY{p}{]}\PY{p}{,}
                    \PY{p}{[}\PY{l+m+mi}{0}\PY{p}{,} \PY{l+m+mi}{0}\PY{p}{,} \PY{l+m+mi}{1}\PY{p}{,} \PY{l+m+mi}{0}\PY{p}{]}\PY{p}{,}
                    \PY{p}{[}\PY{l+m+mi}{0}\PY{p}{,} \PY{l+m+mi}{0}\PY{p}{,} \PY{l+m+mi}{0}\PY{p}{,}\PY{o}{\PYZhy{}}\PY{l+m+mi}{1}\PY{p}{]}\PY{p}{]}\PY{p}{)}

\PY{n}{SWAP\PYZus{}gate} \PY{o}{=} \PY{n}{np}\PY{o}{.}\PY{n}{array}\PY{p}{(}\PY{p}{[}\PY{p}{[}\PY{l+m+mi}{1}\PY{p}{,} \PY{l+m+mi}{0}\PY{p}{,} \PY{l+m+mi}{0}\PY{p}{,} \PY{l+m+mi}{0}\PY{p}{]}\PY{p}{,}       \PY{c+c1}{\PYZsh{} Swap gate}
                      \PY{p}{[}\PY{l+m+mi}{0}\PY{p}{,} \PY{l+m+mi}{0}\PY{p}{,} \PY{l+m+mi}{1}\PY{p}{,} \PY{l+m+mi}{0}\PY{p}{]}\PY{p}{,}
                      \PY{p}{[}\PY{l+m+mi}{0}\PY{p}{,} \PY{l+m+mi}{1}\PY{p}{,} \PY{l+m+mi}{0}\PY{p}{,} \PY{l+m+mi}{0}\PY{p}{]}\PY{p}{,}
                      \PY{p}{[}\PY{l+m+mi}{0}\PY{p}{,} \PY{l+m+mi}{0}\PY{p}{,} \PY{l+m+mi}{0}\PY{p}{,} \PY{l+m+mi}{1}\PY{p}{]}\PY{p}{]}\PY{p}{)}

\PY{n}{TOFF\PYZus{}gate} \PY{o}{=} \PY{n}{np}\PY{o}{.}\PY{n}{array}\PY{p}{(}\PY{p}{[}\PY{p}{[}\PY{l+m+mi}{1}\PY{p}{,} \PY{l+m+mi}{0}\PY{p}{,} \PY{l+m+mi}{0}\PY{p}{,} \PY{l+m+mi}{0}\PY{p}{,} \PY{l+m+mi}{0}\PY{p}{,} \PY{l+m+mi}{0}\PY{p}{,} \PY{l+m+mi}{0}\PY{p}{,} \PY{l+m+mi}{0}\PY{p}{]}\PY{p}{,} \PY{c+c1}{\PYZsh{} Toffoli gate}
                      \PY{p}{[}\PY{l+m+mi}{0}\PY{p}{,} \PY{l+m+mi}{1}\PY{p}{,} \PY{l+m+mi}{0}\PY{p}{,} \PY{l+m+mi}{0}\PY{p}{,} \PY{l+m+mi}{0}\PY{p}{,} \PY{l+m+mi}{0}\PY{p}{,} \PY{l+m+mi}{0}\PY{p}{,} \PY{l+m+mi}{0}\PY{p}{]}\PY{p}{,}
                      \PY{p}{[}\PY{l+m+mi}{0}\PY{p}{,} \PY{l+m+mi}{0}\PY{p}{,} \PY{l+m+mi}{1}\PY{p}{,} \PY{l+m+mi}{0}\PY{p}{,} \PY{l+m+mi}{0}\PY{p}{,} \PY{l+m+mi}{0}\PY{p}{,} \PY{l+m+mi}{0}\PY{p}{,} \PY{l+m+mi}{0}\PY{p}{]}\PY{p}{,}
                      \PY{p}{[}\PY{l+m+mi}{0}\PY{p}{,} \PY{l+m+mi}{0}\PY{p}{,} \PY{l+m+mi}{0}\PY{p}{,} \PY{l+m+mi}{1}\PY{p}{,} \PY{l+m+mi}{0}\PY{p}{,} \PY{l+m+mi}{0}\PY{p}{,} \PY{l+m+mi}{0}\PY{p}{,} \PY{l+m+mi}{0}\PY{p}{]}\PY{p}{,}
                      \PY{p}{[}\PY{l+m+mi}{0}\PY{p}{,} \PY{l+m+mi}{0}\PY{p}{,} \PY{l+m+mi}{0}\PY{p}{,} \PY{l+m+mi}{0}\PY{p}{,} \PY{l+m+mi}{1}\PY{p}{,} \PY{l+m+mi}{0}\PY{p}{,} \PY{l+m+mi}{0}\PY{p}{,} \PY{l+m+mi}{0}\PY{p}{]}\PY{p}{,}
                      \PY{p}{[}\PY{l+m+mi}{0}\PY{p}{,} \PY{l+m+mi}{0}\PY{p}{,} \PY{l+m+mi}{0}\PY{p}{,} \PY{l+m+mi}{0}\PY{p}{,} \PY{l+m+mi}{0}\PY{p}{,} \PY{l+m+mi}{1}\PY{p}{,} \PY{l+m+mi}{0}\PY{p}{,} \PY{l+m+mi}{0}\PY{p}{]}\PY{p}{,}
                      \PY{p}{[}\PY{l+m+mi}{0}\PY{p}{,} \PY{l+m+mi}{0}\PY{p}{,} \PY{l+m+mi}{0}\PY{p}{,} \PY{l+m+mi}{0}\PY{p}{,} \PY{l+m+mi}{0}\PY{p}{,} \PY{l+m+mi}{0}\PY{p}{,} \PY{l+m+mi}{0}\PY{p}{,} \PY{l+m+mi}{1}\PY{p}{]}\PY{p}{,}
                      \PY{p}{[}\PY{l+m+mi}{0}\PY{p}{,} \PY{l+m+mi}{0}\PY{p}{,} \PY{l+m+mi}{0}\PY{p}{,} \PY{l+m+mi}{0}\PY{p}{,} \PY{l+m+mi}{0}\PY{p}{,} \PY{l+m+mi}{0}\PY{p}{,} \PY{l+m+mi}{1}\PY{p}{,} \PY{l+m+mi}{0}\PY{p}{]}\PY{p}{]}\PY{p}{)}
\end{Verbatim}
\end{tcolorbox}

    It's the gate operations that are applied to more than one qubit
(\(M \ge 2\)) that introduce \textbf{entanglement} and enable the
quantum computer to perform interesting calculations.

\subsection{Instruction 3: Move a Qubit to the Top of the
Stack}\label{instruction-3-move-a-qubit-to-the-top-of-the-stack}

Since we only apply gates to the top qubits in the stack, we need an
instruction that moves the \(K\)th qubit from the top to TOS (Top Of
Stack), and shifts the other \(K-1\) qubits down a step. This can be
achieved with numpy's \texttt{swapaxes} function, but first, we reshape
the workspace so that the second to last index corresponds to the qubit
we want to move to TOS.

    \begin{tcolorbox}[breakable, size=fbox, boxrule=1pt, pad at break*=1mm,colback=cellbackground, colframe=cellborder]
\prompt{In}{incolor}{63}{\boxspacing}
\begin{Verbatim}[commandchars=\\\{\}]
\PY{k}{def} \PY{n+nf}{tosQubit}\PY{p}{(}\PY{n}{k}\PY{p}{)}\PY{p}{:}
    \PY{k}{global} \PY{n}{workspace}
    \PY{k}{if} \PY{n}{k} \PY{o}{\PYZgt{}} \PY{l+m+mi}{1}\PY{p}{:}                              \PY{c+c1}{\PYZsh{} if non\PYZhy{}trivial}
        \PY{n}{workspace} \PY{o}{=} \PY{n}{np}\PY{o}{.}\PY{n}{reshape}\PY{p}{(}\PY{n}{workspace}\PY{p}{,}\PY{p}{(}\PY{o}{\PYZhy{}}\PY{l+m+mi}{1}\PY{p}{,}\PY{l+m+mi}{2}\PY{p}{,}\PY{l+m+mi}{2}\PY{o}{*}\PY{o}{*}\PY{p}{(}\PY{n}{k}\PY{o}{\PYZhy{}}\PY{l+m+mi}{1}\PY{p}{)}\PY{p}{)}\PY{p}{)} 
        \PY{n}{workspace} \PY{o}{=} \PY{n}{np}\PY{o}{.}\PY{n}{swapaxes}\PY{p}{(}\PY{n}{workspace}\PY{p}{,}\PY{o}{\PYZhy{}}\PY{l+m+mi}{2}\PY{p}{,}\PY{o}{\PYZhy{}}\PY{l+m+mi}{1}\PY{p}{)}
\end{Verbatim}
\end{tcolorbox}

    \paragraph{\texorpdfstring{Example of
\texttt{tosQubit}}{Example of tosQubit}}\label{example-of-tosqubit}

In the previous example, we used a SWAP gate to switch the positions of
two qubits in the stack. Now, we can do it with the
\texttt{tosQubit} instruction instead:

    \begin{tcolorbox}[breakable, size=fbox, boxrule=1pt, pad at break*=1mm,colback=cellbackground, colframe=cellborder]
\prompt{In}{incolor}{64}{\boxspacing}
\begin{Verbatim}[commandchars=\\\{\}]
\PY{n}{workspace} \PY{o}{=} \PY{n}{np}\PY{o}{.}\PY{n}{array}\PY{p}{(}\PY{p}{[}\PY{p}{[}\PY{l+m+mf}{1.}\PY{p}{]}\PY{p}{]}\PY{p}{)}
\PY{n}{pushQubit}\PY{p}{(}\PY{p}{[}\PY{l+m+mi}{1}\PY{p}{,}\PY{l+m+mi}{0}\PY{p}{]}\PY{p}{)}                                  
\PY{n}{pushQubit}\PY{p}{(}\PY{p}{[}\PY{l+m+mf}{0.6}\PY{p}{,}\PY{l+m+mf}{0.8}\PY{p}{]}\PY{p}{)}                              
\PY{n+nb}{print}\PY{p}{(}\PY{n}{workspace}\PY{p}{)}
\PY{n}{tosQubit}\PY{p}{(}\PY{l+m+mi}{2}\PY{p}{)}
\PY{n+nb}{print}\PY{p}{(}\PY{n}{workspace}\PY{p}{)}
\end{Verbatim}
\end{tcolorbox}

    \begin{Verbatim}[commandchars=\\\{\}]
[[0.6 0.8 0.  0. ]]
[[[0.6 0. ]
  [0.8 0. ]]]
    \end{Verbatim}

    In the second line of the output, the reshaping remains after
\texttt{tosQubit}. If we don't want this and prefer it in the form of a
row vector, we can instead write

    \begin{tcolorbox}[breakable, size=fbox, boxrule=1pt, pad at break*=1mm,colback=cellbackground, colframe=cellborder]
\prompt{In}{incolor}{65}{\boxspacing}
\begin{Verbatim}[commandchars=\\\{\}]
\PY{n+nb}{print}\PY{p}{(}\PY{n}{np}\PY{o}{.}\PY{n}{reshape}\PY{p}{(}\PY{n}{workspace}\PY{p}{,}\PY{p}{(}\PY{l+m+mi}{1}\PY{p}{,}\PY{o}{\PYZhy{}}\PY{l+m+mi}{1}\PY{p}{)}\PY{p}{)}\PY{p}{)}
\end{Verbatim}
\end{tcolorbox}

    \begin{Verbatim}[commandchars=\\\{\}]
[[0.6 0.  0.8 0. ]]
    \end{Verbatim}

    \subsection{Instruction 4: Measure a
Qubit}\label{instruction-4-measure-a-qubit}

When we measure a qubit, it's automatically popped from the stack.
Physicists call this \textbf{collapse}. We don't get a weight as a
result, but rather either zero or one, depending on the combined
probabilities of the small cubes that correspond to zero or one. To know
the probability of a 0, we need to sum up the probabilities for all the
small cubes where the corresponding coordinate is a 0. If \(\alpha\) is
the weight in the small cube, then \(|\alpha|^2\) is the probability for
that cube. We can calculate the total probability using numpy's
\texttt{linalg.norm} function. Then, we draw either a one or a zero
according to these probabilities. That's the final value of the popped
qubit, and this drawing is what's called \textbf{measurement}. After the
measurement, we discard all the small cubes that don't match the
measurement result. That's the actual \textbf{collapse}. The workspace
will then shrink to half its size. We have simply popped away one
dimension of the cube. We also normalize the remaining weights so that
the total probability becomes one.

    \begin{tcolorbox}[breakable, size=fbox, boxrule=1pt, pad at break*=1mm,colback=cellbackground, colframe=cellborder]
\prompt{In}{incolor}{66}{\boxspacing}
\begin{Verbatim}[commandchars=\\\{\}]
\PY{k}{def} \PY{n+nf}{probQubit}\PY{p}{(}\PY{p}{)}\PY{p}{:}
    \PY{k}{global} \PY{n}{workspace}
    \PY{n}{workspace} \PY{o}{=} \PY{n}{np}\PY{o}{.}\PY{n}{reshape}\PY{p}{(}\PY{n}{workspace}\PY{p}{,}\PY{p}{(}\PY{o}{\PYZhy{}}\PY{l+m+mi}{1}\PY{p}{,}\PY{l+m+mi}{2}\PY{p}{)}\PY{p}{)}
    \PY{k}{return} \PY{n}{np}\PY{o}{.}\PY{n}{linalg}\PY{o}{.}\PY{n}{norm}\PY{p}{(}\PY{n}{workspace}\PY{p}{,}\PY{n}{axis}\PY{o}{=}\PY{l+m+mi}{0}\PY{p}{)}\PY{o}{*}\PY{o}{*}\PY{l+m+mi}{2}

\PY{k}{def} \PY{n+nf}{measureQubit}\PY{p}{(}\PY{p}{)}\PY{p}{:}
    \PY{k}{global} \PY{n}{workspace}
    \PY{n}{prob} \PY{o}{=} \PY{n}{probQubit}\PY{p}{(}\PY{p}{)}
    \PY{n}{measurement} \PY{o}{=} \PY{n}{np}\PY{o}{.}\PY{n}{random}\PY{o}{.}\PY{n}{choice}\PY{p}{(}\PY{l+m+mi}{2}\PY{p}{,}\PY{n}{p}\PY{o}{=}\PY{n}{prob}\PY{p}{)}  \PY{c+c1}{\PYZsh{} select 0 or 1 }
    \PY{n}{workspace} \PY{o}{=} {(}\PY{n}{workspace}\PY{p}{[}\PY{p}{:}\PY{p}{,}\PY{p}{[}\PY{n}{measurement}\PY{p}{]}\PY{p}{]}\PY{o}{/}
                 \PY{n}{np}\PY{o}{.}\PY{n}{sqrt}\PY{p}{(}\PY{n}{prob}\PY{p}{[}\PY{n}{measurement}\PY{p}{]}\PY{p}{)}{)}
    \PY{k}{return} \PY{n+nb}{str}\PY{p}{(}\PY{n}{measurement}\PY{p}{)}
\end{Verbatim}
\end{tcolorbox}

    The function \texttt{probQubit} calculates the probability for the qubit
on top of the stack to be zero or one by computing the sum of squares of
the corresponding columns in the workspace. It is called by the function
\texttt{measureQubit}, which randomly draws a zero or one according to
these probabilities using \texttt{random.choice}. Finally, the
corresponding column is extracted and normalized to become the new
workspace.

\paragraph{\texorpdfstring{Example of
\texttt{measureQubit}}{Example of measureQubit}}\label{example-of-measurequbit}

In the next example, we push a qubit that has the value zero with a
probability of \(0.6^2=0.36\) and the value one with a probability of
\(0.8^2=0.64\). The odds \(36:64=9:16\) mean that if we run the program
25 times, we'll get approximately 9 zeros and 16 ones:

    \begin{tcolorbox}[breakable, size=fbox, boxrule=1pt, pad at break*=1mm,colback=cellbackground, colframe=cellborder]
\prompt{In}{incolor}{67}{\boxspacing}
\begin{Verbatim}[commandchars=\\\{\}]
\PY{n}{workspace} \PY{o}{=} \PY{n}{np}\PY{o}{.}\PY{n}{array}\PY{p}{(}\PY{p}{[}\PY{p}{[}\PY{l+m+mf}{1.}\PY{p}{]}\PY{p}{]}\PY{p}{)}
\PY{k}{for} \PY{n}{n} \PY{o+ow}{in} \PY{n+nb}{range}\PY{p}{(}\PY{l+m+mi}{30}\PY{p}{)}\PY{p}{:}
    \PY{n}{pushQubit}\PY{p}{(}\PY{p}{[}\PY{l+m+mf}{0.6}\PY{p}{,}\PY{l+m+mf}{0.8}\PY{p}{]}\PY{p}{)}
    \PY{n+nb}{print}\PY{p}{(}\PY{n}{measureQubit}\PY{p}{(}\PY{p}{)}\PY{p}{,} \PY{n}{end}\PY{o}{=}\PY{l+s+s2}{\PYZdq{}}\PY{l+s+s2}{\PYZdq{}}\PY{p}{)}
\end{Verbatim}
\end{tcolorbox}

    \begin{Verbatim}[commandchars=\\\{\}]
001010110010111001101111110110
    \end{Verbatim}

    To be precise, all possible values for the qubit are calculated, but we
can only read off one value for each qubit per run, because as soon as
we read it, the representation of the bit collapses to that value.

In a slightly more advanced example with three qubits, we combine two
Hadamard gates with a Toffoli gate. This combination computes AND, and
to demonstrate this, we generate a truth table:

    \begin{tcolorbox}[breakable, size=fbox, boxrule=1pt, pad at break*=1mm,colback=cellbackground, colframe=cellborder]
\prompt{In}{incolor}{68}{\boxspacing}
\begin{Verbatim}[commandchars=\\\{\}]
\PY{n}{workspace} \PY{o}{=} \PY{n}{np}\PY{o}{.}\PY{n}{array}\PY{p}{(}\PY{p}{[}\PY{p}{[}\PY{l+m+mf}{1.}\PY{p}{]}\PY{p}{]}\PY{p}{)}
\PY{k}{for} \PY{n}{i} \PY{o+ow}{in} \PY{n+nb}{range}\PY{p}{(}\PY{l+m+mi}{16}\PY{p}{)}\PY{p}{:}
    \PY{n}{pushQubit}\PY{p}{(}\PY{p}{[}\PY{l+m+mi}{1}\PY{p}{,}\PY{l+m+mi}{0}\PY{p}{]}\PY{p}{)}          \PY{c+c1}{\PYZsh{} push a zero qubit}
    \PY{n}{applyGate}\PY{p}{(}\PY{n}{H\PYZus{}gate}\PY{p}{)}         \PY{c+c1}{\PYZsh{} set equal 0 and 1 probability}
    \PY{n}{pushQubit}\PY{p}{(}\PY{p}{[}\PY{l+m+mi}{1}\PY{p}{,}\PY{l+m+mi}{0}\PY{p}{]}\PY{p}{)}          \PY{c+c1}{\PYZsh{} push a 2nd zero qubit}
    \PY{n}{applyGate}\PY{p}{(}\PY{n}{H\PYZus{}gate}\PY{p}{)}         \PY{c+c1}{\PYZsh{} set equal 0 and 1 probability}
    \PY{n}{pushQubit}\PY{p}{(}\PY{p}{[}\PY{l+m+mi}{1}\PY{p}{,}\PY{l+m+mi}{0}\PY{p}{]}\PY{p}{)}          \PY{c+c1}{\PYZsh{} push a dummy zero qubit}
    \PY{n}{applyGate}\PY{p}{(}\PY{n}{TOFF\PYZus{}gate}\PY{p}{)}      \PY{c+c1}{\PYZsh{} compute Q3 = Q1 AND Q2}
    \PY{n}{q3} \PY{o}{=} \PY{n}{measureQubit}\PY{p}{(}\PY{p}{)}       \PY{c+c1}{\PYZsh{} pop qubit 3}
    \PY{n}{q2} \PY{o}{=} \PY{n}{measureQubit}\PY{p}{(}\PY{p}{)}       \PY{c+c1}{\PYZsh{} pop qubit 2}
    \PY{n}{q1} \PY{o}{=} \PY{n}{measureQubit}\PY{p}{(}\PY{p}{)}       \PY{c+c1}{\PYZsh{} pop qubit 1}
    \PY{n+nb}{print}\PY{p}{(}\PY{n}{q1}\PY{o}{+}\PY{n}{q2}\PY{o}{+}\PY{n}{q3}\PY{p}{,}\PY{n}{end}\PY{o}{=}\PY{l+s+s2}{\PYZdq{}}\PY{l+s+s2}{,}\PY{l+s+s2}{\PYZdq{}}\PY{p}{)}
\end{Verbatim}
\end{tcolorbox}

    \begin{Verbatim}[commandchars=\\\{\}]
000,010,000,100,000,100,111,010,100,010,100,000,000,100,010,010,
    \end{Verbatim}

    What's happening here is that even though the code calculates qubit 3
for all values of qubits 1 and 2 each loop iteration, we still need to
rerun the program to get the full table. The big problem with quantum
computers isn't doing calculations; it's getting the results out!

This is it---a quantum computer isn't really any more complicated than
this. The instructions \texttt{pushQubit}, \texttt{applyGate},
\texttt{tosQubit}, and \texttt{measureQubit} can do everything a quantum
computer can do, but of course, they take much (exponentially) more
space and time. In the next section, we'll improve the code and try some
more advanced examples.

\section{Some Code Improvements}\label{some-code-improvements}

We can make some improvements to the code to make it easier to use. Most
importantly, we want to be able to refer to qubits by names. So, we
introduce a name stack, which is a classic stack for the names of the
qubits.

\paragraph{\texorpdfstring{Improved
\texttt{pushQubit}}{Improved pushQubit}}\label{improved-pushqubit}

For \texttt{pushQubit}, we also normalize the weights inside the
procedure so that the user doesn't have to think about it. We take the
opportunity to reset the name stack if we see that the workspace is
empty. When we push the qubit, we ensure that the weights are converted
to the same type as \texttt{workspace} to prevent \texttt{kron} from
using up unnecessary memory.

    \begin{tcolorbox}[breakable, size=fbox, boxrule=1pt, pad at break*=1mm,colback=cellbackground, colframe=cellborder]
\prompt{In}{incolor}{69}{\boxspacing}
\begin{Verbatim}[commandchars=\\\{\}]
\PY{k}{def} \PY{n+nf}{pushQubit}\PY{p}{(}\PY{n}{name}\PY{p}{,}\PY{n}{weights}\PY{p}{)}\PY{p}{:}
    \PY{k}{global} \PY{n}{workspace}
    \PY{k}{global} \PY{n}{namestack}
    \PY{k}{if} \PY{n}{workspace}\PY{o}{.}\PY{n}{shape} \PY{o}{==} \PY{p}{(}\PY{l+m+mi}{1}\PY{p}{,}\PY{l+m+mi}{1}\PY{p}{)}\PY{p}{:}        \PY{c+c1}{\PYZsh{} if workspace empty}
        \PY{n}{namestack} \PY{o}{=} \PY{p}{[}\PY{p}{]}                  \PY{c+c1}{\PYZsh{}   then reset}
    \PY{n}{namestack}\PY{o}{.}\PY{n}{append}\PY{p}{(}\PY{n}{name}\PY{p}{)}              \PY{c+c1}{\PYZsh{} push name}
    \PY{n}{weights} \PY{o}{=} \PY{n}{weights}\PY{o}{/}\PY{n}{np}\PY{o}{.}\PY{n}{linalg}\PY{o}{.}\PY{n}{norm}\PY{p}{(}\PY{n}{weights}\PY{p}{)} \PY{c+c1}{\PYZsh{} normalize}
    \PY{n}{weights} \PY{o}{=} \PY{n}{np}\PY{o}{.}\PY{n}{array}\PY{p}{(}\PY{n}{weights}\PY{p}{,}\PY{n}{dtype}\PY{o}{=}\PY{n}{workspace}\PY{p}{[}\PY{l+m+mi}{0}\PY{p}{,}\PY{l+m+mi}{0}\PY{p}{]}\PY{o}{.}\PY{n}{dtype}\PY{p}{)}
    \PY{n}{workspace} \PY{o}{=} \PY{n}{np}\PY{o}{.}\PY{n}{reshape}\PY{p}{(}\PY{n}{workspace}\PY{p}{,}\PY{p}{(}\PY{l+m+mi}{1}\PY{p}{,}\PY{o}{\PYZhy{}}\PY{l+m+mi}{1}\PY{p}{)}\PY{p}{)} \PY{c+c1}{\PYZsh{} to row vector}
    \PY{n}{workspace} \PY{o}{=} \PY{n}{np}\PY{o}{.}\PY{n}{kron}\PY{p}{(}\PY{n}{workspace}\PY{p}{,}\PY{n}{weights}\PY{p}{)}        
\end{Verbatim}
\end{tcolorbox}

    \paragraph{\texorpdfstring{Example of
\texttt{pushQubit}}{Example of pushQubit}}\label{example-of-pushqubit}

A simple example:

    \begin{tcolorbox}[breakable, size=fbox, boxrule=1pt, pad at break*=1mm,colback=cellbackground, colframe=cellborder]
\prompt{In}{incolor}{70}{\boxspacing}
\begin{Verbatim}[commandchars=\\\{\}]
\PY{n}{workspace} \PY{o}{=} \PY{n}{np}\PY{o}{.}\PY{n}{array}\PY{p}{(}\PY{p}{[}\PY{p}{[}\PY{l+m+mf}{1.}\PY{p}{]}\PY{p}{]}\PY{p}{)}        \PY{c+c1}{\PYZsh{} create empty qubit stack}
\PY{n}{pushQubit}\PY{p}{(}\PY{l+s+s2}{\PYZdq{}}\PY{l+s+s2}{Q1}\PY{l+s+s2}{\PYZdq{}}\PY{p}{,}\PY{p}{[}\PY{l+m+mi}{1}\PY{p}{,}\PY{l+m+mi}{1}\PY{p}{]}\PY{p}{)}               \PY{c+c1}{\PYZsh{} push a qubit}
\PY{n+nb}{print}\PY{p}{(}\PY{n}{np}\PY{o}{.}\PY{n}{reshape}\PY{p}{(}\PY{n}{workspace}\PY{p}{,}\PY{p}{(}\PY{l+m+mi}{1}\PY{p}{,}\PY{o}{\PYZhy{}}\PY{l+m+mi}{1}\PY{p}{)}\PY{p}{)}\PY{p}{)} \PY{c+c1}{\PYZsh{} print workspace as vector}
\PY{n+nb}{print}\PY{p}{(}\PY{n}{namestack}\PY{p}{)}
\PY{n}{pushQubit}\PY{p}{(}\PY{l+s+s2}{\PYZdq{}}\PY{l+s+s2}{Q2}\PY{l+s+s2}{\PYZdq{}}\PY{p}{,}\PY{p}{[}\PY{l+m+mi}{0}\PY{p}{,}\PY{l+m+mi}{1}\PY{p}{]}\PY{p}{)}               \PY{c+c1}{\PYZsh{} push a 2nd qubit}
\PY{n+nb}{print}\PY{p}{(}\PY{n}{np}\PY{o}{.}\PY{n}{reshape}\PY{p}{(}\PY{n}{workspace}\PY{p}{,}\PY{p}{(}\PY{l+m+mi}{1}\PY{p}{,}\PY{o}{\PYZhy{}}\PY{l+m+mi}{1}\PY{p}{)}\PY{p}{)}\PY{p}{)} \PY{c+c1}{\PYZsh{} print workspace as vector}
\PY{n+nb}{print}\PY{p}{(}\PY{n}{namestack}\PY{p}{)}
\end{Verbatim}
\end{tcolorbox}

    \begin{Verbatim}[commandchars=\\\{\}]
[[0.70710678 0.70710678]]
['Q1']
[[0.         0.70710678 0.         0.70710678]]
['Q1', 'Q2']
    \end{Verbatim}

    Here, we've initialized qubit Q1 with weights \texttt{{[}1,1{]}} instead
of first initializing with \texttt{{[}1,0{]}} and then applying a
Hadamard gate. The weights are normalized inside \texttt{pushQubit}, so
the result is identical.

\paragraph{\texorpdfstring{Improved
\texttt{tosQubit}}{Improved tosQubit}}\label{improved-tosqubit}

For \texttt{tosQubit}, we only need to calculate how many qubits need to
be moved, and also rotate the name stack. Negative indices in Python
mean counting from the end of the list, so minus one refers to the last
element in the list.

    \begin{tcolorbox}[breakable, size=fbox, boxrule=1pt, pad at break*=1mm,colback=cellbackground, colframe=cellborder]
\prompt{In}{incolor}{71}{\boxspacing}
\begin{Verbatim}[commandchars=\\\{\}]
\PY{k}{def} \PY{n+nf}{tosQubit}\PY{p}{(}\PY{n}{name}\PY{p}{)}\PY{p}{:}
    \PY{k}{global} \PY{n}{workspace}
    \PY{k}{global} \PY{n}{namestack}
    \PY{n}{k} \PY{o}{=} \PY{n+nb}{len}\PY{p}{(}\PY{n}{namestack}\PY{p}{)}\PY{o}{\PYZhy{}}\PY{n}{namestack}\PY{o}{.}\PY{n}{index}\PY{p}{(}\PY{n}{name}\PY{p}{)}\PY{c+c1}{\PYZsh{} qubit pos}
    \PY{k}{if} \PY{n}{k} \PY{o}{\PYZgt{}} \PY{l+m+mi}{1}\PY{p}{:}                               \PY{c+c1}{\PYZsh{} if non\PYZhy{}trivial}
        \PY{n}{namestack}\PY{o}{.}\PY{n}{append}\PY{p}{(}\PY{n}{namestack}\PY{o}{.}\PY{n}{pop}\PY{p}{(}\PY{o}{\PYZhy{}}\PY{n}{k}\PY{p}{)}\PY{p}{)} \PY{c+c1}{\PYZsh{} rotate name stack}
        \PY{n}{workspace} \PY{o}{=} \PY{n}{np}\PY{o}{.}\PY{n}{reshape}\PY{p}{(}\PY{n}{workspace}\PY{p}{,}\PY{p}{(}\PY{o}{\PYZhy{}}\PY{l+m+mi}{1}\PY{p}{,}\PY{l+m+mi}{2}\PY{p}{,}\PY{l+m+mi}{2}\PY{o}{*}\PY{o}{*}\PY{p}{(}\PY{n}{k}\PY{o}{\PYZhy{}}\PY{l+m+mi}{1}\PY{p}{)}\PY{p}{)}\PY{p}{)} 
        \PY{n}{workspace} \PY{o}{=} \PY{n}{np}\PY{o}{.}\PY{n}{swapaxes}\PY{p}{(}\PY{n}{workspace}\PY{p}{,}\PY{o}{\PYZhy{}}\PY{l+m+mi}{2}\PY{p}{,}\PY{o}{\PYZhy{}}\PY{l+m+mi}{1}\PY{p}{)}
\end{Verbatim}
\end{tcolorbox}

    \paragraph{\texorpdfstring{Example of
\texttt{tosQubit}}{Example of tosQubit}}\label{example-of-tosqubit-2}

We continue the previous example:

    \begin{tcolorbox}[breakable, size=fbox, boxrule=1pt, pad at break*=1mm,colback=cellbackground, colframe=cellborder]
\prompt{In}{incolor}{72}{\boxspacing}
\begin{Verbatim}[commandchars=\\\{\}]
\PY{n+nb}{print}\PY{p}{(}\PY{n}{np}\PY{o}{.}\PY{n}{reshape}\PY{p}{(}\PY{n}{workspace}\PY{p}{,}\PY{p}{(}\PY{l+m+mi}{1}\PY{p}{,}\PY{o}{\PYZhy{}}\PY{l+m+mi}{1}\PY{p}{)}\PY{p}{)}\PY{p}{)}  \PY{c+c1}{\PYZsh{} print workspace as vector}
\PY{n+nb}{print}\PY{p}{(}\PY{n}{namestack}\PY{p}{)}
\PY{n}{tosQubit}\PY{p}{(}\PY{l+s+s2}{\PYZdq{}}\PY{l+s+s2}{Q1}\PY{l+s+s2}{\PYZdq{}}\PY{p}{)}                       \PY{c+c1}{\PYZsh{} swap qubits}
\PY{n+nb}{print}\PY{p}{(}\PY{n}{np}\PY{o}{.}\PY{n}{reshape}\PY{p}{(}\PY{n}{workspace}\PY{p}{,}\PY{p}{(}\PY{l+m+mi}{1}\PY{p}{,}\PY{o}{\PYZhy{}}\PY{l+m+mi}{1}\PY{p}{)}\PY{p}{)}\PY{p}{)}  \PY{c+c1}{\PYZsh{} print workspace as vector}
\PY{n+nb}{print}\PY{p}{(}\PY{n}{namestack}\PY{p}{)}
\end{Verbatim}
\end{tcolorbox}

    \begin{Verbatim}[commandchars=\\\{\}]
[[0.         0.70710678 0.         0.70710678]]
['Q1', 'Q2']
[[0.         0.         0.70710678 0.70710678]]
['Q2', 'Q1']
    \end{Verbatim}

    \paragraph{\texorpdfstring{Improved
\texttt{applyGate}}{Improved applyGate}}\label{improved-applygate}

We allow the procedure to take the names of the qubits that are to be
moved to the top of the stack as arguments, so this move can be built
into the gate operation.

    \begin{tcolorbox}[breakable, size=fbox, boxrule=1pt, pad at break*=1mm,colback=cellbackground, colframe=cellborder]
\prompt{In}{incolor}{73}{\boxspacing}
\begin{Verbatim}[commandchars=\\\{\}]
\PY{k}{def} \PY{n+nf}{applyGate}\PY{p}{(}\PY{n}{gate}\PY{p}{,}\PY{o}{*}\PY{n}{names}\PY{p}{)}\PY{p}{:}
    \PY{k}{global} \PY{n}{workspace}
    \PY{k}{for} \PY{n}{name} \PY{o+ow}{in} \PY{n}{names}\PY{p}{:}              \PY{c+c1}{\PYZsh{} move qubits to TOS}
        \PY{n}{tosQubit}\PY{p}{(}\PY{n}{name}\PY{p}{)}             
    \PY{n}{workspace} \PY{o}{=} \PY{n}{np}\PY{o}{.}\PY{n}{reshape}\PY{p}{(}\PY{n}{workspace}\PY{p}{,}\PY{p}{(}\PY{o}{\PYZhy{}}\PY{l+m+mi}{1}\PY{p}{,}\PY{n}{gate}\PY{o}{.}\PY{n}{shape}\PY{p}{[}\PY{l+m+mi}{0}\PY{p}{]}\PY{p}{)}\PY{p}{)} 
    \PY{n}{np}\PY{o}{.}\PY{n}{matmul}\PY{p}{(}\PY{n}{workspace}\PY{p}{,}\PY{n}{gate}\PY{o}{.}\PY{n}{T}\PY{p}{,}\PY{n}{out}\PY{o}{=}\PY{n}{workspace}\PY{p}{)} 
\end{Verbatim}
\end{tcolorbox}

    \paragraph{\texorpdfstring{Example of
\texttt{applyGate}}{Example of applyGate}}\label{example-of-applygate-2}

Let's continue with the latest example:

    \begin{tcolorbox}[breakable, size=fbox, boxrule=1pt, pad at break*=1mm,colback=cellbackground, colframe=cellborder]
\prompt{In}{incolor}{74}{\boxspacing}
\begin{Verbatim}[commandchars=\\\{\}]
\PY{n+nb}{print}\PY{p}{(}\PY{n}{np}\PY{o}{.}\PY{n}{reshape}\PY{p}{(}\PY{n}{workspace}\PY{p}{,}\PY{p}{(}\PY{l+m+mi}{1}\PY{p}{,}\PY{o}{\PYZhy{}}\PY{l+m+mi}{1}\PY{p}{)}\PY{p}{)}\PY{p}{)} \PY{c+c1}{\PYZsh{} print workspace as vector}
\PY{n+nb}{print}\PY{p}{(}\PY{n}{namestack}\PY{p}{)}
\PY{n}{applyGate}\PY{p}{(}\PY{n}{H\PYZus{}gate}\PY{p}{,}\PY{l+s+s2}{\PYZdq{}}\PY{l+s+s2}{Q2}\PY{l+s+s2}{\PYZdq{}}\PY{p}{)}              \PY{c+c1}{\PYZsh{} H gate on qubit 2}
\PY{n+nb}{print}\PY{p}{(}\PY{n}{np}\PY{o}{.}\PY{n}{reshape}\PY{p}{(}\PY{n}{workspace}\PY{p}{,}\PY{p}{(}\PY{l+m+mi}{1}\PY{p}{,}\PY{o}{\PYZhy{}}\PY{l+m+mi}{1}\PY{p}{)}\PY{p}{)}\PY{p}{)} \PY{c+c1}{\PYZsh{}  turns a 0 qubit to 1}
\PY{n+nb}{print}\PY{p}{(}\PY{n}{namestack}\PY{p}{)}                    \PY{c+c1}{\PYZsh{}  with 50\PYZpc{} probability}
\end{Verbatim}
\end{tcolorbox}

    \begin{Verbatim}[commandchars=\\\{\}]
[[0.         0.         0.70710678 0.70710678]]
['Q2', 'Q1']
[[ 0.5 -0.5  0.5 -0.5]]
['Q1', 'Q2']
    \end{Verbatim}

    \paragraph{\texorpdfstring{Improved
\texttt{measureQubit}}{Improved measureQubit}}\label{improved-measurequbit}

There are no major changes needed for \texttt{measureQubit}, other than
that we now provide the name of the qubit to be measured as an argument.

    \begin{tcolorbox}[breakable, size=fbox, boxrule=1pt, pad at break*=1mm,colback=cellbackground, colframe=cellborder]
\prompt{In}{incolor}{75}{\boxspacing}
\begin{Verbatim}[commandchars=\\\{\}]
\PY{k}{def} \PY{n+nf}{probQubit}\PY{p}{(}\PY{n}{name}\PY{p}{)}\PY{p}{:}                  
    \PY{k}{global} \PY{n}{workspace}                
    \PY{n}{tosQubit}\PY{p}{(}\PY{n}{name}\PY{p}{)}         
    \PY{n}{workspace} \PY{o}{=} \PY{n}{np}\PY{o}{.}\PY{n}{reshape}\PY{p}{(}\PY{n}{workspace}\PY{p}{,}\PY{p}{(}\PY{o}{\PYZhy{}}\PY{l+m+mi}{1}\PY{p}{,}\PY{l+m+mi}{2}\PY{p}{)}\PY{p}{)}  
    \PY{n}{prob} \PY{o}{=} \PY{n}{np}\PY{o}{.}\PY{n}{linalg}\PY{o}{.}\PY{n}{norm}\PY{p}{(}\PY{n}{workspace}\PY{p}{,}\PY{n}{axis}\PY{o}{=}\PY{l+m+mi}{0}\PY{p}{)}\PY{o}{*}\PY{o}{*}\PY{l+m+mi}{2} 
    \PY{k}{return} \PY{n}{prob}\PY{o}{/}\PY{n}{prob}\PY{o}{.}\PY{n}{sum}\PY{p}{(}\PY{p}{)}       \PY{c+c1}{\PYZsh{} make sure sum is one}

\PY{k}{def} \PY{n+nf}{measureQubit}\PY{p}{(}\PY{n}{name}\PY{p}{)}\PY{p}{:} 
    \PY{k}{global} \PY{n}{workspace}
    \PY{k}{global} \PY{n}{namestack}
    \PY{n}{prob} \PY{o}{=} \PY{n}{probQubit}\PY{p}{(}\PY{n}{name}\PY{p}{)}
    \PY{n}{measurement} \PY{o}{=} \PY{n}{np}\PY{o}{.}\PY{n}{random}\PY{o}{.}\PY{n}{choice}\PY{p}{(}\PY{l+m+mi}{2}\PY{p}{,}\PY{n}{p}\PY{o}{=}\PY{n}{prob}\PY{p}{)}      
    \PY{n}{workspace} \PY{o}{=} {(}\PY{n}{workspace}\PY{p}{[}\PY{p}{:}\PY{p}{,}\PY{p}{[}\PY{n}{measurement}\PY{p}{]}\PY{p}{]}\PY{o}{/}
                 \PY{n}{np}\PY{o}{.}\PY{n}{sqrt}\PY{p}{(}\PY{n}{prob}\PY{p}{[}\PY{n}{measurement}\PY{p}{]}\PY{p}{)}{)}
    \PY{n}{namestack}\PY{o}{.}\PY{n}{pop}\PY{p}{(}\PY{p}{)}                   
    \PY{k}{return} \PY{n+nb}{str}\PY{p}{(}\PY{n}{measurement}\PY{p}{)}
\end{Verbatim}
\end{tcolorbox}

    \paragraph{\texorpdfstring{Examples of \texttt{probQubit}and
\texttt{measureQubit}}{Examples of probQubitand measureQubit}}\label{examples-of-probqubitand-measurequbit}

First a simple example:

    \begin{tcolorbox}[breakable, size=fbox, boxrule=1pt, pad at break*=1mm,colback=cellbackground, colframe=cellborder]
\prompt{In}{incolor}{76}{\boxspacing}
\begin{Verbatim}[commandchars=\\\{\}]
\PY{n}{workspace} \PY{o}{=} \PY{n}{np}\PY{o}{.}\PY{n}{array}\PY{p}{(}\PY{p}{[}\PY{p}{[}\PY{l+m+mf}{1.}\PY{p}{]}\PY{p}{]}\PY{p}{)}    
\PY{n}{pushQubit}\PY{p}{(}\PY{l+s+s2}{\PYZdq{}}\PY{l+s+s2}{Q1}\PY{l+s+s2}{\PYZdq{}}\PY{p}{,}\PY{p}{[}\PY{l+m+mi}{1}\PY{p}{,}\PY{l+m+mi}{0}\PY{p}{]}\PY{p}{)}
\PY{n}{applyGate}\PY{p}{(}\PY{n}{H\PYZus{}gate}\PY{p}{,}\PY{l+s+s2}{\PYZdq{}}\PY{l+s+s2}{Q1}\PY{l+s+s2}{\PYZdq{}}\PY{p}{)}
\PY{n+nb}{print}\PY{p}{(}\PY{l+s+s2}{\PYZdq{}}\PY{l+s+s2}{Q1 probabilities:}\PY{l+s+s2}{\PYZdq{}}\PY{p}{,} \PY{n}{probQubit}\PY{p}{(}\PY{l+s+s2}{\PYZdq{}}\PY{l+s+s2}{Q1}\PY{l+s+s2}{\PYZdq{}}\PY{p}{)}\PY{p}{)} \PY{c+c1}{\PYZsh{} peek Q1 prob}
\PY{n}{pushQubit}\PY{p}{(}\PY{l+s+s2}{\PYZdq{}}\PY{l+s+s2}{Q2}\PY{l+s+s2}{\PYZdq{}}\PY{p}{,}\PY{p}{[}\PY{l+m+mf}{0.6}\PY{p}{,}\PY{l+m+mf}{0.8}\PY{p}{]}\PY{p}{)}
\PY{n+nb}{print}\PY{p}{(}\PY{l+s+s2}{\PYZdq{}}\PY{l+s+s2}{Q2 probabilities:}\PY{l+s+s2}{\PYZdq{}}\PY{p}{,} \PY{n}{probQubit}\PY{p}{(}\PY{l+s+s2}{\PYZdq{}}\PY{l+s+s2}{Q2}\PY{l+s+s2}{\PYZdq{}}\PY{p}{)}\PY{p}{)} \PY{c+c1}{\PYZsh{} peek Q2 prob}
\PY{n+nb}{print}\PY{p}{(}\PY{n}{measureQubit}\PY{p}{(}\PY{l+s+s2}{\PYZdq{}}\PY{l+s+s2}{Q1}\PY{l+s+s2}{\PYZdq{}}\PY{p}{)}\PY{p}{,} \PY{n}{measureQubit}\PY{p}{(}\PY{l+s+s2}{\PYZdq{}}\PY{l+s+s2}{Q2}\PY{l+s+s2}{\PYZdq{}}\PY{p}{)}\PY{p}{)}
\end{Verbatim}
\end{tcolorbox}

    \begin{Verbatim}[commandchars=\\\{\}]
Q1 probabilities: [0.5 0.5]
Q2 probabilities: [0.36 0.64]
0 1
    \end{Verbatim}

    The function \texttt{probQubit} is impossible in a physical quantum
computer because the probabilities collapse as soon as you try to
measure them. However, the function is very useful for debugging in a
quantum computer simulator.

Here's a more complex example that demonstrates how a Toffoli gate can
be implemented with Hadamard, T, inverse T, and CNOT gates. How it works
is quite baffling, but the fact that it works is interesting because it
shows that it is \emph{possible}:

    \begin{tcolorbox}[breakable, size=fbox, boxrule=1pt, pad at break*=1mm,colback=cellbackground, colframe=cellborder]
\prompt{In}{incolor}{77}{\boxspacing}
\begin{Verbatim}[commandchars=\\\{\}]
\PY{k}{def} \PY{n+nf}{toffEquiv\PYZus{}gate}\PY{p}{(}\PY{n}{q1}\PY{p}{,}\PY{n}{q2}\PY{p}{,}\PY{n}{q3}\PY{p}{)}\PY{p}{:}           \PY{c+c1}{\PYZsh{} define Toffoli gate}
    \PY{n}{applyGate}\PY{p}{(}\PY{n}{H\PYZus{}gate}\PY{p}{,}\PY{n}{q3}\PY{p}{)}                \PY{c+c1}{\PYZsh{} using H, T, T*, CNOT}
    \PY{n}{applyGate}\PY{p}{(}\PY{n}{CNOT\PYZus{}gate}\PY{p}{,}\PY{n}{q2}\PY{p}{,}\PY{n}{q3}\PY{p}{)}
    \PY{n}{applyGate}\PY{p}{(}\PY{n}{Tinv\PYZus{}gate}\PY{p}{,}\PY{n}{q3}\PY{p}{)}
    \PY{n}{applyGate}\PY{p}{(}\PY{n}{CNOT\PYZus{}gate}\PY{p}{,}\PY{n}{q1}\PY{p}{,}\PY{n}{q3}\PY{p}{)}    
    \PY{n}{applyGate}\PY{p}{(}\PY{n}{T\PYZus{}gate}\PY{p}{,}\PY{n}{q3}\PY{p}{)}
    \PY{n}{applyGate}\PY{p}{(}\PY{n}{CNOT\PYZus{}gate}\PY{p}{,}\PY{n}{q2}\PY{p}{,}\PY{n}{q3}\PY{p}{)}     
    \PY{n}{applyGate}\PY{p}{(}\PY{n}{Tinv\PYZus{}gate}\PY{p}{,}\PY{n}{q3}\PY{p}{)}
    \PY{n}{applyGate}\PY{p}{(}\PY{n}{CNOT\PYZus{}gate}\PY{p}{,}\PY{n}{q1}\PY{p}{,}\PY{n}{q3}\PY{p}{)} 
    \PY{n}{applyGate}\PY{p}{(}\PY{n}{T\PYZus{}gate}\PY{p}{,}\PY{n}{q2}\PY{p}{)}
    \PY{n}{applyGate}\PY{p}{(}\PY{n}{T\PYZus{}gate}\PY{p}{,}\PY{n}{q3}\PY{p}{)}
    \PY{n}{applyGate}\PY{p}{(}\PY{n}{H\PYZus{}gate}\PY{p}{,}\PY{n}{q3}\PY{p}{)}
    \PY{n}{applyGate}\PY{p}{(}\PY{n}{CNOT\PYZus{}gate}\PY{p}{,}\PY{n}{q1}\PY{p}{,}\PY{n}{q2}\PY{p}{)}
    \PY{n}{applyGate}\PY{p}{(}\PY{n}{T\PYZus{}gate}\PY{p}{,}\PY{n}{q1}\PY{p}{)}
    \PY{n}{applyGate}\PY{p}{(}\PY{n}{Tinv\PYZus{}gate}\PY{p}{,}\PY{n}{q2}\PY{p}{)}
    \PY{n}{applyGate}\PY{p}{(}\PY{n}{CNOT\PYZus{}gate}\PY{p}{,}\PY{n}{q1}\PY{p}{,}\PY{n}{q2}\PY{p}{)}
    
\PY{n}{workspace} \PY{o}{=} \PY{n}{np}\PY{o}{.}\PY{n}{array}\PY{p}{(}\PY{p}{[}\PY{p}{[}\PY{l+m+mf}{1.}\PY{o}{+}\PY{l+m+mi}{0}\PY{n}{j}\PY{p}{]}\PY{p}{]}\PY{p}{)}         \PY{c+c1}{\PYZsh{} prep COMPLEX array}
\PY{k}{for} \PY{n}{i} \PY{o+ow}{in} \PY{n+nb}{range}\PY{p}{(}\PY{l+m+mi}{16}\PY{p}{)}\PY{p}{:}                     \PY{c+c1}{\PYZsh{} test function}
    \PY{n}{pushQubit}\PY{p}{(}\PY{l+s+s2}{\PYZdq{}}\PY{l+s+s2}{Q1}\PY{l+s+s2}{\PYZdq{}}\PY{p}{,}\PY{p}{[}\PY{l+m+mi}{1}\PY{p}{,}\PY{l+m+mi}{1}\PY{p}{]}\PY{p}{)}               \PY{c+c1}{\PYZsh{} and print truth table}
    \PY{n}{pushQubit}\PY{p}{(}\PY{l+s+s2}{\PYZdq{}}\PY{l+s+s2}{Q2}\PY{l+s+s2}{\PYZdq{}}\PY{p}{,}\PY{p}{[}\PY{l+m+mi}{1}\PY{p}{,}\PY{l+m+mi}{1}\PY{p}{]}\PY{p}{)}
    \PY{n}{pushQubit}\PY{p}{(}\PY{l+s+s2}{\PYZdq{}}\PY{l+s+s2}{Q3}\PY{l+s+s2}{\PYZdq{}}\PY{p}{,}\PY{p}{[}\PY{l+m+mi}{1}\PY{p}{,}\PY{l+m+mi}{0}\PY{p}{]}\PY{p}{)}
    \PY{n}{toffEquiv\PYZus{}gate}\PY{p}{(}\PY{l+s+s2}{\PYZdq{}}\PY{l+s+s2}{Q1}\PY{l+s+s2}{\PYZdq{}}\PY{p}{,}\PY{l+s+s2}{\PYZdq{}}\PY{l+s+s2}{Q2}\PY{l+s+s2}{\PYZdq{}}\PY{p}{,}\PY{l+s+s2}{\PYZdq{}}\PY{l+s+s2}{Q3}\PY{l+s+s2}{\PYZdq{}}\PY{p}{)}      \PY{c+c1}{\PYZsh{} compute Q3 = Q1 AND Q2}
    \PY{n+nb}{print}\PY{p}{(}\PY{n}{measureQubit}\PY{p}{(}\PY{l+s+s2}{\PYZdq{}}\PY{l+s+s2}{Q1}\PY{l+s+s2}{\PYZdq{}}\PY{p}{)}\PY{o}{+}\PY{n}{measureQubit}\PY{p}{(}\PY{l+s+s2}{\PYZdq{}}\PY{l+s+s2}{Q2}\PY{l+s+s2}{\PYZdq{}}\PY{p}{)}\PY{o}{+}
          \PY{n}{measureQubit}\PY{p}{(}\PY{l+s+s2}{\PYZdq{}}\PY{l+s+s2}{Q3}\PY{l+s+s2}{\PYZdq{}}\PY{p}{)}\PY{p}{,} \PY{n}{end}\PY{o}{=}\PY{l+s+s2}{\PYZdq{}}\PY{l+s+s2}{,}\PY{l+s+s2}{\PYZdq{}}\PY{p}{)}
\end{Verbatim}
\end{tcolorbox}

    \begin{Verbatim}[commandchars=\\\{\}]
000,010,100,010,010,111,010,010,000,100,100,000,111,000,100,111,
    \end{Verbatim}

    It can be shown that at least six CNOT gates are needed to implement the
Toffoli gate. Any circuit that can be built with the logical gates AND,
OR, and NOT---that is, combinational circuits---can also be built with
quantum gates. The finesse of doing it with quantum gates is that you
essentially calculate the function for all possible inputs in a single
run. The big challenge is how to read out the result. We'll take a
closer look at that in the section on Grover's search.

\section{Controlled Gates and Recycling of
Qubits}\label{controlled-gates-and-recycling-of-qubits}

As we saw earlier, the CNOT gate plays a significant role in quantum
computers. The `C' in `CNOT' stands for `Controlled,' meaning that the
gate performs a NOT on the controlled qubit if and only if the control
qubit is one; otherwise, the gate does nothing. It's common for a gate
to have multiple control qubits, and then there's an implicit AND
between them, meaning all control qubits must be one for the operation
to occur. Examples of this in the list of common gates above include,
besides CNOT, `Controlled Z' (CZ) and the Toffoli gate (TOFF).

Controlled gates with many control qubits can always be implemented
using gates with only one or two control qubits. Here's an example
showing how a Toffoli gate with three control qubits can be constructed
using a regular Toffoli gate with two control qubits:

    \begin{tcolorbox}[breakable, size=fbox, boxrule=1pt, pad at break*=1mm,colback=cellbackground, colframe=cellborder]
\prompt{In}{incolor}{78}{\boxspacing}
\begin{Verbatim}[commandchars=\\\{\}]
\PY{k}{def} \PY{n+nf}{TOFF3\PYZus{}gate}\PY{p}{(}\PY{n}{q1}\PY{p}{,}\PY{n}{q2}\PY{p}{,}\PY{n}{q3}\PY{p}{,}\PY{n}{q4}\PY{p}{)}\PY{p}{:}   \PY{c+c1}{\PYZsh{} q4 = q4 XOR (q1 AND q2 AND q3)}
    \PY{n}{pushQubit}\PY{p}{(}\PY{l+s+s2}{\PYZdq{}}\PY{l+s+s2}{temp}\PY{l+s+s2}{\PYZdq{}}\PY{p}{,}\PY{p}{[}\PY{l+m+mi}{1}\PY{p}{,}\PY{l+m+mi}{0}\PY{p}{]}\PY{p}{)}    \PY{c+c1}{\PYZsh{} push a zero temporary qubit}
    \PY{n}{applyGate}\PY{p}{(}\PY{n}{TOFF\PYZus{}gate}\PY{p}{,}\PY{n}{q1}\PY{p}{,}\PY{n}{q2}\PY{p}{,}\PY{l+s+s2}{\PYZdq{}}\PY{l+s+s2}{temp}\PY{l+s+s2}{\PYZdq{}}\PY{p}{)} \PY{c+c1}{\PYZsh{} t = q1 AND q2}
    \PY{n}{applyGate}\PY{p}{(}\PY{n}{TOFF\PYZus{}gate}\PY{p}{,}\PY{l+s+s2}{\PYZdq{}}\PY{l+s+s2}{temp}\PY{l+s+s2}{\PYZdq{}}\PY{p}{,}\PY{n}{q3}\PY{p}{,}\PY{n}{q4}\PY{p}{)} \PY{c+c1}{\PYZsh{} q4 = q4 XOR (t AND q3) }
    \PY{n}{measureQubit}\PY{p}{(}\PY{l+s+s2}{\PYZdq{}}\PY{l+s+s2}{temp}\PY{l+s+s2}{\PYZdq{}}\PY{p}{)}       \PY{c+c1}{\PYZsh{} pop temp qubit \PYZhy{} PROBLEM HERE!}
\end{Verbatim}
\end{tcolorbox}

    This code looks deceptively simple, but there's a catch. We've had to
introduce an `ancilla qubit,' \texttt{temp}. So far, so good, but:
\textbf{You must reset ancilla qubits to definitely be either 0 or 1
before they are popped or reused!} Otherwise, the qubit's entanglement
with other qubits can create problems similar to `dangling pointers' in
classical programming. The standard way to reset an ancilla qubit is to
reverse all the operations performed on it, using the inverse gates. For
instance, if you've applied \(HT\) to a qubit, you should perform
\((HT)^{-1}=T^{-1}H^{-1}=T^{-1}H\) to reset it. This process is known as
\textbf{`uncomputing.'} Since the Toffoli gate is its own inverse,
including the sample run, the correct code becomes

    \begin{tcolorbox}[breakable, size=fbox, boxrule=1pt, pad at break*=1mm,colback=cellbackground, colframe=cellborder]
\prompt{In}{incolor}{79}{\boxspacing}
\begin{Verbatim}[commandchars=\\\{\}]
\PY{k}{def} \PY{n+nf}{TOFF3\PYZus{}gate}\PY{p}{(}\PY{n}{q1}\PY{p}{,}\PY{n}{q2}\PY{p}{,}\PY{n}{q3}\PY{p}{,}\PY{n}{q4}\PY{p}{)}\PY{p}{:}      
    \PY{n}{pushQubit}\PY{p}{(}\PY{l+s+s2}{\PYZdq{}}\PY{l+s+s2}{temp}\PY{l+s+s2}{\PYZdq{}}\PY{p}{,}\PY{p}{[}\PY{l+m+mi}{1}\PY{p}{,}\PY{l+m+mi}{0}\PY{p}{]}\PY{p}{)}    
    \PY{n}{applyGate}\PY{p}{(}\PY{n}{TOFF\PYZus{}gate}\PY{p}{,}\PY{n}{q1}\PY{p}{,}\PY{n}{q2}\PY{p}{,}\PY{l+s+s2}{\PYZdq{}}\PY{l+s+s2}{temp}\PY{l+s+s2}{\PYZdq{}}\PY{p}{)} 
    \PY{n}{applyGate}\PY{p}{(}\PY{n}{TOFF\PYZus{}gate}\PY{p}{,}\PY{l+s+s2}{\PYZdq{}}\PY{l+s+s2}{temp}\PY{l+s+s2}{\PYZdq{}}\PY{p}{,}\PY{n}{q3}\PY{p}{,}\PY{n}{q4}\PY{p}{)} 
    \PY{n}{applyGate}\PY{p}{(}\PY{n}{TOFF\PYZus{}gate}\PY{p}{,}\PY{n}{q1}\PY{p}{,}\PY{n}{q2}\PY{p}{,}\PY{l+s+s2}{\PYZdq{}}\PY{l+s+s2}{temp}\PY{l+s+s2}{\PYZdq{}}\PY{p}{)} \PY{c+c1}{\PYZsh{} restore temp}
    \PY{n}{measureQubit}\PY{p}{(}\PY{l+s+s2}{\PYZdq{}}\PY{l+s+s2}{temp}\PY{l+s+s2}{\PYZdq{}}\PY{p}{)}              \PY{c+c1}{\PYZsh{} t is surely zero}
    
\PY{n}{workspace} \PY{o}{=} \PY{n}{np}\PY{o}{.}\PY{n}{array}\PY{p}{(}\PY{p}{[}\PY{p}{[}\PY{l+m+mf}{1.}\PY{p}{]}\PY{p}{]}\PY{p}{)}          \PY{c+c1}{\PYZsh{} test!}
\PY{k}{for} \PY{n}{i} \PY{o+ow}{in} \PY{n+nb}{range}\PY{p}{(}\PY{l+m+mi}{20}\PY{p}{)}\PY{p}{:}                   \PY{c+c1}{\PYZsh{} generate truth table}
    \PY{n}{pushQubit}\PY{p}{(}\PY{l+s+s2}{\PYZdq{}}\PY{l+s+s2}{Q1}\PY{l+s+s2}{\PYZdq{}}\PY{p}{,}\PY{p}{[}\PY{l+m+mi}{1}\PY{p}{,}\PY{l+m+mi}{1}\PY{p}{]}\PY{p}{)}
    \PY{n}{pushQubit}\PY{p}{(}\PY{l+s+s2}{\PYZdq{}}\PY{l+s+s2}{Q2}\PY{l+s+s2}{\PYZdq{}}\PY{p}{,}\PY{p}{[}\PY{l+m+mi}{1}\PY{p}{,}\PY{l+m+mi}{1}\PY{p}{]}\PY{p}{)}
    \PY{n}{pushQubit}\PY{p}{(}\PY{l+s+s2}{\PYZdq{}}\PY{l+s+s2}{Q3}\PY{l+s+s2}{\PYZdq{}}\PY{p}{,}\PY{p}{[}\PY{l+m+mi}{1}\PY{p}{,}\PY{l+m+mi}{1}\PY{p}{]}\PY{p}{)}
    \PY{n}{pushQubit}\PY{p}{(}\PY{l+s+s2}{\PYZdq{}}\PY{l+s+s2}{Q4}\PY{l+s+s2}{\PYZdq{}}\PY{p}{,}\PY{p}{[}\PY{l+m+mi}{1}\PY{p}{,}\PY{l+m+mi}{0}\PY{p}{]}\PY{p}{)}              \PY{c+c1}{\PYZsh{} Q4 starts at zero so}
    \PY{n}{TOFF3\PYZus{}gate}\PY{p}{(}\PY{l+s+s2}{\PYZdq{}}\PY{l+s+s2}{Q1}\PY{l+s+s2}{\PYZdq{}}\PY{p}{,}\PY{l+s+s2}{\PYZdq{}}\PY{l+s+s2}{Q2}\PY{l+s+s2}{\PYZdq{}}\PY{p}{,}\PY{l+s+s2}{\PYZdq{}}\PY{l+s+s2}{Q3}\PY{l+s+s2}{\PYZdq{}}\PY{p}{,}\PY{l+s+s2}{\PYZdq{}}\PY{l+s+s2}{Q4}\PY{l+s+s2}{\PYZdq{}}\PY{p}{)}    \PY{c+c1}{\PYZsh{} Q4 = AND of Q1 thru Q3}
    \PY{n+nb}{print}\PY{p}{(}\PY{l+s+s2}{\PYZdq{}}\PY{l+s+s2}{\PYZdq{}}\PY{o}{.}\PY{n}{join}\PY{p}{(}\PY{p}{[}\PY{n}{measureQubit}\PY{p}{(}\PY{n}{q}\PY{p}{)} \PY{k}{for} \PY{n}{q} \PY{o+ow}{in}
                   \PY{p}{[}\PY{l+s+s2}{\PYZdq{}}\PY{l+s+s2}{Q1}\PY{l+s+s2}{\PYZdq{}}\PY{p}{,}\PY{l+s+s2}{\PYZdq{}}\PY{l+s+s2}{Q2}\PY{l+s+s2}{\PYZdq{}}\PY{p}{,}\PY{l+s+s2}{\PYZdq{}}\PY{l+s+s2}{Q3}\PY{l+s+s2}{\PYZdq{}}\PY{p}{,}\PY{l+s+s2}{\PYZdq{}}\PY{l+s+s2}{Q4}\PY{l+s+s2}{\PYZdq{}}\PY{p}{]}\PY{p}{]}\PY{p}{)}\PY{p}{,} \PY{n}{end}\PY{o}{=}\PY{l+s+s2}{\PYZdq{}}\PY{l+s+s2}{,}\PY{l+s+s2}{\PYZdq{}}\PY{p}{)}
\end{Verbatim}
\end{tcolorbox}

    \begin{Verbatim}[commandchars=\\\{\}]
0000,1000,1111,0110,1111,1111,0110,1111,1010,0100,0010,0110,1000,
0000,0100,0110,0000,0100,1100,0110,
    \end{Verbatim}

    We can generalize the Toffoli gate to take an arbitrary number of
control qubits in the following way:

    \begin{tcolorbox}[breakable, size=fbox, boxrule=1pt, pad at break*=1mm,colback=cellbackground, colframe=cellborder]
\prompt{In}{incolor}{80}{\boxspacing}
\begin{Verbatim}[commandchars=\\\{\}]
\PY{k}{def} \PY{n+nf}{TOFFn\PYZus{}gate}\PY{p}{(}\PY{n}{ctl}\PY{p}{,}\PY{n}{result}\PY{p}{)}\PY{p}{:}  \PY{c+c1}{\PYZsh{} result = result XOR AND(qubits)}
    \PY{n}{n} \PY{o}{=} \PY{n+nb}{len}\PY{p}{(}\PY{n}{ctl}\PY{p}{)}
    \PY{k}{if} \PY{n}{n} \PY{o}{==} \PY{l+m+mi}{0}\PY{p}{:}
        \PY{n}{applyGate}\PY{p}{(}\PY{n}{X\PYZus{}gate}\PY{p}{,}\PY{n}{result}\PY{p}{)}
    \PY{k}{if} \PY{n}{n} \PY{o}{==} \PY{l+m+mi}{1}\PY{p}{:}
        \PY{n}{applyGate}\PY{p}{(}\PY{n}{CNOT\PYZus{}gate}\PY{p}{,}\PY{n}{ctl}\PY{p}{[}\PY{l+m+mi}{0}\PY{p}{]}\PY{p}{,}\PY{n}{result}\PY{p}{)}
    \PY{k}{elif} \PY{n}{n} \PY{o}{==} \PY{l+m+mi}{2}\PY{p}{:}
        \PY{n}{applyGate}\PY{p}{(}\PY{n}{TOFF\PYZus{}gate}\PY{p}{,}\PY{n}{ctl}\PY{p}{[}\PY{l+m+mi}{0}\PY{p}{]}\PY{p}{,}\PY{n}{ctl}\PY{p}{[}\PY{l+m+mi}{1}\PY{p}{]}\PY{p}{,}\PY{n}{result}\PY{p}{)}
    \PY{k}{elif} \PY{n}{n} \PY{o}{\PYZgt{}} \PY{l+m+mi}{2}\PY{p}{:}
        \PY{n}{k} \PY{o}{=} \PY{l+m+mi}{0}
        \PY{k}{while} \PY{l+s+s2}{\PYZdq{}}\PY{l+s+s2}{temp}\PY{l+s+s2}{\PYZdq{}}\PY{o}{+}\PY{n+nb}{str}\PY{p}{(}\PY{n}{k}\PY{p}{)} \PY{o+ow}{in} \PY{n}{namestack}\PY{p}{:}
            \PY{n}{k} \PY{o}{=} \PY{n}{k} \PY{o}{+} \PY{l+m+mi}{1}
        \PY{n}{temp} \PY{o}{=} \PY{l+s+s2}{\PYZdq{}}\PY{l+s+s2}{temp}\PY{l+s+s2}{\PYZdq{}}\PY{o}{+}\PY{n+nb}{str}\PY{p}{(}\PY{n}{k}\PY{p}{)}      \PY{c+c1}{\PYZsh{} generate unique name}
        \PY{n}{pushQubit}\PY{p}{(}\PY{n}{temp}\PY{p}{,}\PY{p}{[}\PY{l+m+mi}{1}\PY{p}{,}\PY{l+m+mi}{0}\PY{p}{]}\PY{p}{)}     \PY{c+c1}{\PYZsh{} push zero temp qubit}
        \PY{n}{applyGate}\PY{p}{(}\PY{n}{TOFF\PYZus{}gate}\PY{p}{,}\PY{n}{ctl}\PY{p}{[}\PY{l+m+mi}{0}\PY{p}{]}\PY{p}{,}\PY{n}{ctl}\PY{p}{[}\PY{l+m+mi}{1}\PY{p}{]}\PY{p}{,}\PY{n}{temp}\PY{p}{)} \PY{c+c1}{\PYZsh{} apply TOFF}
        \PY{n}{ctl}\PY{o}{.}\PY{n}{append}\PY{p}{(}\PY{n}{temp}\PY{p}{)}          \PY{c+c1}{\PYZsh{} add temp to controls}
        \PY{n}{TOFFn\PYZus{}gate}\PY{p}{(}\PY{n}{ctl}\PY{p}{[}\PY{l+m+mi}{2}\PY{p}{:}\PY{p}{]}\PY{p}{,}\PY{n}{result}\PY{p}{)}\PY{c+c1}{\PYZsh{} recursion}
        \PY{n}{applyGate}\PY{p}{(}\PY{n}{TOFF\PYZus{}gate}\PY{p}{,}\PY{n}{ctl}\PY{p}{[}\PY{l+m+mi}{0}\PY{p}{]}\PY{p}{,}\PY{n}{ctl}\PY{p}{[}\PY{l+m+mi}{1}\PY{p}{]}\PY{p}{,}\PY{n}{temp}\PY{p}{)} \PY{c+c1}{\PYZsh{} uncompute temp}
        \PY{n}{measureQubit}\PY{p}{(}\PY{n}{temp}\PY{p}{)}        \PY{c+c1}{\PYZsh{} pop temp}
        
\PY{n}{workspace} \PY{o}{=} \PY{n}{np}\PY{o}{.}\PY{n}{array}\PY{p}{(}\PY{p}{[}\PY{p}{[}\PY{l+m+mi}{1}\PY{p}{]}\PY{p}{]}\PY{p}{,}\PY{n}{dtype}\PY{o}{=}\PY{n}{np}\PY{o}{.}\PY{n}{single}\PY{p}{)} \PY{c+c1}{\PYZsh{} test!}
\PY{k}{for} \PY{n}{i} \PY{o+ow}{in} \PY{n+nb}{range}\PY{p}{(}\PY{l+m+mi}{20}\PY{p}{)}\PY{p}{:}               \PY{c+c1}{\PYZsh{} generate truth table}
    \PY{n}{pushQubit}\PY{p}{(}\PY{l+s+s2}{\PYZdq{}}\PY{l+s+s2}{Q1}\PY{l+s+s2}{\PYZdq{}}\PY{p}{,}\PY{p}{[}\PY{l+m+mi}{1}\PY{p}{,}\PY{l+m+mi}{1}\PY{p}{]}\PY{p}{)}
    \PY{n}{pushQubit}\PY{p}{(}\PY{l+s+s2}{\PYZdq{}}\PY{l+s+s2}{Q2}\PY{l+s+s2}{\PYZdq{}}\PY{p}{,}\PY{p}{[}\PY{l+m+mi}{1}\PY{p}{,}\PY{l+m+mi}{1}\PY{p}{]}\PY{p}{)}
    \PY{n}{pushQubit}\PY{p}{(}\PY{l+s+s2}{\PYZdq{}}\PY{l+s+s2}{Q3}\PY{l+s+s2}{\PYZdq{}}\PY{p}{,}\PY{p}{[}\PY{l+m+mi}{1}\PY{p}{,}\PY{l+m+mi}{1}\PY{p}{]}\PY{p}{)}
    \PY{n}{pushQubit}\PY{p}{(}\PY{l+s+s2}{\PYZdq{}}\PY{l+s+s2}{Q4}\PY{l+s+s2}{\PYZdq{}}\PY{p}{,}\PY{p}{[}\PY{l+m+mi}{1}\PY{p}{,}\PY{l+m+mi}{0}\PY{p}{]}\PY{p}{)}         \PY{c+c1}{\PYZsh{} Q4 starts at zero, becomes}
    \PY{n}{TOFFn\PYZus{}gate}\PY{p}{(}\PY{p}{[}\PY{l+s+s2}{\PYZdq{}}\PY{l+s+s2}{Q1}\PY{l+s+s2}{\PYZdq{}}\PY{p}{,}\PY{l+s+s2}{\PYZdq{}}\PY{l+s+s2}{Q2}\PY{l+s+s2}{\PYZdq{}}\PY{p}{,}\PY{l+s+s2}{\PYZdq{}}\PY{l+s+s2}{Q3}\PY{l+s+s2}{\PYZdq{}}\PY{p}{]}\PY{p}{,}\PY{l+s+s2}{\PYZdq{}}\PY{l+s+s2}{Q4}\PY{l+s+s2}{\PYZdq{}}\PY{p}{)} \PY{c+c1}{\PYZsh{} AND of Q1 thru Q3}
    \PY{n+nb}{print}\PY{p}{(}\PY{l+s+s2}{\PYZdq{}}\PY{l+s+s2}{\PYZdq{}}\PY{o}{.}\PY{n}{join}\PY{p}{(}\PY{p}{[}\PY{n}{measureQubit}\PY{p}{(}\PY{n}{q}\PY{p}{)} \PY{k}{for} \PY{n}{q} \PY{o+ow}{in}
                   \PY{p}{[}\PY{l+s+s2}{\PYZdq{}}\PY{l+s+s2}{Q1}\PY{l+s+s2}{\PYZdq{}}\PY{p}{,}\PY{l+s+s2}{\PYZdq{}}\PY{l+s+s2}{Q2}\PY{l+s+s2}{\PYZdq{}}\PY{p}{,}\PY{l+s+s2}{\PYZdq{}}\PY{l+s+s2}{Q3}\PY{l+s+s2}{\PYZdq{}}\PY{p}{,}\PY{l+s+s2}{\PYZdq{}}\PY{l+s+s2}{Q4}\PY{l+s+s2}{\PYZdq{}}\PY{p}{]}\PY{p}{]}\PY{p}{)}\PY{p}{,}\PY{n}{end}\PY{o}{=}\PY{l+s+s2}{\PYZdq{}}\PY{l+s+s2}{,}\PY{l+s+s2}{\PYZdq{}}\PY{p}{)}
\end{Verbatim}
\end{tcolorbox}

    \begin{Verbatim}[commandchars=\\\{\}]
1111,1111,1100,0110,0010,0110,0010,0110,0100,0010,1111,0110,0100,
0000,1100,0110,1111,0000,0100,1010,
    \end{Verbatim}

    Such multi-input-controlled gates are very useful. They allow us, for
example, to test equality in a straightforward way by performing an AND
on the control qubits. A downside is that they use many ancilla qubits
(try writing it without them!), and the number of qubits is a critical
resource.

Characteristic of multi-input-controlled gates is that their matrix
representation is almost an identity matrix, that is, a matrix that is
zero everywhere except on the diagonal, where there are ones. They
differ only in the lower right corner, in the last two rows and columns.
We can take advantage of this for an efficient implementation of
controlled gates that avoids using ancilla qubits, like so:

    \begin{tcolorbox}[breakable, size=fbox, boxrule=1pt, pad at break*=1mm,colback=cellbackground, colframe=cellborder]
\prompt{In}{incolor}{81}{\boxspacing}
\begin{Verbatim}[commandchars=\\\{\}]
\PY{k}{def} \PY{n+nf}{applyGate}\PY{p}{(}\PY{n}{gate}\PY{p}{,}\PY{o}{*}\PY{n}{names}\PY{p}{)}\PY{p}{:}
    \PY{k}{global} \PY{n}{workspace}
    \PY{k}{if} \PY{n+nb}{list}\PY{p}{(}\PY{n}{names}\PY{p}{)} \PY{o}{!=} \PY{n}{namestack}\PY{p}{[}\PY{o}{\PYZhy{}}\PY{n+nb}{len}\PY{p}{(}\PY{n}{names}\PY{p}{)}\PY{p}{:}\PY{p}{]}\PY{p}{:} \PY{c+c1}{\PYZsh{} reorder stack}
        \PY{k}{for} \PY{n}{name} \PY{o+ow}{in} \PY{n}{names}\PY{p}{:}                     \PY{c+c1}{\PYZsh{} if necessary}
            \PY{n}{tosQubit}\PY{p}{(}\PY{n}{name}\PY{p}{)} 
    \PY{n}{workspace} \PY{o}{=} \PY{n}{np}\PY{o}{.}\PY{n}{reshape}\PY{p}{(}\PY{n}{workspace}\PY{p}{,}\PY{p}{(}\PY{o}{\PYZhy{}}\PY{l+m+mi}{1}\PY{p}{,}\PY{l+m+mi}{2}\PY{o}{*}\PY{o}{*}\PY{p}{(}\PY{n+nb}{len}\PY{p}{(}\PY{n}{names}\PY{p}{)}\PY{p}{)}\PY{p}{)}\PY{p}{)} 
    \PY{n}{subworkspace} \PY{o}{=} \PY{n}{workspace}\PY{p}{[}\PY{p}{:}\PY{p}{,}\PY{o}{\PYZhy{}}\PY{n}{gate}\PY{o}{.}\PY{n}{shape}\PY{p}{[}\PY{l+m+mi}{0}\PY{p}{]}\PY{p}{:}\PY{p}{]} 
    \PY{n}{np}\PY{o}{.}\PY{n}{matmul}\PY{p}{(}\PY{n}{subworkspace}\PY{p}{,}\PY{n}{gate}\PY{o}{.}\PY{n}{T}\PY{p}{,}\PY{n}{out}\PY{o}{=}\PY{n}{subworkspace}\PY{p}{)} 
\end{Verbatim}
\end{tcolorbox}

    An additional optimization here is to avoid moving qubits if the stack
is already in order. Above we have used the number of rows in the
\texttt{gate} matrix to determine which \texttt{qubits} are the control
qubits. Only a small part of the \texttt{workspace} is involved in the
multiplication. Now we can define the Toffoli gates this elegantly:

    \begin{tcolorbox}[breakable, size=fbox, boxrule=1pt, pad at break*=1mm,colback=cellbackground, colframe=cellborder]
\prompt{In}{incolor}{82}{\boxspacing}
\begin{Verbatim}[commandchars=\\\{\}]
\PY{k}{def} \PY{n+nf}{TOFF3\PYZus{}gate}\PY{p}{(}\PY{n}{q1}\PY{p}{,}\PY{n}{q2}\PY{p}{,}\PY{n}{q3}\PY{p}{,}\PY{n}{q4}\PY{p}{)}\PY{p}{:}       
    \PY{n}{applyGate}\PY{p}{(}\PY{n}{X\PYZus{}gate}\PY{p}{,}\PY{n}{q1}\PY{p}{,}\PY{n}{q2}\PY{p}{,}\PY{n}{q3}\PY{p}{,}\PY{n}{q4}\PY{p}{)}

\PY{k}{def} \PY{n+nf}{TOFFn\PYZus{}gate}\PY{p}{(}\PY{n}{ctl}\PY{p}{,}\PY{n}{result}\PY{p}{)}\PY{p}{:}  
    \PY{n}{applyGate}\PY{p}{(}\PY{n}{X\PYZus{}gate}\PY{p}{,}\PY{o}{*}\PY{n}{ctl}\PY{p}{,}\PY{n}{result}\PY{p}{)}

\PY{n}{workspace} \PY{o}{=} \PY{n}{np}\PY{o}{.}\PY{n}{array}\PY{p}{(}\PY{p}{[}\PY{p}{[}\PY{l+m+mi}{1}\PY{p}{]}\PY{p}{]}\PY{p}{,}\PY{n}{dtype}\PY{o}{=}\PY{n}{np}\PY{o}{.}\PY{n}{single}\PY{p}{)} 
\PY{k}{for} \PY{n}{i} \PY{o+ow}{in} \PY{n+nb}{range}\PY{p}{(}\PY{l+m+mi}{20}\PY{p}{)}\PY{p}{:}      
    \PY{n}{pushQubit}\PY{p}{(}\PY{l+s+s2}{\PYZdq{}}\PY{l+s+s2}{Q1}\PY{l+s+s2}{\PYZdq{}}\PY{p}{,}\PY{p}{[}\PY{l+m+mi}{1}\PY{p}{,}\PY{l+m+mi}{1}\PY{p}{]}\PY{p}{)}
    \PY{n}{pushQubit}\PY{p}{(}\PY{l+s+s2}{\PYZdq{}}\PY{l+s+s2}{Q2}\PY{l+s+s2}{\PYZdq{}}\PY{p}{,}\PY{p}{[}\PY{l+m+mi}{1}\PY{p}{,}\PY{l+m+mi}{1}\PY{p}{]}\PY{p}{)}
    \PY{n}{pushQubit}\PY{p}{(}\PY{l+s+s2}{\PYZdq{}}\PY{l+s+s2}{Q3}\PY{l+s+s2}{\PYZdq{}}\PY{p}{,}\PY{p}{[}\PY{l+m+mi}{1}\PY{p}{,}\PY{l+m+mi}{1}\PY{p}{]}\PY{p}{)}
    \PY{n}{pushQubit}\PY{p}{(}\PY{l+s+s2}{\PYZdq{}}\PY{l+s+s2}{Q4}\PY{l+s+s2}{\PYZdq{}}\PY{p}{,}\PY{p}{[}\PY{l+m+mi}{1}\PY{p}{,}\PY{l+m+mi}{0}\PY{p}{]}\PY{p}{)} 
    \PY{n}{TOFF3\PYZus{}gate}\PY{p}{(}\PY{l+s+s2}{\PYZdq{}}\PY{l+s+s2}{Q1}\PY{l+s+s2}{\PYZdq{}}\PY{p}{,}\PY{l+s+s2}{\PYZdq{}}\PY{l+s+s2}{Q2}\PY{l+s+s2}{\PYZdq{}}\PY{p}{,}\PY{l+s+s2}{\PYZdq{}}\PY{l+s+s2}{Q3}\PY{l+s+s2}{\PYZdq{}}\PY{p}{,}\PY{l+s+s2}{\PYZdq{}}\PY{l+s+s2}{Q4}\PY{l+s+s2}{\PYZdq{}}\PY{p}{)} 
    \PY{n+nb}{print}\PY{p}{(}\PY{l+s+s2}{\PYZdq{}}\PY{l+s+s2}{\PYZdq{}}\PY{o}{.}\PY{n}{join}\PY{p}{(}\PY{p}{[}\PY{n}{measureQubit}\PY{p}{(}\PY{n}{q}\PY{p}{)} \PY{k}{for} \PY{n}{q} \PY{o+ow}{in}
          \PY{p}{[}\PY{l+s+s2}{\PYZdq{}}\PY{l+s+s2}{Q1}\PY{l+s+s2}{\PYZdq{}}\PY{p}{,}\PY{l+s+s2}{\PYZdq{}}\PY{l+s+s2}{Q2}\PY{l+s+s2}{\PYZdq{}}\PY{p}{,}\PY{l+s+s2}{\PYZdq{}}\PY{l+s+s2}{Q3}\PY{l+s+s2}{\PYZdq{}}\PY{p}{,}\PY{l+s+s2}{\PYZdq{}}\PY{l+s+s2}{Q4}\PY{l+s+s2}{\PYZdq{}}\PY{p}{]}\PY{p}{]}\PY{p}{)}\PY{p}{,}\PY{n}{end}\PY{o}{=}\PY{l+s+s2}{\PYZdq{}}\PY{l+s+s2}{/}\PY{l+s+s2}{\PYZdq{}}\PY{p}{)}
    \PY{n}{pushQubit}\PY{p}{(}\PY{l+s+s2}{\PYZdq{}}\PY{l+s+s2}{Q1}\PY{l+s+s2}{\PYZdq{}}\PY{p}{,}\PY{p}{[}\PY{l+m+mi}{1}\PY{p}{,}\PY{l+m+mi}{1}\PY{p}{]}\PY{p}{)}
    \PY{n}{pushQubit}\PY{p}{(}\PY{l+s+s2}{\PYZdq{}}\PY{l+s+s2}{Q2}\PY{l+s+s2}{\PYZdq{}}\PY{p}{,}\PY{p}{[}\PY{l+m+mi}{1}\PY{p}{,}\PY{l+m+mi}{1}\PY{p}{]}\PY{p}{)}
    \PY{n}{pushQubit}\PY{p}{(}\PY{l+s+s2}{\PYZdq{}}\PY{l+s+s2}{Q3}\PY{l+s+s2}{\PYZdq{}}\PY{p}{,}\PY{p}{[}\PY{l+m+mi}{1}\PY{p}{,}\PY{l+m+mi}{1}\PY{p}{]}\PY{p}{)}
    \PY{n}{pushQubit}\PY{p}{(}\PY{l+s+s2}{\PYZdq{}}\PY{l+s+s2}{Q4}\PY{l+s+s2}{\PYZdq{}}\PY{p}{,}\PY{p}{[}\PY{l+m+mi}{1}\PY{p}{,}\PY{l+m+mi}{0}\PY{p}{]}\PY{p}{)}   
    \PY{n}{TOFFn\PYZus{}gate}\PY{p}{(}\PY{p}{[}\PY{l+s+s2}{\PYZdq{}}\PY{l+s+s2}{Q1}\PY{l+s+s2}{\PYZdq{}}\PY{p}{,}\PY{l+s+s2}{\PYZdq{}}\PY{l+s+s2}{Q2}\PY{l+s+s2}{\PYZdq{}}\PY{p}{,}\PY{l+s+s2}{\PYZdq{}}\PY{l+s+s2}{Q3}\PY{l+s+s2}{\PYZdq{}}\PY{p}{]}\PY{p}{,}\PY{l+s+s2}{\PYZdq{}}\PY{l+s+s2}{Q4}\PY{l+s+s2}{\PYZdq{}}\PY{p}{)} 
    \PY{n+nb}{print}\PY{p}{(}\PY{l+s+s2}{\PYZdq{}}\PY{l+s+s2}{\PYZdq{}}\PY{o}{.}\PY{n}{join}\PY{p}{(}\PY{p}{[}\PY{n}{measureQubit}\PY{p}{(}\PY{n}{q}\PY{p}{)} \PY{k}{for} \PY{n}{q} \PY{o+ow}{in}
          \PY{p}{[}\PY{l+s+s2}{\PYZdq{}}\PY{l+s+s2}{Q1}\PY{l+s+s2}{\PYZdq{}}\PY{p}{,}\PY{l+s+s2}{\PYZdq{}}\PY{l+s+s2}{Q2}\PY{l+s+s2}{\PYZdq{}}\PY{p}{,}\PY{l+s+s2}{\PYZdq{}}\PY{l+s+s2}{Q3}\PY{l+s+s2}{\PYZdq{}}\PY{p}{,}\PY{l+s+s2}{\PYZdq{}}\PY{l+s+s2}{Q4}\PY{l+s+s2}{\PYZdq{}}\PY{p}{]}\PY{p}{]}\PY{p}{)}\PY{p}{,}\PY{n}{end}\PY{o}{=}\PY{l+s+s2}{\PYZdq{}}\PY{l+s+s2}{,}\PY{l+s+s2}{\PYZdq{}}\PY{p}{)}
\end{Verbatim}
\end{tcolorbox}

    \begin{Verbatim}[commandchars=\\\{\}]
0010/1000,0000/1000,0100/0000,0110/0110,0010/1100,0000/0100,1100/
1000,0000/0010,1100/1000,0110/1010,1010/1000,0100/1111,1111/0010,
0000/1000,0100/0100,1000/1010,1111/0000,0010/0010,0010/0000,1100/
0000,
    \end{Verbatim}

    \section{A Slightly Larger Example: Grover's
Search}\label{a-slightly-larger-example-grovers-search}

Grover's search is a neat and relatively simple example of how a quantum
computer with enough qubits could be used to speed up a calculation. It
also shows that quite sophisticated techniques are required even to
solve relatively simple problems with a quantum computer.

Suppose we've implemented a function \(f(x)\) with quantum gates, where
\(x\) is represented by \(N\) qubits. Let's also assume that there's
exactly one \(x_0\) such that \(f(x_0)=1\) and \(f(x)=0\) for all
\(x \ne x_0\). Then, we can use Grover's search to find \(x_0\) in time
\(O(2^{N/2})\), while a naive search would require time \(O(2^N)\), thus
a quadratic speedup.

\subsection{Subroutines}\label{subroutines}

First, let's look at a function that returns 1 if all qubits are zero,
and 0 otherwise:

    \begin{tcolorbox}[breakable, size=fbox, boxrule=1pt, pad at break*=1mm,colback=cellbackground, colframe=cellborder]
\prompt{In}{incolor}{83}{\boxspacing}
\begin{Verbatim}[commandchars=\\\{\}]
\PY{k}{def} \PY{n+nf}{zero\PYZus{}booleanOracle}\PY{p}{(}\PY{n}{qubits}\PY{p}{,}\PY{n}{result}\PY{p}{)}\PY{p}{:} \PY{c+c1}{\PYZsh{} all qubits zero?}
    \PY{c+c1}{\PYZsh{} if all qubits==0 return 1 else return 0     }
    \PY{k}{for} \PY{n}{qubit} \PY{o+ow}{in} \PY{n}{qubits}\PY{p}{:}               \PY{c+c1}{\PYZsh{} negate all inputs}
        \PY{n}{applyGate}\PY{p}{(}\PY{n}{X\PYZus{}gate}\PY{p}{,}\PY{n}{qubit}\PY{p}{)}       
    \PY{n}{TOFFn\PYZus{}gate}\PY{p}{(}\PY{n}{qubits}\PY{p}{,}\PY{n}{result}\PY{p}{)}          \PY{c+c1}{\PYZsh{} compute AND}
    \PY{k}{for} \PY{n}{qubit} \PY{o+ow}{in} \PY{n}{qubits}\PY{p}{:}               \PY{c+c1}{\PYZsh{} restore inputs}
        \PY{n}{applyGate}\PY{p}{(}\PY{n}{X\PYZus{}gate}\PY{p}{,}\PY{n}{qubit}\PY{p}{)} 
\end{Verbatim}
\end{tcolorbox}

    It simply inverts the qubits with X gates, and then performs an AND
using the generalized Toffoli gate. The result is delivered through the
ancilla qubit \texttt{result}. Afterward, the input qubits are reset.
Such a function is called a \textbf{Boolean oracle}.

For Grover's algorithm, a variant of the above procedure is used that
doesn't have a result qubit but `returns' the value 1 through a
\textbf{sign change} in the contents of the corresponding small cubes.
Physicists call this sign change a `phase change.' When returning the
value 0, no sign change is made. Such a function is called a
\textbf{phase oracle}. A trick to achieve the sign change is to apply
Hadamard gates to the controlled bit before and after the Toffoli gate.
Which bit is controlled doesn't matter. Such a gate is exactly the same
as a controlled Z gate, so we can write

    \begin{tcolorbox}[breakable, size=fbox, boxrule=1pt, pad at break*=1mm,colback=cellbackground, colframe=cellborder]
\begin{Verbatim}[commandchars=\\\{\}]
\PY{k}{def} \PY{n+nf}{zero\PYZus{}phaseOracle}\PY{p}{(}\PY{n}{qubits}\PY{p}{)}\PY{p}{:}         \PY{c+c1}{\PYZsh{} all qubits zero?}
    \PY{c+c1}{\PYZsh{} if all qubits==0 return \PYZhy{}weight else return weight    }
    \PY{k}{for} \PY{n}{qubit} \PY{o+ow}{in} \PY{n}{qubits}\PY{p}{:}              \PY{c+c1}{\PYZsh{} negate all inputs}
        \PY{n}{applyGate}\PY{p}{(}\PY{n}{X\PYZus{}gate}\PY{p}{,}\PY{n}{qubit}\PY{p}{)}
    \PY{n}{applyGate}\PY{p}{(}\PY{n}{Z\PYZus{}gate}\PY{p}{,}\PY{o}{*}\PY{n}{namestack}\PY{p}{)}      \PY{c+c1}{\PYZsh{} controlled Z gate}
    \PY{k}{for} \PY{n}{qubit} \PY{o+ow}{in} \PY{n}{qubits}\PY{p}{:}              \PY{c+c1}{\PYZsh{} restore inputs}
        \PY{n}{applyGate}\PY{p}{(}\PY{n}{X\PYZus{}gate}\PY{p}{,}\PY{n}{qubit}\PY{p}{)} 
\end{Verbatim}
\end{tcolorbox}

    Note that beccause of the symmetry of the controlled Z gate, any order
of the qubits is OK. Applying it in the \texttt{namestack} order has the
advantage that the qubits don't need to be moved around.

Finally, we come to our `Black box' function. This also needs to be
expressed as a phase oracle, that is, it should give a sign change for
the solutions to the function. As an example, right here, we use the
function \[f(x)=\Big\{
\begin{matrix}
	1 & \text{if all qubits except qubit 1 are one}, \\ 0 & \text{otherwise.}\hfill
\end{matrix}
\]

The implementation is similar to what we did for
\texttt{zero\_phaseOracle}, but here we only negate qubit 1. Given six
qubits, the procedure should negate the weights if the input is 111101
in binary (a programmer might think of this number as -3), otherwise,
leave the weights unchanged.

    \begin{tcolorbox}[breakable, size=fbox, boxrule=1pt, pad at break*=1mm,colback=cellbackground, colframe=cellborder]
\prompt{In}{incolor}{91}{\boxspacing}
\begin{Verbatim}[commandchars=\\\{\}]
\PY{k}{def} \PY{n+nf}{sample\PYZus{}phaseOracle}\PY{p}{(}\PY{n}{qubits}\PY{p}{)}\PY{p}{:}       \PY{c+c1}{\PYZsh{} sample function}
    \PY{c+c1}{\PYZsh{} if all f(x)==1 return \PYZhy{}weight else return weight }
    \PY{n}{applyGate}\PY{p}{(}\PY{n}{X\PYZus{}gate}\PY{p}{,}\PY{n}{qubits}\PY{p}{[}\PY{l+m+mi}{1}\PY{p}{]}\PY{p}{)}       \PY{c+c1}{\PYZsh{} negate qubit 1}
    \PY{n}{applyGate}\PY{p}{(}\PY{n}{Z\PYZus{}gate}\PY{p}{,}\PY{o}{*}\PY{n}{namestack}\PY{p}{)}      \PY{c+c1}{\PYZsh{} controlled Z gate}
    \PY{n}{applyGate}\PY{p}{(}\PY{n}{X\PYZus{}gate}\PY{p}{,}\PY{n}{qubits}\PY{p}{[}\PY{l+m+mi}{1}\PY{p}{]}\PY{p}{)}       \PY{c+c1}{\PYZsh{} restore qubit 1}
\end{Verbatim}
\end{tcolorbox}

    If we wanted to check for -7, we could add one line for negating
qubit 3 by \texttt{applygate(X\_gate,qubits{[}3{]})} before the
application of the Z-gate, and another line for restoring it after the
Z-gate. The sample function would then recognize binary 110101.

\subsection{The Main Loop}\label{the-main-loop}

Now we're ready to implement the main loop of Grover's search. It
alternates between running \texttt{sample\_phaseOracle} and
\texttt{zero\_phaseOracle}, applying Hadamard gates between them. It's
really not obvious why this works, but a brief explanation is in the
next section. With each iteration, the probabilities of the qubits being
`right' gradually improve. So the weight for the small cube that is the
solution increases with each iteration. This is called \textbf{amplitude
amplification}.

    \begin{tcolorbox}[breakable, size=fbox, boxrule=1pt, pad at break*=1mm,colback=cellbackground, colframe=cellborder]
\prompt{In}{incolor}{89}{\boxspacing}
\begin{Verbatim}[commandchars=\\\{\}]
\PY{k}{def} \PY{n+nf}{groverSearch}\PY{p}{(}\PY{n}{n}\PY{p}{,} \PY{n}{printProb}\PY{o}{=}\PY{k+kc}{True}\PY{p}{)}\PY{p}{:}
    \PY{n}{optimalTurns} \PY{o}{=} \PY{n+nb}{int}\PY{p}{(}\PY{n}{np}\PY{o}{.}\PY{n}{pi}\PY{o}{/}\PY{l+m+mi}{4}\PY{o}{*}\PY{n}{np}\PY{o}{.}\PY{n}{sqrt}\PY{p}{(}\PY{l+m+mi}{2}\PY{o}{*}\PY{o}{*}\PY{n}{n}\PY{p}{)}\PY{o}{\PYZhy{}}\PY{l+m+mi}{1}\PY{o}{/}\PY{l+m+mi}{2}\PY{p}{)} \PY{c+c1}{\PYZsh{} iterations}
    \PY{n}{qubits} \PY{o}{=} \PY{n+nb}{list}\PY{p}{(}\PY{n+nb}{range}\PY{p}{(}\PY{n}{n}\PY{p}{)}\PY{p}{)}             \PY{c+c1}{\PYZsh{} generate qubit names}
    \PY{k}{for} \PY{n}{qubit} \PY{o+ow}{in} \PY{n}{qubits}\PY{p}{:}                \PY{c+c1}{\PYZsh{} initialize qubits}
        \PY{n}{pushQubit}\PY{p}{(}\PY{n}{qubit}\PY{p}{,}\PY{p}{[}\PY{l+m+mi}{1}\PY{p}{,}\PY{l+m+mi}{1}\PY{p}{]}\PY{p}{)}
    \PY{k}{for} \PY{n}{k} \PY{o+ow}{in} \PY{n+nb}{range}\PY{p}{(}\PY{n}{optimalTurns}\PY{p}{)}\PY{p}{:}       \PY{c+c1}{\PYZsh{} Grover iterations:}
        \PY{n}{sample\PYZus{}phaseOracle}\PY{p}{(}\PY{n}{qubits}\PY{p}{)}      \PY{c+c1}{\PYZsh{} apply phase oracle}
        \PY{k}{for} \PY{n}{qubit} \PY{o+ow}{in} \PY{n}{qubits}\PY{p}{:}            \PY{c+c1}{\PYZsh{} H\PYZhy{}gate all qubits}
            \PY{n}{applyGate}\PY{p}{(}\PY{n}{H\PYZus{}gate}\PY{p}{,}\PY{n}{qubit}\PY{p}{)}
        \PY{n}{zero\PYZus{}phaseOracle}\PY{p}{(}\PY{n}{qubits}\PY{p}{)}        \PY{c+c1}{\PYZsh{} apply 0 phase oracle}
        \PY{k}{for} \PY{n}{qubit} \PY{o+ow}{in} \PY{n}{qubits}\PY{p}{:}            \PY{c+c1}{\PYZsh{} H\PYZhy{}gate all qubits}
            \PY{n}{applyGate}\PY{p}{(}\PY{n}{H\PYZus{}gate}\PY{p}{,}\PY{n}{qubit}\PY{p}{)}
        \PY{k}{if} \PY{n}{printProb}\PY{p}{:}                   \PY{c+c1}{\PYZsh{} peek probabilities}
            \PY{n+nb}{print}\PY{p}{(}\PY{n}{probQubit}\PY{p}{(}\PY{n}{qubits}\PY{p}{[}\PY{l+m+mi}{0}\PY{p}{]}\PY{p}{)}\PY{p}{)} \PY{c+c1}{\PYZsh{} to show convergence}
    \PY{k}{for} \PY{n}{qubit} \PY{o+ow}{in} \PY{n+nb}{reversed}\PY{p}{(}\PY{n}{qubits}\PY{p}{)}\PY{p}{:}      \PY{c+c1}{\PYZsh{} print result}
        \PY{n+nb}{print}\PY{p}{(}\PY{n}{measureQubit}\PY{p}{(}\PY{n}{qubit}\PY{p}{)}\PY{p}{,}\PY{n}{end}\PY{o}{=}\PY{l+s+s2}{\PYZdq{}}\PY{l+s+s2}{\PYZdq{}}\PY{p}{)}
\end{Verbatim}
\end{tcolorbox}

    To illustrate the convergence, we print out the bit probabilities for
qubit 0 in each iteration. Here's what a run looks like:

    \begin{tcolorbox}[breakable, size=fbox, boxrule=1pt, pad at break*=1mm,colback=cellbackground, colframe=cellborder]
\prompt{In}{incolor}{90}{\boxspacing}
\begin{Verbatim}[commandchars=\\\{\}]
\PY{n}{workspace} \PY{o}{=} \PY{n}{np}\PY{o}{.}\PY{n}{array}\PY{p}{(}\PY{p}{[}\PY{p}{[}\PY{l+m+mf}{1.}\PY{p}{]}\PY{p}{]}\PY{p}{)}            \PY{c+c1}{\PYZsh{} initialize workspace}
\PY{n}{groverSearch}\PY{p}{(}\PY{l+m+mi}{6}\PY{p}{)}                         \PY{c+c1}{\PYZsh{} (only reals used here)}
\end{Verbatim}
\end{tcolorbox}

    \begin{Verbatim}[commandchars=\\\{\}]
[0.43945313 0.56054687]
[0.33325958 0.66674042]
[0.20755294 0.79244706]
[0.09326882 0.90673118]
[0.01853182 0.98146818]
111101
    \end{Verbatim}

    It's clear that the probability of a zero approaches zero and for a one
approaches one, so a correct result is likely, but still not entirely
certain. One must test it and redo the search if the result turns out to
be wrong. Grover's search doesn't offer an exponential speedup but
`only' a quadratic one. However, it can be shown that this is the best
that can be achieved with a quantum computer.

Here again, it's clear that the main problem is not calculating the
function, as this is done in one sweep for all conceivable arguments by
\texttt{sample\_phaseOracle}, but in extracting the solution after the
calculation.

\subsection{Why Grover's Search
Works}\label{why-grovers-search-works}

Let's see the workspace as a row vector of \(M=2^N\) small cubes again.
The sought solution \(x_0\) to \(f(x)=1\) can be described with a
`one-hot' encoding as an \(M\)-dimensional unit vector
\[x_0 = (0, 0, ..., 1, ..., 0),\] where only one coordinate (= small
cube weight) differs from zero. Let the vector \(y\) be another
\(M\)-dimensional unit vector with a zero where \(x_0\) has a one, while
the other coordinates are \(1/\sqrt{M-1}\), so that
\[y = \frac{1}{\sqrt{M-1}}(1, 1, ..., 0, ..., 1).\] This vector
represents all the small cubes that are not solutions. Since the scalar
product \(x \cdot y = 0\), the vectors \(x\) and \(y\) must be
orthogonal. We introduce the vector
\[z = \frac{1}{\sqrt{M}} (1, 1, ..., 1, ..., 1)= \frac{1}{\sqrt{M}}x + \frac{\sqrt{M-1}}{\sqrt{M}}y,\]
which is a linear combination of \(x\) and \(y\).
       
\begin{figure}[!hbt] 
	\centering{\includegraphics[width=0.7 \textwidth]{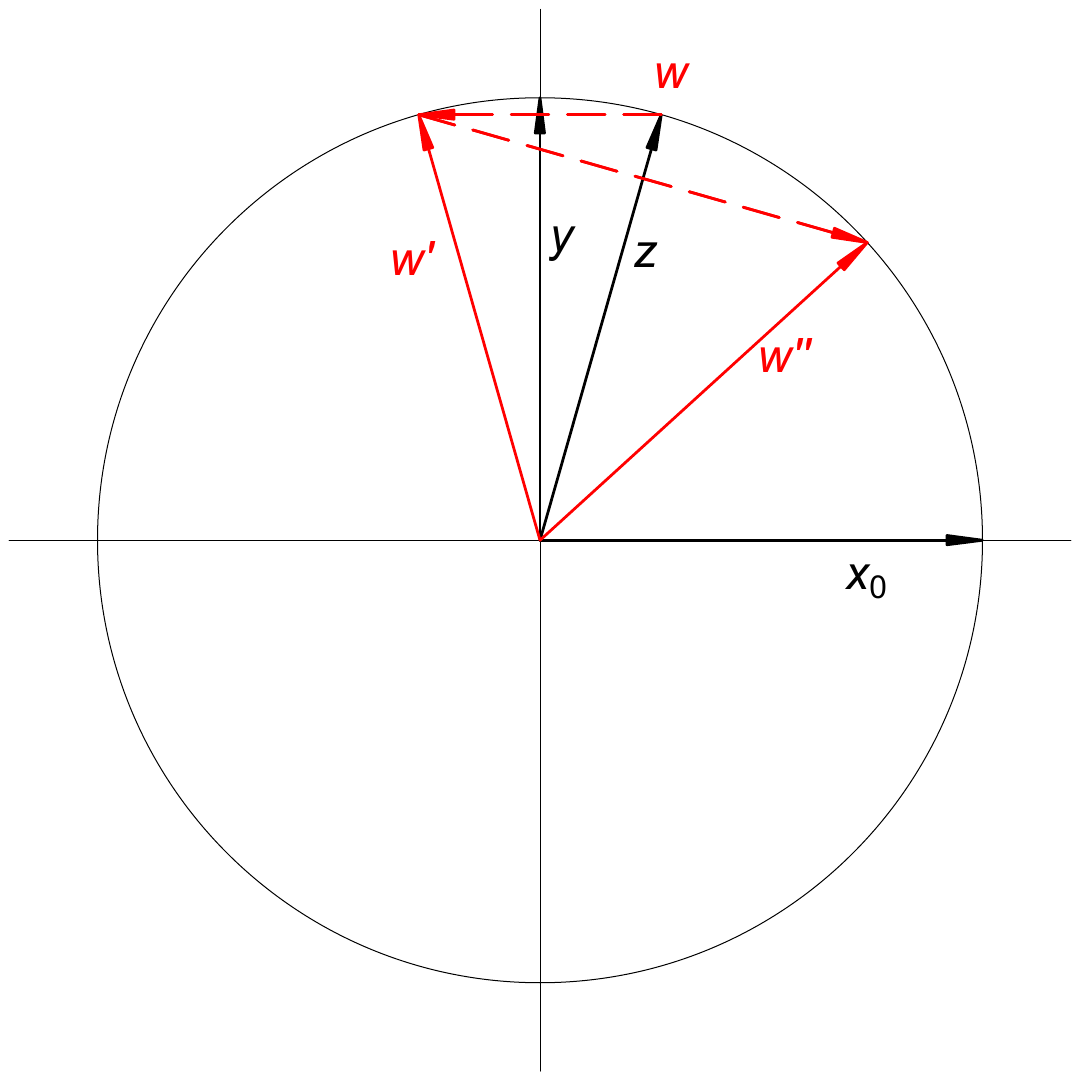}}
	\caption{Grover's search gradually makes the vector $w$ approach $x_0$ by first reflecting
		$w$ in $y$ and then in $z$. The reflections are performed by the phase oracles.}
\end{figure}

    We let the vector \(w\) represent the approximation to the solution
during Grover's search and initialize it to \(z\). The first call to the
phase oracle \texttt{sample\_phaseOracle} reflects \(w\) across the
\(y\)-axis (Fig.~3), to become \(w'\). It's relatively
straightforward to show that the second call, along with the Hadamard
gates, then reflects \(w'\) in \(z\), resulting in \(w''\). This vector
is closer to \(x_0\) than \(w\), so we replace \(w\) with \(w''\) and
repeat the procedure. It can be calculated that the probability of
hitting \(x_0\) is highest after
\[k = \Bigg\lfloor\frac{\pi}{4}\sqrt{M}-\frac{1}{2}\Bigg\rfloor\]
iterations. See references {[}2-3{]} for details.

\section{Quantum Computer Simulation with a
GPU}\label{quantum-computer-simulation-with-a-gpu}

Looking at the code above, it's clear that it would be beneficial to
move \texttt{workspace} to a GPU. The operations performed on the
workspace are perfect for the GPU, and there isn't much communication
needed between the CPU and the GPU.

We start by importing the PyTorch package. For resource requirements and
installation instructions, see \href{https://pytorch.org/}{the PyTorch
documentation}. PyTorch can also run on a CPU, so you don't necessarily
need an NVidia graphics card to try it out. The code to initialize
PyTorch is simple:

    \begin{tcolorbox}[breakable, size=fbox, boxrule=1pt, pad at break*=1mm,colback=cellbackground, colframe=cellborder]
\prompt{In}{incolor}{41}{\boxspacing}
\begin{Verbatim}[commandchars=\\\{\}]
\PY{k+kn}{import} \PY{n+nn}{torch} \PY{k}{as} \PY{n+nn}{pt}                    
\PY{n}{pt}\PY{o}{.}\PY{n}{autograd}\PY{o}{.}\PY{n}{set\PYZus{}grad\PYZus{}enabled}\PY{p}{(}\PY{k+kc}{False}\PY{p}{)}\PY{c+c1}{\PYZsh{} disable autogradients}
\PY{k}{if} \PY{n}{pt}\PY{o}{.}\PY{n}{cuda}\PY{o}{.}\PY{n}{is\PYZus{}available}\PY{p}{(}\PY{p}{)}\PY{p}{:}         \PY{c+c1}{\PYZsh{} check if GPU is available}
    \PY{n+nb}{print}\PY{p}{(}\PY{l+s+s2}{\PYZdq{}}\PY{l+s+s2}{GPU available}\PY{l+s+s2}{\PYZdq{}}\PY{p}{)}
\PY{k}{else}\PY{p}{:}
    \PY{n+nb}{print}\PY{p}{(}\PY{l+s+s2}{\PYZdq{}}\PY{l+s+s2}{Sorry, only CPU available}\PY{l+s+s2}{\PYZdq{}}\PY{p}{)}
\end{Verbatim}
\end{tcolorbox}

    \begin{Verbatim}[commandchars=\\\{\}]
GPU available
    \end{Verbatim}

    Below are the adjustments needed for the quantum computer simulator to
use PyTorch, both on a CPU and a GPU with CUDA. The major difference
from the previous code is that the workspace now lives entirely in the
GPU's memory. Consequently, many calls to numpy have been replaced with
PyTorch calls. Only the lines whose comments begin with \texttt{\#!} are
new or changed.

    \begin{tcolorbox}[breakable, size=fbox, boxrule=1pt, pad at break*=1mm,colback=cellbackground, colframe=cellborder]
\prompt{In}{incolor}{42}{\boxspacing}
\begin{Verbatim}[commandchars=\\\{\}]
\PY{k}{def} \PY{n+nf}{pushQubit}\PY{p}{(}\PY{n}{name}\PY{p}{,}\PY{n}{weights}\PY{p}{)}\PY{p}{:}
    \PY{k}{global} \PY{n}{workspace}
    \PY{k}{global} \PY{n}{namestack} 
    \PY{k}{if} \PY{p}{(}\PY{n}{workspace}\PY{o}{.}\PY{n}{shape}\PY{p}{[}\PY{l+m+mi}{0}\PY{p}{]}\PY{p}{,}\PY{n}{workspace}\PY{o}{.}\PY{n}{shape}\PY{p}{[}\PY{l+m+mi}{1}\PY{p}{]}\PY{p}{)} \PY{o}{==} \PY{p}{(}\PY{l+m+mi}{1}\PY{p}{,}\PY{l+m+mi}{1}\PY{p}{)}\PY{p}{:} \PY{c+c1}{\PYZsh{}!}
        \PY{n}{namestack} \PY{o}{=} \PY{p}{[}\PY{p}{]}             \PY{c+c1}{\PYZsh{} reset if workspace empty}
    \PY{n}{namestack}\PY{o}{.}\PY{n}{append}\PY{p}{(}\PY{n}{name}\PY{p}{)}
    \PY{n}{weights} \PY{o}{=} \PY{n}{weights}\PY{o}{/}\PY{n}{np}\PY{o}{.}\PY{n}{linalg}\PY{o}{.}\PY{n}{norm}\PY{p}{(}\PY{n}{weights}\PY{p}{)} \PY{c+c1}{\PYZsh{} normalize}
    \PY{n}{weights} \PY{o}{=} \PY{n}{pt}\PY{o}{.}\PY{n}{tensor}\PY{p}{(}\PY{n}{weights}\PY{p}{,}\PY{n}{device}\PY{o}{=}\PY{n}{workspace}\PY{o}{.}\PY{n}{device}\PY{p}{,} \PY{c+c1}{\PYZsh{}! }
                        \PY{n}{dtype}\PY{o}{=}\PY{n}{workspace}\PY{p}{[}\PY{l+m+mi}{0}\PY{p}{,}\PY{l+m+mi}{0}\PY{p}{]}\PY{o}{.}\PY{n}{dtype}\PY{p}{)}      \PY{c+c1}{\PYZsh{}! }
    \PY{n}{workspace} \PY{o}{=} \PY{n}{pt}\PY{o}{.}\PY{n}{reshape}\PY{p}{(}\PY{n}{workspace}\PY{p}{,}\PY{p}{(}\PY{l+m+mi}{1}\PY{p}{,}\PY{o}{\PYZhy{}}\PY{l+m+mi}{1}\PY{p}{)}\PY{p}{)}             \PY{c+c1}{\PYZsh{}! }
    \PY{n}{workspace} \PY{o}{=} \PY{n}{pt}\PY{o}{.}\PY{n}{kron}\PY{p}{(}\PY{n}{workspace}\PY{p}{,}\PY{n}{weights}\PY{p}{)}               \PY{c+c1}{\PYZsh{}! }

\PY{k}{def} \PY{n+nf}{tosQubit}\PY{p}{(}\PY{n}{name}\PY{p}{)}\PY{p}{:}
    \PY{k}{global} \PY{n}{workspace}
    \PY{k}{global} \PY{n}{namestack}
    \PY{n}{k} \PY{o}{=} \PY{n+nb}{len}\PY{p}{(}\PY{n}{namestack}\PY{p}{)}\PY{o}{\PYZhy{}}\PY{n}{namestack}\PY{o}{.}\PY{n}{index}\PY{p}{(}\PY{n}{name}\PY{p}{)} \PY{c+c1}{\PYZsh{} position of qubit}
    \PY{k}{if} \PY{n}{k} \PY{o}{\PYZgt{}} \PY{l+m+mi}{1}\PY{p}{:}                                \PY{c+c1}{\PYZsh{} if non\PYZhy{}trivial}
        \PY{n}{namestack}\PY{o}{.}\PY{n}{append}\PY{p}{(}\PY{n}{namestack}\PY{o}{.}\PY{n}{pop}\PY{p}{(}\PY{o}{\PYZhy{}}\PY{n}{k}\PY{p}{)}\PY{p}{)} 
        \PY{n}{workspace} \PY{o}{=} \PY{n}{pt}\PY{o}{.}\PY{n}{reshape}\PY{p}{(}\PY{n}{workspace}\PY{p}{,}\PY{p}{(}\PY{o}{\PYZhy{}}\PY{l+m+mi}{1}\PY{p}{,}\PY{l+m+mi}{2}\PY{p}{,}\PY{l+m+mi}{2}\PY{o}{*}\PY{o}{*}\PY{p}{(}\PY{n}{k}\PY{o}{\PYZhy{}}\PY{l+m+mi}{1}\PY{p}{)}\PY{p}{)}\PY{p}{)} \PY{c+c1}{\PYZsh{}! }
        \PY{n}{workspace} \PY{o}{=} \PY{n}{pt}\PY{o}{.}\PY{n}{swapaxes}\PY{p}{(}\PY{n}{workspace}\PY{p}{,}\PY{o}{\PYZhy{}}\PY{l+m+mi}{2}\PY{p}{,}\PY{o}{\PYZhy{}}\PY{l+m+mi}{1}\PY{p}{)}          \PY{c+c1}{\PYZsh{}! }

\PY{k}{def} \PY{n+nf}{applyGate}\PY{p}{(}\PY{n}{gate}\PY{p}{,}\PY{o}{*}\PY{n}{names}\PY{p}{)}\PY{p}{:}
    \PY{k}{global} \PY{n}{workspace}
    \PY{k}{if} \PY{n+nb}{list}\PY{p}{(}\PY{n}{names}\PY{p}{)} \PY{o}{!=} \PY{n}{namestack}\PY{p}{[}\PY{o}{\PYZhy{}}\PY{n+nb}{len}\PY{p}{(}\PY{n}{names}\PY{p}{)}\PY{p}{:}\PY{p}{]}\PY{p}{:} \PY{c+c1}{\PYZsh{} reorder stack}
        \PY{k}{for} \PY{n}{name} \PY{o+ow}{in} \PY{n}{names}\PY{p}{:}                     \PY{c+c1}{\PYZsh{} if necessary}
            \PY{n}{tosQubit}\PY{p}{(}\PY{n}{name}\PY{p}{)}    
    \PY{n}{workspace} \PY{o}{=} \PY{n}{pt}\PY{o}{.}\PY{n}{reshape}\PY{p}{(}\PY{n}{workspace}\PY{p}{,}\PY{p}{(}\PY{o}{\PYZhy{}}\PY{l+m+mi}{1}\PY{p}{,}\PY{l+m+mi}{2}\PY{o}{*}\PY{o}{*}\PY{n+nb}{len}\PY{p}{(}\PY{n}{names}\PY{p}{)}\PY{p}{)}\PY{p}{)} \PY{c+c1}{\PYZsh{}! }
    \PY{n}{subworkspace} \PY{o}{=} \PY{n}{workspace}\PY{p}{[}\PY{p}{:}\PY{p}{,}\PY{o}{\PYZhy{}}\PY{n}{gate}\PY{o}{.}\PY{n}{shape}\PY{p}{[}\PY{l+m+mi}{0}\PY{p}{]}\PY{p}{:}\PY{p}{]}  
    \PY{n}{gate} \PY{o}{=} \PY{n}{pt}\PY{o}{.}\PY{n}{tensor}\PY{p}{(}\PY{n}{gate}\PY{o}{.}\PY{n}{T}\PY{p}{,}\PY{n}{device}\PY{o}{=}\PY{n}{workspace}\PY{o}{.}\PY{n}{device}\PY{p}{,}     \PY{c+c1}{\PYZsh{}! }
                     \PY{n}{dtype}\PY{o}{=}\PY{n}{workspace}\PY{p}{[}\PY{l+m+mi}{0}\PY{p}{,}\PY{l+m+mi}{0}\PY{p}{]}\PY{o}{.}\PY{n}{dtype}\PY{p}{)}         \PY{c+c1}{\PYZsh{}! }
    \PY{k}{if} \PY{n}{workspace}\PY{o}{.}\PY{n}{device}\PY{o}{.}\PY{n}{type} \PY{o}{==} \PY{l+s+s1}{\PYZsq{}}\PY{l+s+s1}{cuda}\PY{l+s+s1}{\PYZsq{}}\PY{p}{:}                  \PY{c+c1}{\PYZsh{}!}
        \PY{n}{pt}\PY{o}{.}\PY{n}{matmul}\PY{p}{(}\PY{n}{subworkspace}\PY{p}{,}\PY{n}{gate}\PY{p}{,}\PY{n}{out}\PY{o}{=}\PY{n}{subworkspace}\PY{p}{)}    \PY{c+c1}{\PYZsh{}! }
    \PY{k}{else}\PY{p}{:} \PY{c+c1}{\PYZsh{}! workaround for issue \PYZsh{}114350 in torch.matmul}
        \PY{n}{subworkspace}\PY{p}{[}\PY{p}{:}\PY{p}{,}\PY{p}{:}\PY{p}{]}\PY{o}{=}\PY{n}{pt}\PY{o}{.}\PY{n}{matmul}\PY{p}{(}\PY{n}{subworkspace}\PY{p}{,}\PY{n}{gate}\PY{p}{)}   \PY{c+c1}{\PYZsh{}!}

\PY{k}{def} \PY{n+nf}{probQubit}\PY{p}{(}\PY{n}{name}\PY{p}{)}\PY{p}{:}             \PY{c+c1}{\PYZsh{} Check probabilities}
    \PY{k}{global} \PY{n}{workspace}             \PY{c+c1}{\PYZsh{}   of qubit being 0 or 1}
    \PY{n}{tosQubit}\PY{p}{(}\PY{n}{name}\PY{p}{)}               \PY{c+c1}{\PYZsh{} qubit to TOS}
    \PY{n}{workspace} \PY{o}{=} \PY{n}{pt}\PY{o}{.}\PY{n}{reshape}\PY{p}{(}\PY{n}{workspace}\PY{p}{,}\PY{p}{(}\PY{o}{\PYZhy{}}\PY{l+m+mi}{1}\PY{p}{,}\PY{l+m+mi}{2}\PY{p}{)}\PY{p}{)} \PY{c+c1}{\PYZsh{}! to 2 cols}
    \PY{n}{prob} \PY{o}{=} \PY{n}{pt}\PY{o}{.}\PY{n}{linalg}\PY{o}{.}\PY{n}{norm}\PY{p}{(}\PY{n}{workspace}\PY{p}{,}\PY{n}{axis}\PY{o}{=}\PY{l+m+mi}{0}\PY{p}{)}\PY{o}{*}\PY{o}{*}\PY{l+m+mi}{2} \PY{c+c1}{\PYZsh{}! compute prob}
    \PY{n}{prob} \PY{o}{=} \PY{n}{pt}\PY{o}{.}\PY{n}{Tensor}\PY{o}{.}\PY{n}{cpu}\PY{p}{(}\PY{n}{prob}\PY{p}{)}\PY{o}{.}\PY{n}{numpy}\PY{p}{(}\PY{p}{)} \PY{c+c1}{\PYZsh{}! convert to numpy}
    \PY{k}{return} \PY{n}{prob}\PY{o}{/}\PY{n}{prob}\PY{o}{.}\PY{n}{sum}\PY{p}{(}\PY{p}{)}       \PY{c+c1}{\PYZsh{} make sure sum is one}

\PY{k}{def} \PY{n+nf}{measureQubit}\PY{p}{(}\PY{n}{name}\PY{p}{)}\PY{p}{:}          \PY{c+c1}{\PYZsh{} Measure and pop qubit}
    \PY{k}{global} \PY{n}{workspace}
    \PY{k}{global} \PY{n}{namestack}
    \PY{n}{prob} \PY{o}{=} \PY{n}{probQubit}\PY{p}{(}\PY{n}{name}\PY{p}{)}        \PY{c+c1}{\PYZsh{} compute probabilities}
    \PY{n}{measurement} \PY{o}{=} \PY{n}{np}\PY{o}{.}\PY{n}{random}\PY{o}{.}\PY{n}{choice}\PY{p}{(}\PY{l+m+mi}{2}\PY{p}{,}\PY{n}{p}\PY{o}{=}\PY{n}{prob}\PY{p}{)} \PY{c+c1}{\PYZsh{} 0 or 1}
    \PY{n}{workspace} \PY{o}{=} {(}\PY{n}{workspace}\PY{p}{[}\PY{p}{:}\PY{p}{,}\PY{p}{[}\PY{n}{measurement}\PY{p}{]}\PY{p}{]}\PY{o}{/} \PY{c+c1}{\PYZsh{} extract col}
                 \PY{n}{np}\PY{o}{.}\PY{n}{sqrt}\PY{p}{(}\PY{n}{prob}\PY{p}{[}\PY{n}{measurement}\PY{p}{]}\PY{p}{)}{)}
    \PY{n}{namestack}\PY{o}{.}\PY{n}{pop}\PY{p}{(}\PY{p}{)}               \PY{c+c1}{\PYZsh{} pop stacks}
    \PY{k}{return} \PY{n}{measurement}
\end{Verbatim}
\end{tcolorbox}

    Now, we can test run Grover's search with PyTorch on both the GPU and
the CPU:

    \begin{tcolorbox}[breakable, size=fbox, boxrule=1pt, pad at break*=1mm,colback=cellbackground, colframe=cellborder]
\prompt{In}{incolor}{43}{\boxspacing}
\begin{Verbatim}[commandchars=\\\{\}]
\PY{k+kn}{import} \PY{n+nn}{time}
\PY{n}{workspace} \PY{o}{=} \PY{n}{pt}\PY{o}{.}\PY{n}{tensor}\PY{p}{(}\PY{p}{[}\PY{p}{[}\PY{l+m+mf}{1.}\PY{p}{]}\PY{p}{]}\PY{p}{,}\PY{n}{device}\PY{o}{=}\PY{n}{pt}\PY{o}{.}\PY{n}{device}\PY{p}{(}\PY{l+s+s1}{\PYZsq{}}\PY{l+s+s1}{cuda}\PY{l+s+s1}{\PYZsq{}}\PY{p}{)}\PY{p}{,}
                             \PY{n}{dtype}\PY{o}{=}\PY{n}{pt}\PY{o}{.}\PY{n}{float32}\PY{p}{)}
\PY{n}{t} \PY{o}{=} \PY{n}{time}\PY{o}{.}\PY{n}{process\PYZus{}time}\PY{p}{(}\PY{p}{)}             \PY{c+c1}{\PYZsh{} with GPU}
\PY{n}{groverSearch}\PY{p}{(}\PY{l+m+mi}{16}\PY{p}{,} \PY{n}{printProb}\PY{o}{=}\PY{k+kc}{False}\PY{p}{)}   \PY{c+c1}{\PYZsh{} skip prob printouts}
\PY{n+nb}{print}\PY{p}{(}\PY{l+s+s2}{\PYZdq{}}\PY{l+s+se}{\PYZbs{}n}\PY{l+s+s2}{With GPU:}\PY{l+s+s2}{\PYZdq{}}\PY{p}{,} \PY{n}{time}\PY{o}{.}\PY{n}{process\PYZus{}time}\PY{p}{(}\PY{p}{)} \PY{o}{\PYZhy{}} \PY{n}{t}\PY{p}{,} \PY{l+s+s2}{\PYZdq{}}\PY{l+s+s2}{s}\PY{l+s+s2}{\PYZdq{}}\PY{p}{)}

\PY{n}{workspace} \PY{o}{=} \PY{n}{pt}\PY{o}{.}\PY{n}{tensor}\PY{p}{(}\PY{p}{[}\PY{p}{[}\PY{l+m+mf}{1.}\PY{p}{]}\PY{p}{]}\PY{p}{,}\PY{n}{device}\PY{o}{=}\PY{n}{pt}\PY{o}{.}\PY{n}{device}\PY{p}{(}\PY{l+s+s1}{\PYZsq{}}\PY{l+s+s1}{cpu}\PY{l+s+s1}{\PYZsq{}}\PY{p}{)}\PY{p}{,}
                             \PY{n}{dtype}\PY{o}{=}\PY{n}{pt}\PY{o}{.}\PY{n}{float32}\PY{p}{)}
\PY{n}{t} \PY{o}{=} \PY{n}{time}\PY{o}{.}\PY{n}{process\PYZus{}time}\PY{p}{(}\PY{p}{)}             \PY{c+c1}{\PYZsh{} with CPU}
\PY{n}{groverSearch}\PY{p}{(}\PY{l+m+mi}{16}\PY{p}{,} \PY{n}{printProb}\PY{o}{=}\PY{k+kc}{False}\PY{p}{)}   \PY{c+c1}{\PYZsh{} skip prob printouts}
\PY{n+nb}{print}\PY{p}{(}\PY{l+s+s2}{\PYZdq{}}\PY{l+s+se}{\PYZbs{}n}\PY{l+s+s2}{With CPU (single\PYZhy{}core):}\PY{l+s+s2}{\PYZdq{}}\PY{p}{,} \PY{n}{time}\PY{o}{.}\PY{n}{process\PYZus{}time}\PY{p}{(}\PY{p}{)} \PY{o}{\PYZhy{}} \PY{n}{t}\PY{p}{,} \PY{l+s+s2}{\PYZdq{}}\PY{l+s+s2}{s}\PY{l+s+s2}{\PYZdq{}}\PY{p}{)}
\end{Verbatim}
\end{tcolorbox}

    \begin{Verbatim}[commandchars=\\\{\}]
1111111111111101
With GPU: 4.09375 s
1111111111111101
With CPU (single-core): 44.90625 s
    \end{Verbatim}

    With 25 qubits and using the float32 data type, the search takes 1200
seconds on my computer with an 8GB RTX2070 graphics card.

It doesn't seem unlikely that the interface to a future QPU (`Quantum
Processing Unit') will resemble the interface to a GPU, perhaps by
selecting
\texttt{device=pt.device(\textquotesingle{}qpu\textquotesingle{})}.

\section{Conclusions}\label{conclusions}

Alright, so you've got yourself a quantum computer, kind of. What's
next? There's a bunch of
stuff you can do to make it better. Take the implementation, for
instance. Efficiency can be improved, especially in
reducing how often you shuffle things around on the stack when
it's already almost set up for the next gate application. Imagine a
debugger that shows you the odds of your qubits as you step through
the code - wouldn't that be neat?

Then, think about implementing an adder module using as few ancillary
qubits as possible. What about multiplication? Or raising numbers to a
power? Shor's quantum algorithm for prime factorization needs modular
exponentiation. Give that a whirl, or dive into
other famous quantum algorithms. Like the Harrow–Hassidim–Loyd (HHL)
algorithm solving linear equations, or the Quantum Approximate
Optimization Algorithm (QAOA) for tackling combinatorial optimization
problems. And hey, wouldn't it be awesome to have a compiler that
takes high-level concepts and compiles them into gate operations? The
sky's the limit!

As it turns out, a quantum computer is not really that complicated, but
extracting results from the computations can be difficult due to all the
physical restrictions. It may require quite clever algorithms, and it
seems likely that in the future, general quantum algorithms from
libraries will be used rather than developing a new quantum program for
each domain-specific problem.

Quantum computers are currently programmed at the gate level, which can be painful. Probably, some specification or
programming language at higher levels of abstraction will be
developed for quantum computers, perhaps a language similar to VHDL.

Using a quantum computer is somewhat reminiscent of the situation with
GPUs when they were new. One day, quantum computers may be used similarly.

It will take a while before we have practically useful quantum computers, as there are many technical
hurdles. I'm not holding my breath while waiting, but development
is rapid, and breakthroughs can happen anytime. Meanwhile, it's well
worth getting into the technology and development of tools, for
instance, by programming your own quantum computer simulator. I
recommend the very well-written, concise, and freely available references
{[}2-3{]}.

\section{Acknowledgments}\label{acknowledgments}

This article was inspired by a very stimulating and well-organized
`Quantum Autumn School' held in October 2023 by the
\href{https://enccs.se/}{EuroCC National Competence Centre Sweden
(ENCCS)}, \href{https://www.chalmers.se/en/centres/wacqt/}{Wallenberg
Centre of Quantum Technologies (WACQT)}, and
\href{https://nordiquest.net/}{Nordic-Estonian Quantum Computing
e-Infrastructure Quest (NordIQuEst)}. I extend a big thank you to
Dr.~Lars Rasmusson at RISE for many entertaining discussions about
quantum computers and quantum computing, and to Dr.~Sverker Janson at RISE for support.

\section{References}\label{references}

{[}1{]} \emph{List of QC simulators}. Quantum Information Portal and
Wiki. \url{https://quantiki.org/wiki/list-qc-simulators} (accessed 2023-11-20)

\noindent{[}2{]} Aaronson, S.: \emph{Introduction to Quantum Information
Science.} Lecture notes. 2018. \url{https://www.scottaaronson.com/qclec.pdf}
(accessed 2023-11-19)

\noindent{[}3{]} \emph{Azure Quantum documentation.} Microsoft, Inc.
\url{https://learn.microsoft.com/en-us/azure/quantum/} (accessed 2023-11-19)

\section*{Appendix A: A Complete
Implementation}\label{appendix-a-a-complete-implementation}

Below is the complete code for a full implementation. The simulator
consists of 40 lines of code and Grover's search of 23 lines.

    \begin{tcolorbox}[breakable, size=fbox, boxrule=1pt, pad at break*=1mm,colback=cellbackground, colframe=cellborder]
\prompt{In}{incolor}{44}{\boxspacing}
\begin{Verbatim}[commandchars=\\\{\}]
\PY{c+c1}{\PYZsh{} DIY quantum computer simulator}
\PY{c+c1}{\PYZsh{} Martin Nilsson, RISE, 2023\PYZhy{}11\PYZhy{}26}

\PY{k+kn}{import} \PY{n+nn}{numpy} \PY{k}{as} \PY{n+nn}{np}

\PY{k}{def} \PY{n+nf}{pushQubit}\PY{p}{(}\PY{n}{name}\PY{p}{,}\PY{n}{weights}\PY{p}{)}\PY{p}{:}
    \PY{k}{global} \PY{n}{workspace}
    \PY{k}{global} \PY{n}{namestack}
    \PY{k}{if} \PY{n}{workspace}\PY{o}{.}\PY{n}{shape} \PY{o}{==} \PY{p}{(}\PY{l+m+mi}{1}\PY{p}{,}\PY{l+m+mi}{1}\PY{p}{)}\PY{p}{:} \PY{n}{namestack} \PY{o}{=} \PY{p}{[}\PY{p}{]}
    \PY{n}{namestack}\PY{o}{.}\PY{n}{append}\PY{p}{(}\PY{n}{name}\PY{p}{)}
    \PY{n}{weights} \PY{o}{=} \PY{n}{np}\PY{o}{.}\PY{n}{array}\PY{p}{(}\PY{n}{weights}\PY{o}{/}\PY{n}{np}\PY{o}{.}\PY{n}{linalg}\PY{o}{.}\PY{n}{norm}\PY{p}{(}\PY{n}{weights}\PY{p}{)}\PY{p}{,}
                       \PY{n}{dtype}\PY{o}{=}\PY{n}{workspace}\PY{p}{[}\PY{l+m+mi}{0}\PY{p}{,}\PY{l+m+mi}{0}\PY{p}{]}\PY{o}{.}\PY{n}{dtype}\PY{p}{)}
    \PY{n}{workspace} \PY{o}{=} \PY{n}{np}\PY{o}{.}\PY{n}{kron}\PY{p}{(}\PY{n}{np}\PY{o}{.}\PY{n}{reshape}\PY{p}{(}\PY{n}{workspace}\PY{p}{,}\PY{p}{(}\PY{l+m+mi}{1}\PY{p}{,}\PY{o}{\PYZhy{}}\PY{l+m+mi}{1}\PY{p}{)}\PY{p}{)}\PY{p}{,}\PY{n}{weights}\PY{p}{)}
    
\PY{k}{def} \PY{n+nf}{tosQubit}\PY{p}{(}\PY{n}{name}\PY{p}{)}\PY{p}{:}
    \PY{k}{global} \PY{n}{workspace}
    \PY{k}{global} \PY{n}{namestack}
    \PY{n}{k} \PY{o}{=} \PY{n+nb}{len}\PY{p}{(}\PY{n}{namestack}\PY{p}{)}\PY{o}{\PYZhy{}}\PY{n}{namestack}\PY{o}{.}\PY{n}{index}\PY{p}{(}\PY{n}{name}\PY{p}{)}
    \PY{k}{if} \PY{n}{k} \PY{o}{\PYZgt{}} \PY{l+m+mi}{1}\PY{p}{:}
        \PY{n}{namestack}\PY{o}{.}\PY{n}{append}\PY{p}{(}\PY{n}{namestack}\PY{o}{.}\PY{n}{pop}\PY{p}{(}\PY{o}{\PYZhy{}}\PY{n}{k}\PY{p}{)}\PY{p}{)}
        \PY{n}{workspace} \PY{o}{=} \PY{n}{np}\PY{o}{.}\PY{n}{reshape}\PY{p}{(}\PY{n}{workspace}\PY{p}{,}\PY{p}{(}\PY{o}{\PYZhy{}}\PY{l+m+mi}{1}\PY{p}{,}\PY{l+m+mi}{2}\PY{p}{,}\PY{l+m+mi}{2}\PY{o}{*}\PY{o}{*}\PY{p}{(}\PY{n}{k}\PY{o}{\PYZhy{}}\PY{l+m+mi}{1}\PY{p}{)}\PY{p}{)}\PY{p}{)}
        \PY{n}{workspace} \PY{o}{=} \PY{n}{np}\PY{o}{.}\PY{n}{swapaxes}\PY{p}{(}\PY{n}{workspace}\PY{p}{,}\PY{o}{\PYZhy{}}\PY{l+m+mi}{2}\PY{p}{,}\PY{o}{\PYZhy{}}\PY{l+m+mi}{1}\PY{p}{)}
        
\PY{k}{def} \PY{n+nf}{applyGate}\PY{p}{(}\PY{n}{gate}\PY{p}{,}\PY{o}{*}\PY{n}{names}\PY{p}{)}\PY{p}{:}
    \PY{k}{global} \PY{n}{workspace}
    \PY{k}{if} \PY{n+nb}{list}\PY{p}{(}\PY{n}{names}\PY{p}{)} \PY{o}{!=} \PY{n}{namestack}\PY{p}{[}\PY{o}{\PYZhy{}}\PY{n+nb}{len}\PY{p}{(}\PY{n}{names}\PY{p}{)}\PY{p}{:}\PY{p}{]}\PY{p}{:}                            
        \PY{p}{[}\PY{n}{tosQubit}\PY{p}{(}\PY{n}{name}\PY{p}{)} \PY{k}{for} \PY{n}{name} \PY{o+ow}{in} \PY{n}{names}\PY{p}{]}
    \PY{n}{workspace} \PY{o}{=} \PY{n}{np}\PY{o}{.}\PY{n}{reshape}\PY{p}{(}\PY{n}{workspace}\PY{p}{,}\PY{p}{(}\PY{o}{\PYZhy{}}\PY{l+m+mi}{1}\PY{p}{,}\PY{l+m+mi}{2}\PY{o}{*}\PY{o}{*}\PY{p}{(}\PY{n+nb}{len}\PY{p}{(}\PY{n}{names}\PY{p}{)}\PY{p}{)}\PY{p}{)}\PY{p}{)}
    \PY{n}{subworkspace} \PY{o}{=} \PY{n}{workspace}\PY{p}{[}\PY{p}{:}\PY{p}{,}\PY{o}{\PYZhy{}}\PY{n}{gate}\PY{o}{.}\PY{n}{shape}\PY{p}{[}\PY{l+m+mi}{0}\PY{p}{]}\PY{p}{:}\PY{p}{]}
    \PY{n}{np}\PY{o}{.}\PY{n}{matmul}\PY{p}{(}\PY{n}{subworkspace}\PY{p}{,}\PY{n}{gate}\PY{o}{.}\PY{n}{T}\PY{p}{,}\PY{n}{out}\PY{o}{=}\PY{n}{subworkspace}\PY{p}{)}
        
\PY{k}{def} \PY{n+nf}{probQubit}\PY{p}{(}\PY{n}{name}\PY{p}{)}\PY{p}{:}
    \PY{k}{global} \PY{n}{workspace}
    \PY{n}{tosQubit}\PY{p}{(}\PY{n}{name}\PY{p}{)}
    \PY{n}{workspace} \PY{o}{=} \PY{n}{np}\PY{o}{.}\PY{n}{reshape}\PY{p}{(}\PY{n}{workspace}\PY{p}{,}\PY{p}{(}\PY{o}{\PYZhy{}}\PY{l+m+mi}{1}\PY{p}{,}\PY{l+m+mi}{2}\PY{p}{)}\PY{p}{)}
    \PY{n}{prob} \PY{o}{=} \PY{n}{np}\PY{o}{.}\PY{n}{linalg}\PY{o}{.}\PY{n}{norm}\PY{p}{(}\PY{n}{workspace}\PY{p}{,}\PY{n}{axis}\PY{o}{=}\PY{l+m+mi}{0}\PY{p}{)}\PY{o}{*}\PY{o}{*}\PY{l+m+mi}{2}
    \PY{k}{return} \PY{n}{prob}\PY{o}{/}\PY{n}{prob}\PY{o}{.}\PY{n}{sum}\PY{p}{(}\PY{p}{)}

\PY{k}{def} \PY{n+nf}{measureQubit}\PY{p}{(}\PY{n}{name}\PY{p}{)}\PY{p}{:}
    \PY{k}{global} \PY{n}{workspace}
    \PY{k}{global} \PY{n}{namestack}
    \PY{n}{prob} \PY{o}{=} \PY{n}{probQubit}\PY{p}{(}\PY{n}{name}\PY{p}{)}
    \PY{n}{measurement} \PY{o}{=} \PY{n}{np}\PY{o}{.}\PY{n}{random}\PY{o}{.}\PY{n}{choice}\PY{p}{(}\PY{l+m+mi}{2}\PY{p}{,}\PY{n}{p}\PY{o}{=}\PY{n}{prob}\PY{p}{)}
    \PY{n}{workspace} \PY{o}{=} {(}\PY{n}{workspace}\PY{p}{[}\PY{p}{:}\PY{p}{,}\PY{p}{[}\PY{n}{measurement}\PY{p}{]}\PY{p}{]}\PY{o}{/}
                 \PY{n}{np}\PY{o}{.}\PY{n}{sqrt}\PY{p}{(}\PY{n}{prob}\PY{p}{[}\PY{n}{measurement}\PY{p}{]}\PY{p}{)}{)}
    \PY{n}{namestack}\PY{o}{.}\PY{n}{pop}\PY{p}{(}\PY{p}{)}
    \PY{k}{return} \PY{n+nb}{str}\PY{p}{(}\PY{n}{measurement}\PY{p}{)}

\PY{c+c1}{\PYZsh{} \PYZhy{}\PYZhy{}\PYZhy{}\PYZhy{}\PYZhy{}\PYZhy{}\PYZhy{}\PYZhy{}\PYZhy{}\PYZhy{} Grover search example}

\PY{n}{X\PYZus{}gate} \PY{o}{=} \PY{n}{np}\PY{o}{.}\PY{n}{array}\PY{p}{(}\PY{p}{[}\PY{p}{[}\PY{l+m+mi}{0}\PY{p}{,} \PY{l+m+mi}{1}\PY{p}{]}\PY{p}{,}\PY{p}{[}\PY{l+m+mi}{1}\PY{p}{,} \PY{l+m+mi}{0}\PY{p}{]}\PY{p}{]}\PY{p}{)}
\PY{n}{H\PYZus{}gate} \PY{o}{=} \PY{n}{np}\PY{o}{.}\PY{n}{array}\PY{p}{(}\PY{p}{[}\PY{p}{[}\PY{l+m+mi}{1}\PY{p}{,} \PY{l+m+mi}{1}\PY{p}{]}\PY{p}{,}\PY{p}{[}\PY{l+m+mi}{1}\PY{p}{,}\PY{o}{\PYZhy{}}\PY{l+m+mi}{1}\PY{p}{]}\PY{p}{]}\PY{p}{)}\PY{o}{*}\PY{n}{np}\PY{o}{.}\PY{n}{sqrt}\PY{p}{(}\PY{l+m+mi}{1}\PY{o}{/}\PY{l+m+mi}{2}\PY{p}{)}
\PY{n}{Z\PYZus{}gate} \PY{o}{=} \PY{n}{H\PYZus{}gate} \PY{o}{@} \PY{n}{X\PYZus{}gate} \PY{o}{@} \PY{n}{H\PYZus{}gate}

\PY{k}{def} \PY{n+nf}{sample\PYZus{}phaseOracle}\PY{p}{(}\PY{n}{qubits}\PY{p}{)}\PY{p}{:}
    \PY{n}{applyGate}\PY{p}{(}\PY{n}{X\PYZus{}gate}\PY{p}{,}\PY{n}{qubits}\PY{p}{[}\PY{l+m+mi}{1}\PY{p}{]}\PY{p}{)}
    \PY{n}{applyGate}\PY{p}{(}\PY{n}{Z\PYZus{}gate}\PY{p}{,}\PY{o}{*}\PY{n}{namestack}\PY{p}{)}
    \PY{n}{applyGate}\PY{p}{(}\PY{n}{X\PYZus{}gate}\PY{p}{,}\PY{n}{qubits}\PY{p}{[}\PY{l+m+mi}{1}\PY{p}{]}\PY{p}{)}
    
\PY{k}{def} \PY{n+nf}{zero\PYZus{}phaseOracle}\PY{p}{(}\PY{n}{qubits}\PY{p}{)}\PY{p}{:}
    \PY{p}{[}\PY{n}{applyGate}\PY{p}{(}\PY{n}{X\PYZus{}gate}\PY{p}{,}\PY{n}{q}\PY{p}{)} \PY{k}{for} \PY{n}{q} \PY{o+ow}{in} \PY{n}{qubits}\PY{p}{]}
    \PY{n}{applyGate}\PY{p}{(}\PY{n}{Z\PYZus{}gate}\PY{p}{,}\PY{o}{*}\PY{n}{namestack}\PY{p}{)}
    \PY{p}{[}\PY{n}{applyGate}\PY{p}{(}\PY{n}{X\PYZus{}gate}\PY{p}{,}\PY{n}{q}\PY{p}{)} \PY{k}{for} \PY{n}{q} \PY{o+ow}{in} \PY{n}{qubits}\PY{p}{]}
        
\PY{k}{def} \PY{n+nf}{groverSearch}\PY{p}{(}\PY{n}{n}\PY{p}{,} \PY{n}{printProb}\PY{o}{=}\PY{k+kc}{True}\PY{p}{)}\PY{p}{:}
    \PY{n}{qubits} \PY{o}{=} \PY{n+nb}{list}\PY{p}{(}\PY{n+nb}{range}\PY{p}{(}\PY{n}{n}\PY{p}{)}\PY{p}{)}
    \PY{p}{[}\PY{n}{pushQubit}\PY{p}{(}\PY{n}{q}\PY{p}{,}\PY{p}{[}\PY{l+m+mi}{1}\PY{p}{,}\PY{l+m+mi}{1}\PY{p}{]}\PY{p}{)} \PY{k}{for} \PY{n}{q} \PY{o+ow}{in} \PY{n}{qubits}\PY{p}{]}
    \PY{k}{for} \PY{n}{k} \PY{o+ow}{in} \PY{n+nb}{range}\PY{p}{(}\PY{n+nb}{int}\PY{p}{(}\PY{n}{np}\PY{o}{.}\PY{n}{pi}\PY{o}{/}\PY{l+m+mi}{4}\PY{o}{*}\PY{n}{np}\PY{o}{.}\PY{n}{sqrt}\PY{p}{(}\PY{l+m+mi}{2}\PY{o}{*}\PY{o}{*}\PY{n}{n}\PY{p}{)}\PY{o}{\PYZhy{}}\PY{l+m+mi}{1}\PY{o}{/}\PY{l+m+mi}{2}\PY{p}{)}\PY{p}{)}\PY{p}{:}
        \PY{n}{sample\PYZus{}phaseOracle}\PY{p}{(}\PY{n}{qubits}\PY{p}{)}
        \PY{p}{[}\PY{n}{applyGate}\PY{p}{(}\PY{n}{H\PYZus{}gate}\PY{p}{,}\PY{n}{q}\PY{p}{)} \PY{k}{for} \PY{n}{q} \PY{o+ow}{in} \PY{n}{qubits}\PY{p}{]}
        \PY{n}{zero\PYZus{}phaseOracle}\PY{p}{(}\PY{n}{qubits}\PY{p}{)}
        \PY{p}{[}\PY{n}{applyGate}\PY{p}{(}\PY{n}{H\PYZus{}gate}\PY{p}{,}\PY{n}{q}\PY{p}{)} \PY{k}{for} \PY{n}{q} \PY{o+ow}{in} \PY{n}{qubits}\PY{p}{]}
        \PY{k}{if} \PY{n}{printProb}\PY{p}{:} \PY{n+nb}{print}\PY{p}{(}\PY{n}{probQubit}\PY{p}{(}\PY{n}{qubits}\PY{p}{[}\PY{l+m+mi}{0}\PY{p}{]}\PY{p}{)}\PY{p}{)}                    
    \PY{p}{[}\PY{n+nb}{print}\PY{p}{(}\PY{n}{measureQubit}\PY{p}{(}\PY{n}{q}\PY{p}{)}\PY{p}{,}\PY{n}{end}\PY{o}{=}\PY{l+s+s2}{\PYZdq{}}\PY{l+s+s2}{\PYZdq{}}\PY{p}{)} \PY{k}{for} \PY{n}{q} \PY{o+ow}{in} \PY{n+nb}{reversed}\PY{p}{(}\PY{n}{qubits}\PY{p}{)}\PY{p}{]}
        
\PY{n}{workspace} \PY{o}{=} \PY{n}{np}\PY{o}{.}\PY{n}{array}\PY{p}{(}\PY{p}{[}\PY{p}{[}\PY{l+m+mf}{1.}\PY{p}{]}\PY{p}{]}\PY{p}{)}
\PY{n}{groverSearch}\PY{p}{(}\PY{l+m+mi}{8}\PY{p}{)}
\end{Verbatim}
\end{tcolorbox}

    \begin{Verbatim}[commandchars=\\\{\}]
[0.48449707 0.51550293]
[0.45445636 0.54554364]
[0.41174808 0.58825192]
[0.35903106 0.64096894]
[0.29958726 0.70041274]
[0.2371174 0.7628826]
[0.17551059 0.82448941]
[0.11860222 0.88139778]
[0.06993516 0.93006484]
[0.03253923 0.96746077]
[0.00874254 0.99125746]
[2.65827874e-05 9.99973417e-01]
11111101
    \end{Verbatim}

    \section*{Appendix B: A Minimalist Quantum
Computer}\label{appendix-b-a-minimalist-quantum-computer}

A complete implementation in just twelve lines, just to show it's
possible! Grover's search here is 30 lines.

    \begin{tcolorbox}[breakable, size=fbox, boxrule=1pt, pad at break*=1mm,colback=cellbackground, colframe=cellborder]
\prompt{In}{incolor}{45}{\boxspacing}
\begin{Verbatim}[commandchars=\\\{\}]
\PY{c+c1}{\PYZsh{} Minimalist quantum computer}
\PY{c+c1}{\PYZsh{} Martin Nilsson, RISE, 2023\PYZhy{}11\PYZhy{}24}

\PY{k+kn}{import} \PY{n+nn}{numpy} \PY{k}{as} \PY{n+nn}{np}

\PY{k}{def} \PY{n+nf}{pushQubit}\PY{p}{(}\PY{n}{q}\PY{p}{,}\PY{n}{w}\PY{p}{)}\PY{p}{:}
    \PY{k}{return} \PY{n}{np}\PY{o}{.}\PY{n}{kron}\PY{p}{(}\PY{n}{np}\PY{o}{.}\PY{n}{reshape}\PY{p}{(}\PY{n}{w}\PY{p}{,}\PY{p}{(}\PY{l+m+mi}{1}\PY{p}{,}\PY{o}{\PYZhy{}}\PY{l+m+mi}{1}\PY{p}{)}\PY{p}{)}\PY{p}{,}\PY{n}{q}\PY{p}{)}

\PY{k}{def} \PY{n+nf}{applyGate}\PY{p}{(}\PY{n}{g}\PY{p}{,}\PY{n}{w}\PY{p}{)}\PY{p}{:}
    \PY{k}{return} \PY{n}{np}\PY{o}{.}\PY{n}{matmul}\PY{p}{(}\PY{n}{np}\PY{o}{.}\PY{n}{reshape}\PY{p}{(}\PY{n}{w}\PY{p}{,}\PY{p}{(}\PY{o}{\PYZhy{}}\PY{l+m+mi}{1}\PY{p}{,}\PY{n}{g}\PY{o}{.}\PY{n}{shape}\PY{p}{[}\PY{l+m+mi}{0}\PY{p}{]}\PY{p}{)}\PY{p}{)}\PY{p}{,}\PY{n}{g}\PY{o}{.}\PY{n}{T}\PY{p}{)}

\PY{k}{def} \PY{n+nf}{tosQubit}\PY{p}{(}\PY{n}{k}\PY{p}{,}\PY{n}{w}\PY{p}{)}\PY{p}{:}
    \PY{k}{return} \PY{n}{np}\PY{o}{.}\PY{n}{swapaxes}\PY{p}{(}\PY{n}{np}\PY{o}{.}\PY{n}{reshape}\PY{p}{(}\PY{n}{w}\PY{p}{,}\PY{p}{(}\PY{o}{\PYZhy{}}\PY{l+m+mi}{1}\PY{p}{,}\PY{l+m+mi}{2}\PY{p}{,}\PY{l+m+mi}{2}\PY{o}{*}\PY{o}{*}\PY{p}{(}\PY{n}{k}\PY{o}{\PYZhy{}}\PY{l+m+mi}{1}\PY{p}{)}\PY{p}{)}\PY{p}{)}\PY{p}{,}\PY{o}{\PYZhy{}}\PY{l+m+mi}{2}\PY{p}{,}\PY{o}{\PYZhy{}}\PY{l+m+mi}{1}\PY{p}{)}

\PY{k}{def} \PY{n+nf}{measureQubit}\PY{p}{(}\PY{n}{w}\PY{p}{)}\PY{p}{:}
    \PY{n}{w} \PY{o}{=} \PY{n}{np}\PY{o}{.}\PY{n}{reshape}\PY{p}{(}\PY{n}{w}\PY{p}{,}\PY{p}{(}\PY{o}{\PYZhy{}}\PY{l+m+mi}{1}\PY{p}{,}\PY{l+m+mi}{2}\PY{p}{)}\PY{p}{)}
    \PY{n}{p} \PY{o}{=} \PY{n}{np}\PY{o}{.}\PY{n}{linalg}\PY{o}{.}\PY{n}{norm}\PY{p}{(}\PY{n}{w}\PY{p}{,}\PY{n}{axis}\PY{o}{=}\PY{l+m+mi}{0}\PY{p}{)}
    \PY{n}{m} \PY{o}{=} \PY{n}{np}\PY{o}{.}\PY{n}{random}\PY{o}{.}\PY{n}{choice}\PY{p}{(}\PY{l+m+mi}{2}\PY{p}{,}\PY{n}{p}\PY{o}{=}\PY{n}{p}\PY{o}{*}\PY{o}{*}\PY{l+m+mi}{2}\PY{p}{)}
    \PY{k}{return} \PY{p}{(}\PY{n}{m}\PY{p}{,}\PY{n}{w}\PY{p}{[}\PY{p}{:}\PY{p}{,}\PY{p}{[}\PY{n}{m}\PY{p}{]}\PY{p}{]}\PY{o}{/}\PY{n}{p}\PY{p}{[}\PY{n}{m}\PY{p}{]}\PY{p}{)}

\PY{c+c1}{\PYZsh{} \PYZhy{}\PYZhy{}\PYZhy{}\PYZhy{}\PYZhy{}\PYZhy{}\PYZhy{}\PYZhy{}\PYZhy{}\PYZhy{} Grover search example}

\PY{k}{def} \PY{n+nf}{sample\PYZus{}phaseOracle}\PY{p}{(}\PY{n}{w}\PY{p}{)}\PY{p}{:}
    \PY{n}{w} \PY{o}{=} \PY{n}{applyGate}\PY{p}{(}\PY{n}{X\PYZus{}gate}\PY{p}{,}\PY{n}{tosQubit}\PY{p}{(}\PY{l+m+mi}{2}\PY{p}{,}\PY{n}{w}\PY{p}{)}\PY{p}{)}
    \PY{n}{w} \PY{o}{=} \PY{n}{applyGate}\PY{p}{(}\PY{n}{CZn\PYZus{}gate}\PY{p}{,}\PY{n}{tosQubit}\PY{p}{(}\PY{l+m+mi}{2}\PY{p}{,}\PY{n}{w}\PY{p}{)}\PY{p}{)}
    \PY{n}{w} \PY{o}{=} \PY{n}{applyGate}\PY{p}{(}\PY{n}{X\PYZus{}gate}\PY{p}{,}\PY{n}{tosQubit}\PY{p}{(}\PY{l+m+mi}{2}\PY{p}{,}\PY{n}{w}\PY{p}{)}\PY{p}{)}
    \PY{k}{return} \PY{n}{tosQubit}\PY{p}{(}\PY{l+m+mi}{2}\PY{p}{,}\PY{n}{w}\PY{p}{)}

\PY{k}{def} \PY{n+nf}{zero\PYZus{}phaseOracle}\PY{p}{(}\PY{n}{w}\PY{p}{)}\PY{p}{:}
    \PY{k}{for} \PY{n}{i} \PY{o+ow}{in} \PY{n+nb}{range}\PY{p}{(}\PY{n}{n}\PY{p}{)}\PY{p}{:}
        \PY{n}{w} \PY{o}{=} \PY{n}{applyGate}\PY{p}{(}\PY{n}{X\PYZus{}gate}\PY{p}{,}\PY{n}{tosQubit}\PY{p}{(}\PY{n}{n}\PY{p}{,}\PY{n}{w}\PY{p}{)}\PY{p}{)}
    \PY{n}{w}  \PY{o}{=} \PY{n}{applyGate}\PY{p}{(}\PY{n}{CZn\PYZus{}gate}\PY{p}{,}\PY{n}{w}\PY{p}{)}
    \PY{k}{for} \PY{n}{i} \PY{o+ow}{in} \PY{n+nb}{range}\PY{p}{(}\PY{n}{n}\PY{p}{)}\PY{p}{:}
        \PY{n}{w} \PY{o}{=} \PY{n}{applyGate}\PY{p}{(}\PY{n}{X\PYZus{}gate}\PY{p}{,}\PY{n}{tosQubit}\PY{p}{(}\PY{n}{n}\PY{p}{,}\PY{n}{w}\PY{p}{)}\PY{p}{)}
    \PY{k}{return} \PY{n}{w}

\PY{k}{def} \PY{n+nf}{groverSearch}\PY{p}{(}\PY{n}{w}\PY{p}{)}\PY{p}{:}
    \PY{k}{for} \PY{n}{i} \PY{o+ow}{in} \PY{n+nb}{range}\PY{p}{(}\PY{n}{n}\PY{p}{)}\PY{p}{:}
        \PY{n}{w} \PY{o}{=} \PY{n}{pushQubit}\PY{p}{(}\PY{n}{H\PYZus{}gate}\PY{o}{@}\PY{p}{[}\PY{l+m+mi}{1}\PY{p}{,}\PY{l+m+mi}{0}\PY{p}{]}\PY{p}{,}\PY{n}{w}\PY{p}{)}
    \PY{k}{for} \PY{n}{k} \PY{o+ow}{in} \PY{n+nb}{range}\PY{p}{(}\PY{n+nb}{int}\PY{p}{(}\PY{n}{np}\PY{o}{.}\PY{n}{pi}\PY{o}{/}\PY{l+m+mi}{4}\PY{o}{*}\PY{n}{np}\PY{o}{.}\PY{n}{sqrt}\PY{p}{(}\PY{l+m+mi}{2}\PY{o}{*}\PY{o}{*}\PY{n}{n}\PY{p}{)}\PY{o}{\PYZhy{}}\PY{l+m+mi}{1}\PY{o}{/}\PY{l+m+mi}{2}\PY{p}{)}\PY{p}{)}\PY{p}{:}
        \PY{n}{w} \PY{o}{=} \PY{n}{sample\PYZus{}phaseOracle}\PY{p}{(}\PY{n}{w}\PY{p}{)}
        \PY{k}{for} \PY{n}{i} \PY{o+ow}{in} \PY{n+nb}{range}\PY{p}{(}\PY{n}{n}\PY{p}{)}\PY{p}{:}
            \PY{n}{w} \PY{o}{=} \PY{n}{applyGate}\PY{p}{(}\PY{n}{H\PYZus{}gate}\PY{p}{,}\PY{n}{tosQubit}\PY{p}{(}\PY{n}{n}\PY{p}{,}\PY{n}{w}\PY{p}{)}\PY{p}{)}
        \PY{n}{w} \PY{o}{=} \PY{n}{zero\PYZus{}phaseOracle}\PY{p}{(}\PY{n}{w}\PY{p}{)}
        \PY{k}{for} \PY{n}{i} \PY{o+ow}{in} \PY{n+nb}{range}\PY{p}{(}\PY{n}{n}\PY{p}{)}\PY{p}{:}
            \PY{n}{w} \PY{o}{=} \PY{n}{applyGate}\PY{p}{(}\PY{n}{H\PYZus{}gate}\PY{p}{,}\PY{n}{tosQubit}\PY{p}{(}\PY{n}{n}\PY{p}{,}\PY{n}{w}\PY{p}{)}\PY{p}{)}
    \PY{k}{for} \PY{n}{i} \PY{o+ow}{in} \PY{n+nb}{range}\PY{p}{(}\PY{n}{n}\PY{p}{)}\PY{p}{:}
        \PY{p}{(}\PY{n}{m}\PY{p}{,}\PY{n}{w}\PY{p}{)} \PY{o}{=} \PY{n}{measureQubit}\PY{p}{(}\PY{n}{tosQubit}\PY{p}{(}\PY{n}{n}\PY{o}{\PYZhy{}}\PY{n}{i}\PY{p}{,}\PY{n}{w}\PY{p}{)}\PY{p}{)}
        \PY{n+nb}{print}\PY{p}{(}\PY{n}{m}\PY{p}{,}\PY{n}{end}\PY{o}{=}\PY{l+s+s2}{\PYZdq{}}\PY{l+s+s2}{\PYZdq{}}\PY{p}{)}

\PY{n}{n} \PY{o}{=} \PY{l+m+mi}{10}
\PY{n}{X\PYZus{}gate} \PY{o}{=} \PY{n}{np}\PY{o}{.}\PY{n}{array}\PY{p}{(}\PY{p}{[}\PY{p}{[}\PY{l+m+mi}{0}\PY{p}{,} \PY{l+m+mi}{1}\PY{p}{]}\PY{p}{,}\PY{p}{[}\PY{l+m+mi}{1}\PY{p}{,} \PY{l+m+mi}{0}\PY{p}{]}\PY{p}{]}\PY{p}{)}
\PY{n}{H\PYZus{}gate} \PY{o}{=} \PY{n}{np}\PY{o}{.}\PY{n}{array}\PY{p}{(}\PY{p}{[}\PY{p}{[}\PY{l+m+mi}{1}\PY{p}{,} \PY{l+m+mi}{1}\PY{p}{]}\PY{p}{,}\PY{p}{[}\PY{l+m+mi}{1}\PY{p}{,}\PY{o}{\PYZhy{}}\PY{l+m+mi}{1}\PY{p}{]}\PY{p}{]}\PY{p}{)}\PY{o}{*}\PY{n}{np}\PY{o}{.}\PY{n}{sqrt}\PY{p}{(}\PY{l+m+mi}{1}\PY{o}{/}\PY{l+m+mi}{2}\PY{p}{)}
\PY{n}{CZn\PYZus{}gate} \PY{o}{=} \PY{n}{np}\PY{o}{.}\PY{n}{diag}\PY{p}{(}\PY{n+nb}{list}\PY{p}{(}\PY{n+nb}{reversed}\PY{p}{(}\PY{l+m+mi}{2}\PY{o}{*}\PY{n}{np}\PY{o}{.}\PY{n}{sign}\PY{p}{(}\PY{n+nb}{range}\PY{p}{(}\PY{l+m+mi}{2}\PY{o}{*}\PY{o}{*}\PY{n}{n}\PY{p}{)}\PY{p}{)}\PY{o}{\PYZhy{}}\PY{l+m+mi}{1}\PY{p}{)}\PY{p}{)}\PY{p}{)}

\PY{n}{groverSearch}\PY{p}{(}\PY{n}{np}\PY{o}{.}\PY{n}{array}\PY{p}{(}\PY{p}{[}\PY{p}{[}\PY{l+m+mf}{1.}\PY{p}{]}\PY{p}{]}\PY{p}{)}\PY{p}{)}
\end{Verbatim}
\end{tcolorbox}

    \begin{Verbatim}[commandchars=\\\{\}]
1111111101
    \end{Verbatim}


\end{document}